\documentclass[aps,prb,showpacs,preprintnumbers,twocolumn,superscriptaddress,longbibliography]{revtex4-2}
\usepackage{amsmath,amssymb}
\usepackage{tipa}
\usepackage{upgreek}
\usepackage{physics}
\usepackage{tikz}
\usepackage{comment}
\usepackage{bbm}
\usepackage{multirow}
\usepackage{mathrsfs}
\usepackage{braket}
\usepackage{enumitem}
\usepackage{mathbbol}
\usepackage{booktabs}
\usepackage[normalem]{ulem}
\usepackage{color}
\usepackage{tabularx}

\usepackage{array, makecell}
\usepackage[final,colorlinks,bookmarks=true,citecolor=blue,linkcolor=red,urlcolor=blue]{hyperref}
\usepackage{graphicx}
\usepackage{bm}
\usepackage[toc,page]{appendix}

\newcommand{\nocontentsline}[3]{}
\newcommand{\tocless}[3]{\bgroup\let\addcontentsline=\nocontentsline#1{#2\label{#3}}\egroup}

\addcontentsline{toc}{chapter}{Appendix}
\addtocontents{toc}{\protect\setcounter{tocdepth}{0}}
\addtocontents{ptc}{\protect\setcounter{tocdepth}{2}}

\renewcommand{\vec}[1]{\bm{#1}}
\usepackage{pifont}

\newcommand{\go}{g^{(1)}} 
\newcommand{\gt}{g^{(2)}} 
\newcommand{\gn}{g^{(n)}} 

\newcommand{\Go}{\underline{g}^{(1)}} 

\newcommand{\Gt}{\underline{g}^{(2)}} 

\newcommand{\Gn}{\underline{g}^{(n)}} 

\newcommand{\Gti}{[\underline{g}^{(2)}]^{-1}} 

\newcommand{\Goi}{[\underline{g}^{(1)}]^{-1}} 

\allowdisplaybreaks
\usepackage{titlesec}
\begin{document}

\titlespacing\section{0pt}{12pt plus 6pt minus 2pt}{6pt plus 2pt minus 2pt}
\titlespacing\subsection{0pt}{12pt plus 4pt minus 2pt}{4pt plus 2pt minus 2pt}
\titlespacing\subsubsection{0pt}{12pt plus 4pt minus 2pt}{4pt plus 2pt minus 2pt}

\title{Interacting topological quantum chemistry of Mott atomic limits} 
	
\author{Martina~O.\,Soldini}

\author{Nikita Astrakhantsev}
\affiliation{University of Zurich, Winterthurerstrasse 190, 8057 Zurich, Switzerland}

\author{Mikel Iraola}
\affiliation{Donostia International Physics Center, 20018 Donostia-San Sebastian, Spain}
\affiliation{Department of Physics, University of the Basque Country UPV/EHU, 48080 Bilbao, Spain}

\author{Apoorv Tiwari}
\affiliation{Department of Physics, KTH Royal Institute of Technology, Roslagstullsbacken 21, 114 21 Stockholm, Sweden}
	
\author{Mark~H.\,Fischer}
\affiliation{University of Zurich, Winterthurerstrasse 190, 8057 Zurich, Switzerland}

\author{Roser Valent\'{i}}
\affiliation{Institut für Theoretische Physik, Goethe-Universität Frankfurt, 60438 Frankfurt am Main, Germany}

\author{Maia~G.\,Vergniory}
\affiliation{Donostia International Physics Center, 20018 Donostia-San Sebastian, Spain}
\affiliation{Max Planck Institute for Chemical Physics of Solids, 01187 Dresden, Germany}

\author{Glenn Wagner}

\affiliation{University of Zurich, Winterthurerstrasse 190, 8057 Zurich, Switzerland}

\author{Titus Neupert}
\affiliation{University of Zurich, Winterthurerstrasse 190, 8057 Zurich, Switzerland}

\begin{abstract}
Topological quantum chemistry (TQC) is a successful framework for identifying (noninteracting) topological materials. Based on the symmetry eigenvalues of Bloch eigenstates at maximal momenta, which are attainable from first principles calculations, a band structure can either be classified as an \emph{atomic limit}, in other words adiabatically connected to independent electronic orbitals on the respective crystal lattice, or it is topological.
For interacting systems, there is no single-particle band structure and hence, the TQC machinery grinds to a halt. We develop a framework analogous to TQC, but employing $n$-particle Green's function to classify interacting systems. Fundamentally, we define a class of interacting reference states that generalize the notion of atomic limits, which we call \emph{Mott atomic limits}, and are symmetry protected topological states. Our formalism allows to fully classify these reference states (with $n=2$), which can themselves represent symmetry protected topological states.
We present a comprehensive classification of such states in one-dimension and provide numerical results on model systems. 
With this, we establish Mott atomic limit states as a generalization of the atomic limits to interacting systems. 
\end{abstract}

\setcounter{page}{1}
\maketitle

\tocless\section{Introduction}{sec:Intro}
 	
Topological quantum matter harbors universal and robust physical phenomena that are appealing for fundamental research as well as for applications~\cite{IQHE_PhysRevLett.45.494, KaneMele_PhysRevLett.95.226801, KaneMele_PhysRevLett.95.146802,RevModPhys.82.3045,Sato_2017_review_TopologicalSC}. The experimental identification of topological materials can be challenging, which may explain why topological insulators have only been discovered following theoretical predictions~\cite{BHZ, PhysRevLett.98.106803}, despite decades of semiconductor research~\cite{2007Sci...318..766K, 2008Natur.452..970H}. 
Since these discoveries, the notion of topological states has been considerably refined, starting from the 10-fold way classification~\cite{Schnyder2009, Ryu_2010}, via the inclusion of spatial symmetries in topological crystalline insulators~\cite{Fu_PhysRevLett.106.106802}, to the concepts of fragile~\cite{PhysRevLett.121.126402, PhysRevLett.123.186401,PhysRevB.102.115135} and delicate~\cite{DelicateTopology_Nelson} topology. These concepts have also been extended to other types of systems, beyond electronic materials, such as magnons~\cite{PhysRevLett.122.187203,magnons} and optical excitations~\cite{OpticalInsulatorPhysRevResearch.4.023011}, to name a few.
In parallel, theoretical methods have been developed to predict such topological phases in real materials. Aided by density functional theory calculations, so-called symmetry indicators~\cite{Po_2017_SimmetryIndicators,Po_2020} and the more comprehensive framework of topological quantum chemistry (TQC) has allowed to identify large numbers of candidate topological materials~\cite{Bradlyn2017, catalogueTQC, Cano, MagneticTQC, PhysRevX.7.041069}.

The principle underlying TQC is to relate \emph{atomic limits} (ALs), defined by placing electrons on (maximal) Wyckoff positions of a crystallographic space group, to representations of a set of electronic bands in momentum space~\cite{PhysRevB.26.3010, PhysRevB.59.5998,Bradlyn2017,Cano}. Independent of these considerations, one can obtain the representation of a set of bands for a material of interest from first principles band structure calculations. One says that ALs \textit{induce} band representations. The efficient identification of band representations is obtained by listing the irreducible representations (irreps) of Bloch states at maximal momentum points in the Brillouin zone. Composing multiple ALs in the real space unit cell of a crystal corresponds to adding band representations in momentum space with positive integer coefficients. The generators of this space of band representations are called elementary band representations. Importantly, the AL-generated band representations do not span the space of all the possible band structures.

The key statement of TQC is enclosed in the following: if the representation of the bands of a given material does not admit a decomposition in terms of elementary band representations (with positive integer coefficients), the material realizes a topological state. Note that the converse is not true~\cite{PhysRevB.102.115135,PhysRevB.105.094518,Cano21}:
A non-magnetic system in space group $C_1$ may for instance be a three-dimensional (3D) topological insulator protected by time-reversal symmetry (TRS), but in this low-symmetry space group there are no non-trivial irreps that could be used to differentiate between distinct band representations. 

The TQC framework is fundamentally limited in its formulation to noninteracting systems and can thus only distinguish topological and trivial systems well described in the single-particle approximation. Topological insulators are protected by their gap and therefore, weak interactions will not destroy these phases and TQC might be applicable even with interactions. However, interactions can lead to entirely new topological phases that are not adiabatically connected to any noninteracting limit~\cite{Rachel_2018,Wen2010,Chen2012}. Fundamentally interacting systems, such as Mott insulators like the $3d$ transition metal oxides (NiO~\cite{NiO}, MnO~\cite{MnO}, FeO, CeO~\cite{PhysRevB.97.035107}) or certain sulfides (NiS$_2$~\cite{NiS2}), are beyond the TQC classification scheme.

Here, we develop an interacting TQC (iTQC) formalism, which provides an extension of TQC that allows to also treat fundamentally interacting states~\footnote{A previously proposed interacting TQC extension is limited to states adiabatically connected to noninteracting ones~\cite{Iraola,Lessnich_2021}.}. To that end, we start by introducing a class of interacting reference states that extend the notion of ALs. Instead of the (single-particle) Bloch Hamiltonian and its `bands' in momentum space, we consider the $n$-particle Green's functions $\gn(\omega)$, with $\omega$ the Matsubara frequency, and we extend the concept of band representation to also be applicable in the context of these $n$-particle correlation functions. Specifically, we re-interpret the $n$-particle correlation functions as matrices in the space of $n$-particle excitations (hence, we denote them by an underscored symbol) and consider the correlation functions at zero frequency, where $\Gn \equiv \Gn(\omega=0)$ are Hermitian, thereby admitting a spectral ordering. Correlation functions are well defined for interacting states, and they relate to experimental observables in some instances. They are also numerically accessible within advanced computational methods for modeling correlated quantum materials, such as Quantum Monte Carlo (QMC)~\cite{becca_sorella_2017} and coupled cluster theory~\cite{CC}.

What type of topological phases can we expect to discover with this approach?
First, as with TQC, the topological properties must be indicated or protected by spatial symmetries. Second, when working with $n$th-order Green's functions, the phases must be discoverable from $n$-body correlation functions. This is not the case for intrinsic topological order, as found in the fractional quantum Hall effect and various types of gapped spin liquids: discriminating these phases requires the measurement of correlation functions of extensive order $N$, where $N$ indicates the number of particles, that scales at least linearly with the system size~\cite{TO}.

In the presence of interactions, superconductivity and spontaneous symmetry breaking are abundant, and can be characterized by local order parameters. However, featureless insulating phases, which are not adiabatically connected to free-fermion insulators, also exists and are known as symmetry-protected topological states (SPTs)~\cite{doi:10.1146/annurev-conmatphys-031214-014740}. These are presumably not as abundant and certainly very hard to discern and discover.

The reference states proposed later on in this paper are realizations of certain classes of `crystalline' and `point-group' SPTs (cSPTs and pgSPT, resp.)~\cite{PhysRevB.96.205106,PhysRevX.7.011020,doi:10.1126/sciadv.aax2007,NatCommSong2020,PhysRevResearch.4.033081}, in which a symmetry of the space group or point group acts as internal symmetry protecting the phase. Importantly, the reference states we propose capture a large set of cSPT classes, beyond the ones accessible through noninteracting states.
Our formulation is suitable for identifying large classes of SPTs, naturally including all phases discoverable by TQC. Beyond TQC, there are fermionic SPT phases that are intrinsically interacting~\cite{iSPT}. Our method is in particular susceptible to bosonic SPT phases of spins, which arise as effective descriptions of localized electrons and more broadly topological Mott insulators~\cite{Pesin2010,TMIPhysRevLett.113.106401,PhysRevB.97.125142,TMI2022}.
In the following, we focus on the “Hubbard” class of models for Mott insulators as one well-known example where such states arise. However, we expect our approach to be general enough to be extended beyond this class of models.

\bigskip\tocless\section{Summary of results}{sec:Summary of results}

\begin{figure*}[t]
    \centering
    \includegraphics[width=\textwidth]{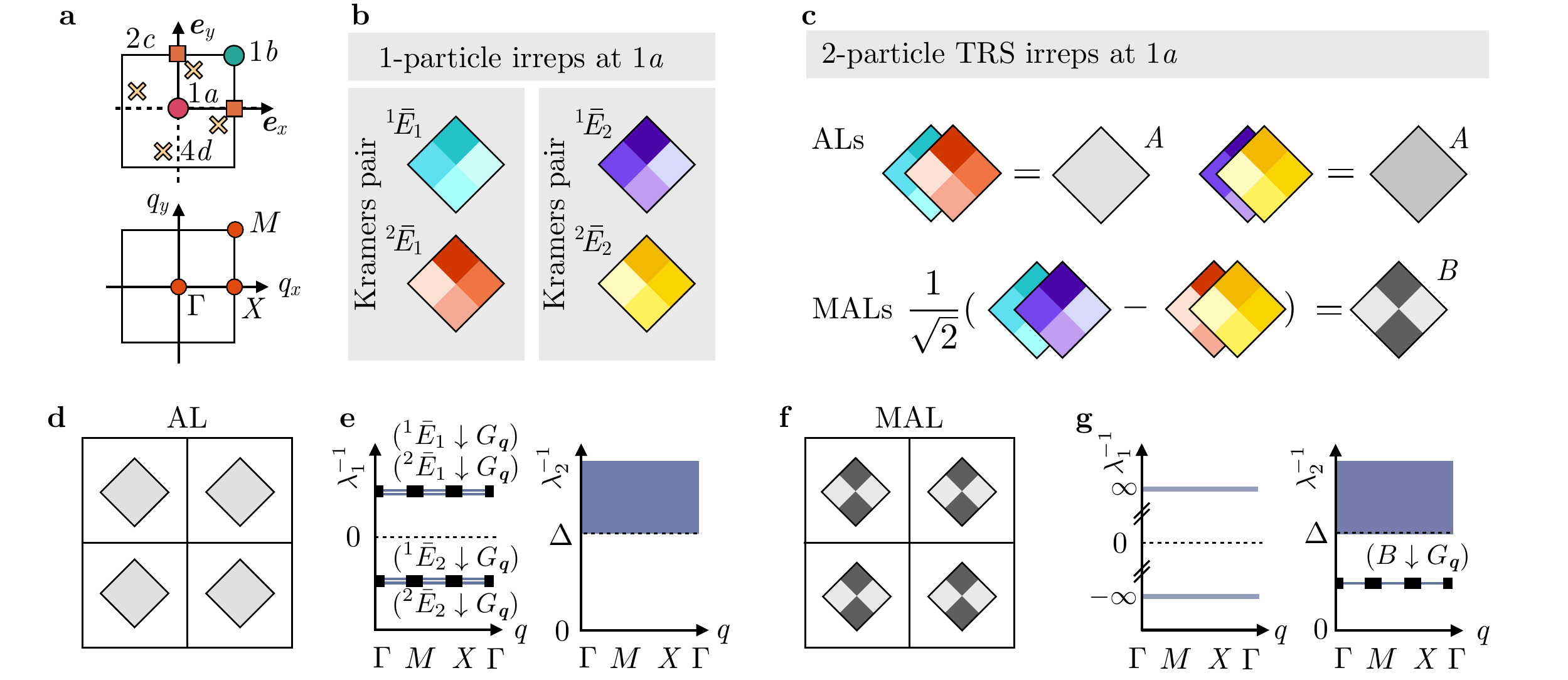}
      \caption{\textbf{Mott atomic limits and induced bands representations.} \textbf{a} Unit cell and first Brillouin zone of the square lattice with the Wyckoff positions and high-symmetry points marked, respectively. \textbf{b} Two sets of single-particle irreps compatible with the site symmetry group $C_4^D$ of Wyckoff position $1a$ : The two blocks represent each two TRS-related states (Kramers pairs) and the colors indicate non-trivial rotation eigenvalues under $C_4$. \textbf{c} Examples of time-reversal symmetric two-particle irreps constructed out of the single-particle irreps in \textbf{b}. In the first row we show two example of ALs: in this case, the two states of a Kramers pair from a Slater determinant and consequently, the state transforms in the $A$ representation of the point-group. In the second row, we show an example of an MAL, where the state transforms in the $B$ representation of the point-group.
    \textbf{d} and \textbf{f} show examples of two-particle irreps placed in each unit cell at Wyckoff position $1a$ with \textbf{d} an AL and \textbf{f} an MAL state on the square lattice. \textbf{e} and \textbf{g} schematically show the respective inverse spectra of the single-particle and two-particle Green's functions, marked by $\lambda_1^{-1}$ and $\lambda_2^{-1}$.}
    \label{fig:IntroPlot}
\end{figure*}

The iTQC framework, summarized in Fig.~\ref{fig:IntroPlot}, closely follows the basic ideas of TQC, which is the classification of ALs. An AL consists of atomic orbitals placed at lattice sites corresponding to some Wyckoff positions of the lattice. We first define a reference class of many-body states in Sec.~\ref{sec:MAL}, which we dub $n$-Mott atomic limits ($n$-MALs) which consist of entangled clusters of $n$ electrons placed at some Wyckoff positions of the lattice. These entangled clusters can be constructed in such a way as to satisfy TRS as well as the lattice symmetries, but to transform non-trivially under the latter. More accurately, these reference states realize cSPTs connected to zero-dimensional (0D) block states, as discussed in Sec.~\ref{subsec:MALs and SPTs}. Note that, by contrast, to obtain a (noninteracting) AL state, which satisfies TRS, we must always combine Kramers pairs of orbitals at the same site which transform trivially under any spatial symmetry. These ideas are summarized in Figs.~\ref{fig:IntroPlot}\textcolor{red}{a}--\textcolor{red}{c} for the case $n=2$.

We use Green's functions as our main tool for diagnosing 2-MALs (see Sec.~\ref{Sec:Greens Functions}). While the single-particle Green's function is related to the Bloch Hamiltonian for a noninteracting system and thus, TQC can also be viewed as a classification scheme based on the single-particle Green's function, a many-body state is not fully specified by its single-particle Green's function. Consequently, we turn to the two-particle Green's function $\Gt$, for which we derive a crucial property: For an AL state, there is a lower bound on the eigenvalues of $\Gt$ which is given by the two-particle gap. Therefore, any eigenvalue of $\Gt$, which appears below this bound, originates from an interacting ground state. Our classification scheme thus focuses on this part of the spectrum of $\Gt$.
Figures~\ref{fig:IntroPlot}\textcolor{red}{d}--\textcolor{red}{f} show a comparison of the classification based on TQC and iTQC.

The iTQC framework then allows to classify all the possible band representations of a $\Gn$ Green's function induced by \mbox{$n$-MAL} ground states as discussed in Sec.~\ref{Sec:GF classification}.  It is advantageous to consider an idealized limit, where the many-body Hamiltonian is spectrally flattened, in analogy with the band flattening procedure in the single-particle case. In this limit, we can derive analytical results for the $\Gt$ band structure. In particular, for an AL state, $\Gt$ only contains eigenvalues above the aforementioned bound (Fig.~\ref{fig:IntroPlot}\textcolor{red}{e}). However, for a $2$-MAL state constructed as a product state of a single cluster operator acting on each unit cell, $\Gt$ contains a single eigenvalue below the bound, at each value of the momentum, whose eigenvector corresponds to the $2$-MAL cluster operator out of which the $2$-MAL state is formed (Fig.~\ref{fig:IntroPlot}\textcolor{red}{g}). We show that stacking several $2$-MAL operators in each unit cell just leads to a direct sum of the $\Gt$-bands below the bound. If we consider all the possible $2$-MAL operators compatible with a given space group and the $\Gt$ band structures induced by them, then we can check whether a given $\Gt$ band structure can be constructed out of a sum of $2$-MAL band structures. Thus, iTQC provides a new definition of topological states: 
\emph{If the band representation associated with the interaction-driven spectrum of a Green's function cannot be induced from ALs and \mbox{$2$-MALs}, the state is either (i) an SPT that cannot be induced from 0-dimensional blocks or (ii) it is a many-body state that is dominated by many-body correlations that involve more than two electrons.}

This statement represents an extension of the TQC framework that in principle can be naturally applied to~$n\geq2$. 

For concreteness, we discuss the full classification of \mbox{$2$-MALs} in one-dimension (1D) (Sec.~\ref{sec:1D}), which gives an explicit example of the procedure outlined above. In addition, simple two-dimensional (2D) examples where the classification can be done by hand are discussed in App.~\ref{app:2D_Examples}. Finally, we apply our formalism in Sec.~\ref{sec:numerics} to four illustrative models whose ground states depart from ideal AL and $2$-MAL states: (i) the Hubbard square~\cite{FMIPhysRevLett.105.166402}, (ii) the Hubbard diamond chain~\cite{Iraola}, (iii) the Hubbard checkerboard lattice~\cite{FMIPhysRevLett.105.166402} and (iv) the Hubbard model on a star of David. In each case, we use exact diagonalization or QMC to diagnose AL and $2$-MAL states, and the transitions between them. 

In this work, we restrict ourselves to the discussion of $n=1$ and $n=2$, while the explicit treatment of higher values of $n$ is left to future work.
In addition, the ideas developed here are in principle easily extended to different types of $n$-particle correlation functions, alongside an appropriate class of reference states, while in the following we concentrate on the anomalous retarded particle-particle Green's function, for $n=2$.

Further details can be found in the Supplementary Material~\cite{SI} (see also Refs.~\cite{CrystTables,PhysRevB.56.15001,BradleyBook2,CracknellBook} therein).

\bigskip\tocless\section{Mott atomic limits}{sec:MAL}

\bigskip\tocless\subsection{\textit{n}-MALs}{sec:n-MALs}

As the overarching motivation of this paper is to develop a framework useful for making predictions about real crystalline materials, we focus on the case of TRS systems of itinerant electrons with spin-orbit coupling. We consider systems at zero temperature, gapped, short-range entangled and with a unique ground state. This class of systems, encompassed by the SPT phases, has proven to be a promising domain for studying topological phenomena, and allows for a detailed characterization by first principles calculations. We will not consider any sublattice or chiral symmetry, as these are generically absent in real crystals. 

With these constraints, we single out a class of interacting many-body states that our formalism is able to capture and classify, and that we consider as reference states of the SPT type. These states are the natural extension of the concept of ALs: while ALs are product states of exponentially localized single-particle states distributed on the lattice, $n$-MALs are product states of interacting clusters of $n$-electrons placed on the crystal.

To set the notation, we consider a lattice with space group $G$ with exponentially localized Wannier orbitals placed at some Wyckoff positions of the lattice.
We denote by $\vec{x}_{a}$ the positions of the sites in the orbit of a Wyckoff position of multiplicity $m$, where $a=1, \dots, m$. The set of Wannier orbitals placed on the site at $\vec{x}_{a}$ must transform under a representation $\rho$ of its site-symmetry group $G_{\vec{x}_{a}}$.

With these notions, we define $\hat{c}^{\dagger}_{\vec{r}, \alpha}$ ($\hat{c}_{\vec{r}, \alpha}$) as the creation (annihilation) operator of an electron placed at the unit cell $\vec{r}$ and created (annihilated) in the single-particle state of an orbital whose quantum numbers are described by the compactified index $\alpha = (W, a, \rho, i)$. For an orbital located at $\vec{x}_a$, $W$ indicates the label of the Wyckoff position of the site at $\vec{x}_a$, $a=1,\dots,m$ specifies at which $\vec{x}_a$ the electron is placed, $\rho$ labels the representation of $G_{\vec{x}_a}$ of the orbital, and $i=1,\cdots, \text{dim}(\rho)$ enumerates the various orbitals placed at $\vec{x}_{a}$.

Noninteracting ALs are constructed as Slater determinants of exponentially localized single-electron states $\hat{c}_{\bm{r},\alpha}$, viz.
\begin{equation} \label{eq:decomp}
    |\textrm{AL}\rangle=\prod_{\bm{r}}\prod_{\alpha \in \textrm{occ.}}\hat{c}_{\bm{r},\alpha}^\dagger |0\rangle,
\end{equation}
with $\alpha$ ranging over the occupied orbitals in each unit cell.
In some instances, we may refer to \mbox{$n$-ALs} as states of the form~\eqref{eq:decomp} where each creation operator in~\eqref{eq:decomp} is replaced by a product of $n$~creation operators with different quantum numbers.
A state is called \textit{topological} if it is not possible to define exponentially localized Wannier orbitals compatible with the space group,
such that the decomposition~\eqref{eq:decomp} holds. An elementary example are Chern bands in 2D.
Therefore, in TQC, non-trivial topology refers to an obstruction in going from a filled band description in momentum space to a real space description in terms of localized orbitals.

We now introduce $n$-MALs, which are quantum many-body states of several electrons that are also exponentially localized, but may not be adiabatically connected to a single Slater determinant, or single reference state in quantum chemistry terms, without the breaking of a relevant spatial symmetry. The \mbox{$n$-MALs} are obtained by placing localized interacting clusters of electrons on (maximal) Wyckoff positions of the crystalline lattice, in analogy with the construction of ALs. Hence, the wave function of an \mbox{$n$-MAL} is
\begin{equation}
    |n\textrm{-MAL}\rangle=\prod_{\bm{r}} \prod_{\xi \in \text{occ.}}\hat{O}_{\bm{r},\xi}^\dagger|0\rangle,
    \label{eq:decomp_MAL}
\end{equation}
where each $\hat{O}_{\bm{r},\xi}^\dagger$ is now a $n$-particle ``cluster" operator, consisting of linear combinations of products of $n$ single-particle operators creating electrons centered at the unit cell with lattice vector $\vec{r}$, and the index $\xi$ ranges over the $n$-particle operators that are occupied in the unit cell.

To get an insight on the fundamental distinctions between ALs and $n$-MALs, note that TRS constrains all \mbox{$n$-MALs} to transform as real-valued 1D irreps of the site-symmetry group of their site, leaving $\pm1$ as possible eigenvalues for any spatial symmetry. Conversely, TRS single Slater determinant states always transform with eigenvalue $+1$: They are composed of products of Kramers pairs of electrons which contribute complex conjugate eigenvalues $\lambda$ and $\lambda^\star$, such that $\lambda\lambda^\star=+1$ is the symmetry eigenvalue of the whole pair~\cite{FMIPhysRevLett.105.166402}.
Hence, \mbox{$n$-MAL}s of type~\eqref{eq:decomp_MAL} are characterized by transformation rules under symmetry action that in principle can be distinct from the set of all the possible representation realized by TRS ALs. 

Some of the states described by~\eqref{eq:decomp_MAL} can be adiabatically connected to ALs, while others form an adiabatically disconnected class of states. With the iTQC framework, we aim to identify these classes by means of the Green's function band representation, as we will discuss later.

Some prominent examples of \mbox{$n$-MALs} are (i)~valence bond states~\cite{ANDERSON_RVB}, (ii)~coupled cluster wave functions~\cite{CC} (iii)~cluster Mott insulators with star of David order, as shown in this work.

\bigskip\tocless\subsection{2-MALs}{sec:2-MAL}
In practice, in the present paper we will specialize to the case $n=2$. In the following, we denote \mbox{$2$-MALs} as MALs for compactness of notation, unless otherwise stated, and maintain the label \mbox{$n$-MALs} for the general case.

A MAL can be in general written as 
\begin{equation}\label{eq:2-MAL wf}
    |\textrm{MAL}\rangle =\prod_{\bm{r}, \xi}\hat{O}^{\dagger}_{\bm{r}, \xi}|0\rangle, \
    \hat{O}^{\dagger}_{\bm{r}, \xi} = \sum_{\alpha, \beta, \vec{u}}M^{\xi}_{\alpha\beta\vec{u}} \hat{c}^{\dagger}_{\bm{r}, \alpha} \hat{c}^{\dagger}_{\bm{r}+\vec{u}, \beta},
\end{equation}
where the coefficients $M^{\xi}_{\alpha\beta\vec{u}}$ are constrained by TRS and the spatial crystalline symmetries of the relevant space group, and $\xi$ labels the type of MAL cluster operator. We assume that distinct cluster operators do not overlap, and therefore $\hat{O}^{\dagger}_{\vec{r}, \xi}$ commute pairwise~\footnote{A more general notion of an MAL is possible, where the cluster operators overlap. Such states include resonant valence bond states as the $S'$-Mott phase of the Hubbard ladder \cite{1DHubbardLadderTsuchiizu_2002}. We leave this extension to future work.},~\cite{1DHubbardLadderTsuchiizu_2002}. In Eq.~\eqref{eq:2-MAL wf}, we introduce the lattice vector $\vec{u}$ to take into account the spatial separation between pairs of electrons.
In momentum space, the MAL operator depends on a single momentum $\vec{q}$, and reads
\begin{equation}\label{eq:MAL operator in k space}
\begin{split}
    \hat{O}^{\dagger}_{\vec{q}, \xi} &= \frac{1}{\sqrt{N}} \sum_{\bm{r}} e^{-\mathrm{i} \Vec{q}\cdot \Vec{r}} \hat{O}^{\dagger}_{\vec{r}, \xi}\\
    &= \frac{1}{\sqrt{N}} \sum_{\Vec{k}, \alpha, \beta, \vec{u}} e^{\mathrm{i}(-\vec{k}+\vec{q})\cdot \vec{u}} M^{\xi}_{\alpha\beta \vec{u}} \hat{c}^{\dagger}_{\bm{k},  \alpha} \hat{c}^{\dagger}_{-\bm{k}+\bm{q}, \beta}.
\end{split}
\end{equation}
Note that the definition in Eq.~\eqref{eq:2-MAL wf} also includes ALs with an even number of electrons.
The MAL operators transform under two-particle representations $\rho$ of the space group $G$. We discuss the explicit form of $\rho$ and how to systematically construct MAL operators compatible with a certain space group $G$ in App.~\ref{App: MAL operators}.

As an explicit example of MALs, we consider a 1D lattice. Depending on the spatial embedding in a 3D system, the 1D system may be considered in the presence of mirror symmetry or inversion symmetry ($\mathcal{I}$)~\footnote{We are interested in symmetries which map the 1D system back to itself, but reverse its one spatial coordinate axis. Another symmetry which does this is two-fold rotation about an axis perpendicular to the 1D system. We do not separately discuss it here as it has the same representation as mirror symmetry.}, and in the following we only focus on the latter.
Let us consider the case of Wyckoff position $2c$, which has two-fold multiplicity, equipped with a Kramers pair of orbitals (Fig.~\ref{fig:1D_unitcell_irreps}). We denote the creation operator at the unit cell coordinate $r=0,\cdots L-1$ as $\hat{c}^\dagger_{r, a, \sigma}$, where $\sigma \in \{\uparrow,\downarrow\}$ is the spin, and $a\in\{1,2\}$ labels the two $2c$ sites~\footnote{The set of orbitals placed at each $2c$ Wyckoff position transforms in the $\bar{A}\bar{A}$ physical representation of the double site symmetry group $1$. According to the notation introduced above, the $\alpha$ index labeling these operators reads
$\hat{c}^{\dagger}_{r,\alpha}:=\hat{c}^\dagger_{r, 2c, a, \bar{A}\bar{A},\sigma}$, with $a\in\{1, 2\}$ and $\sigma \in \{\uparrow, \downarrow\}$, where we use a spin label to indicate a realization of TRS states. In the main text, we suppress the $2c$ and $\bar{A}\bar{A}$ labels for the sake of simplicity.}.
The antiunitary TRS acts as
\begin{equation}
    \mathcal{T}\hat{c}^\dagger_{r, a, \sigma}\mathcal{T}^{-1}
    =
    \mathrm{i}
    \sum_{\sigma'=\uparrow, \downarrow}
    \sigma^{(2)}_{\sigma' \sigma}\hat{c}^\dagger_{r, a, \sigma'},
\end{equation}
where we denote with $\sigma^{(i)}$, $i=1,2,3$ the three Pauli matrices, and inversion acts as
\begin{align}
    \mathcal{I}\hat{c}^\dagger_{r, a, \sigma}\mathcal{I}^{-1}&=
    \sum_{a'=1,2}
    \sigma^{(1)}_{a' a}\hat{c}^\dagger_{-r, a', \sigma},
\end{align}
where we understand $r\mod L$.
An example of a two-electron cluster operator, constructed out of the available single-electron creation operators, is
\begin{equation}
    \hat{O}^{\dagger}_{r, \mathrm{MAL}}
    =\frac{1}{\sqrt{2}}(\hat{c}^\dagger_{r, 1, \uparrow}\hat{c}^\dagger_{r,1, \downarrow}-\hat{c}^\dagger_{r,2,\uparrow}\hat{c}^\dagger_{r,2,\downarrow})
\end{equation}
which obeys
\begin{equation}\label{eq:transform MAL I}
    \mathcal{I}\hat{O}^{\dagger}_{r, \mathrm{MAL}}\mathcal{I}^{-1}=-\hat{O}^{\dagger}_{-r, \mathrm{MAL}}.
\end{equation}
The operator in Eq.~\eqref{eq:transform MAL I} creates a two-electron cluster that transforms with inversion eigenvalue $(-1)$, when we choose $r=0$ as inversion center. It is easy to prove that the ground state
\begin{equation}\label{eq:MAL example}
    |\mathrm{MAL}\rangle=
    \prod_{r=0}^{L-1}\hat{O}^{\dagger}_{r, \mathrm{MAL}}|0\rangle
\end{equation}
cannot be connected adiabatically to any 2-AL ground state: For odd $L$, it has inversion eigenvalue $(-1)$, while any TRS AL state has inversion eigenvalue $(+1)$~\cite{FMIPhysRevLett.105.166402}. As no TRS AL behaves the same way at the same filling, the state in Eq.~\eqref{eq:MAL example} has to be considered as a representative of a distinct phase. 

\bigskip\tocless\subsection{MALs and SPTs}{subsec:MALs and SPTs}
\begin{figure}[t]
    \centering
    \includegraphics[width=0.48\textwidth]{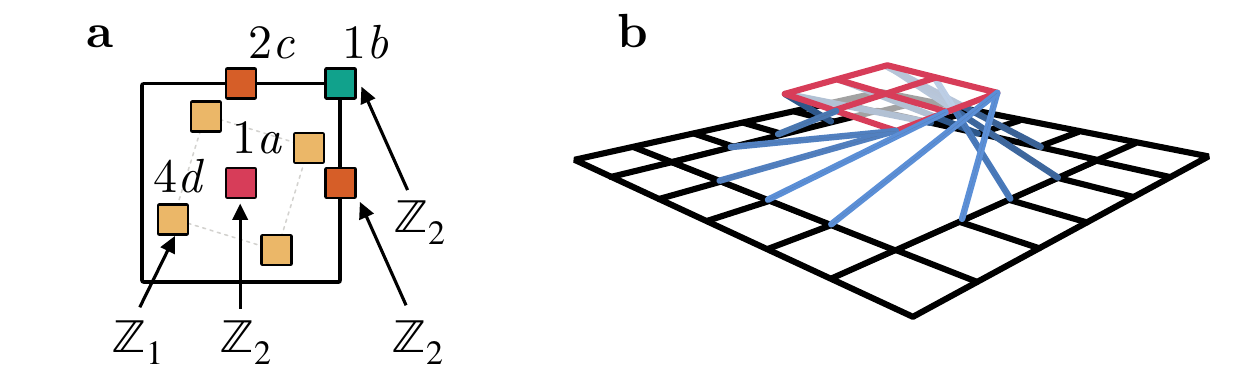}
    \caption{\textbf{Crystalline SPTs.} \textbf{a} Classification of 0D-block cSPTs in a $C_4$ symmetric unit cell with TRS $\mathcal{T}^2 =+1$. \textbf{b} Illustration of the action of a partial symmetry operation applied to a subsystem in the square lattice. The pink region indicates the subsystem and the blue lines indicate the new lattice connections after the partial $C_4$ rotation is applied to the subsystem.}
    \label{fig:SPTs}
\end{figure}
We now discuss how MALs realize SPT phases. Crystalline SPTs and pgSPTs~\cite{PhysRevB.96.205106,PhysRevX.7.011020,doi:10.1126/sciadv.aax2007,NatCommSong2020,PhysRevResearch.4.033081} are SPT phases whose protecting symmetries are crystalline space group or point group symmetries, which act as internal onsite operations on portions of the unit cell, called `blocks' in this context~\footnote{Although it is enough to consider the more general case of cSPTs, here we also explicitly refer to pgSPTs as the results presented later on in the section were derived in the context of pgSPTs. For the latter, the classification is based on a point group rather than the full space group. Although less refined, the pgSPT case is sufficient to convey the key concepts of the cSPT classification.}. A block $b$ of the unit cell is left invariant under a subset of the point group, $G_b \subset G$, i.e., elements of $G_b$ act as on-site or internal symmetries on the block. Let us first recall a few properties of the pgSPTs, following Ref.~\cite{PhysRevB.96.205106}. A possible construction scheme for such phases consists in decorating different $d_b$-dimensional unit cell blocks $b$ ($d_b=0, 1, 2$ for 3D systems) with $d_b$-dimensional SPTs.
In Ref.~\cite{PhysRevB.96.205106}, Eq.~(1) defines the `block states' as
    \begin{equation}\label{eq:block wf SPT}
        \ket{\Psi} = \bigotimes_{b \in B} \ket{\psi_b} ,
    \end{equation}
where each factor $\ket{\psi_b}$ corresponds to an SPT wavefunction in $d_b$-dimensions defined on block $b$, whose protecting symmetry belongs to $G_b$. For the case of a 0D-block, one says that $\ket{\psi_b}$ has a `$G_b$ charge', meaning that it transforms under a 1D $G_b$ irrep, and different irreps correspond to distinguished 0D-block SPTs.
The classifications of pgSPTs with point group $G$ in $d$-dimensions and bosonic degrees of freedom are provided by the cobordism classification, and can be decomposed as follows (Eq.~(2) in Ref.~\cite{PhysRevB.96.205106})
\begin{equation}\label{eq:block decompos SPT class}
        \mathcal{C}(G) = \mathcal{C}_0(G) \times \cdots \times \mathcal{C}_{d-1}(G),
\end{equation}
where $\mathcal{C}_{d_b}(G)$ is the classification of SPTs built only out of $d_b$ blocks.
For fermionic degree of freedom the factorization of Eq.~\eqref{eq:block decompos SPT class} does not hold, a fact that we can ignore in this work, as we argue below.

From Eqs.~\eqref{eq:2-MAL wf} and~\eqref{eq:block wf SPT}, we see that MALs are   0D-block state cSPTs. Our interest here are TRS fermionic MALs that conserve particle number. Particle number conservation $U(1)$ and TRS $\mathcal{T}$ thus have to be imposed in addition to the symmetry $G$. MALs have even fermion parity, otherwise they would not have a unique TRS ground state. Therefore, time-reversal $\mathcal{T}^2=(-1)^F=+1$ ($F$ being the fermion parity) is represented as in bosonic states. MALs thus follow the bosonic $\mathcal{C}_0(G)$ classification, supplemented by the TRS constraint and particle number conservation.
 The relevant symmetry group for classifying the 0D-block SPTs is thus $G_0= (U(1)\rtimes \mathbb{Z}_{4}^{TF})/\mathbb{Z}_2^F \times \tilde{G}$, where $\tilde{G}$ is the unitary onsite-symmetry of the block, and $\mathbb{Z}_2^F$ the fermion parity. For concreteness, we exemplify this for the case $\tilde{G}=\mathbb{Z}_2$ with generator $S$ that could originate from mirror, two-fold rotation or inversion symmetry and contrast it to $\tilde{G}=\mathbb{Z}_1$. Table~\ref{tab:SPT 0D classif} lists the classification for both cases, differentiating whether the system is interacting or not.  The noninteracting case corresponds to ALs, where $\mathbb{Z}$ simply counts the number of occupied Kramers pairs. Importantly, interactions allow for an additional $\mathbb{Z}_2$-grading, which is realized by MALs and allows for 0D-block states odd under $S$. 
We illustrate in Fig.~\ref{fig:SPTs}\textcolor{red}{a} how these blocks can be used to build pgSPTs in wallpaper group $p4$, that has four-fold rotation symmetry and translation symmetry.  
At the Wyckoff positions $1a$ and $1b$ we can place MALs with $C_4$ eigenvalues $\pm 1$ and at the Wyckoff position $2c$ we can place MALs with $C_2$ eigenvalues $\pm 1$. This results in a $\mathbb{Z}_2\times \mathbb{Z}_2 \times \mathbb{Z}_2$ group of pgSPT phases beyond those that have an AL representation.

\begin{table}[t]
    \centering
    \begin{tabular}{c@{\hskip 12 pt}@{\hskip 12 pt}c@{\hskip 12 pt}c@{\hskip 10 pt}c}
                \hline\hline
                &&& \\[-6pt]
                $\frac{U(1) \rtimes \mathbb{Z}_4^{T F}}{\mathbb{Z}_2^{F}}\times$ & $\mathbb{Z}_1$ & $\mathbb{Z}_2, \ [S, T]=0$ &  $\mathbb{Z}_2, \ \{S, T\}=0$ \\[2pt]
                \hline
                &&& \\[-6pt]
                Noninteracting & $\mathbb{Z}$ & $\mathbb{Z}\times\mathbb{Z}$ & $\mathbb{Z}$  \\[2pt]
                &&& \\[-6pt]
                Interacting & $\mathbb{Z}$ & $\mathbb{Z}\times\mathbb{Z}\times\mathbb{Z}_2$ & $\mathbb{Z}\times\mathbb{Z}_2$  \\[2pt]
                \hline\hline
    \end{tabular}
    \caption{\textbf{Classification of 0D-block SPTs.} Classification of SPT phases constructed from 0D-block states, where the symmetries of the systems include charge conservation, $U(1)$, and TRS $\mathbb{Z}_{4}^{T F}$, generated by $\mathcal{T}=T\mathcal{K}$, where $T$ is the unitary part of TRS and $\mathcal{T}^2=(-1)^F$, $F$ the fermion number, and $\mathbb{Z}_2^F$ is the fermion parity. The three columns correspond to the case of no additional internal symmetry (indicated by $\mathbb{Z}_1$), and a two-fold graded internal symmetry that commutes (indicated by $\mathbb{Z}_2$, $[S, T]=0$) or anti-commutes (indicated by $\mathbb{Z}_2$, $\{S, T\}=0$) with the unitary part of TRS.}
    \label{tab:SPT 0D classif}
\end{table}

A practical question is, given a (non-fixed-point) correlated many-body state, how one computes topological invariants that allow to place the state in this classification. These SPT invariants can be obtained as the $U(1)$ phases of a quantum system defined on a spacetime manifold potentially equipped with a background symmetry bundle and having a topology which probes the relevant spatial symmetries. Thus far, such invariants have been extracted from the ground states of interacting models for only a handful of examples by applying partial symmetry operations~\cite{PhysRevB.95.205139, PhysRevLett.118.216402}, by gauging the protecting symmetry~\cite{PhysRevX.8.011040}, and by computing a partial transpose operation~\cite{PhysRevB.98.035151}. Concretely, the invariants are obtained as the ground state expectation value of a partial point group operation, i.\,e., a point group operation applied to a subsystem. Since one needs to perform such an operation on a subregion of the space that is large compared with the correlation length of the system, it typically involves taking the expectation value of an operator that implements $\mathcal O(L^{d})$ swap operations where $L>
\xi$ (the correlation length) and $d$ is the dimension of space. Figure~\ref{fig:SPTs}\textcolor{red}{c} shows an example of how a partial symmetry operation acts for the square lattice case: first, the $C_4$ rotation is applied to a subsystem of size $L^d$, comparable with the total system size, while leaving the rest of the system unchanged. The expectation value of this operation on the initial ground state allows to extract the topological invariant.

Depending on the form in which the state is represented, the computation of such an expectation value can be of vastly different numerical complexity. In 1D systems, if the ground state is known in a matrix-product state (MPS) form, it is possible to efficiently evaluate the partial symmetry action on the ground state~\cite{PhysRevB.86.125441}. On the other hand, diagrammatic  QMC calculations as well as higher-dimensional tensor networks are typically not suited for computing such high-order operator expectation values. 

\bigskip\tocless\section{Green's functions}{Sec:Greens Functions}

\bigskip\tocless\subsection{Single-particle retarded Green's function}{sec:G1}
In this section, we review a few properties of the single-particle Green's functions which will be relevant in connection to higher-order Green's functions.

Consider in real time the retarded Green's function with an electron created at time $t=0$ and another annihilated at time $t$
\begin{equation}
\label{eq:retarded G1 tau}
\go_{\alpha \beta}(t, \vec{k})=\mathrm{i} \Theta(t)\left\langle \{ \hat{c}_{\vec{k}, \alpha}(t), \hat{c}_{\vec{k}, \beta}^{\dagger}(0)\} \right\rangle_{\mathrm{GS}},
\end{equation}
where the brackets $\langle...\rangle_{\mathrm{GS}}$ indicate the expectation value over the ground state (assumed to be unique and gapped), and $\vec{k}, \alpha, \beta$ describe the quantum numbers of the created and annihilated electrons.
Equivalently, we can write the single-particle Green's function in terms of Matsubara frequency $\omega$. Focusing on the case of zero frequency for reasons that will become clear in the following discussion, by Fourier transforming Eq.~\eqref{eq:retarded G1 tau} to Matsubara frequency and setting $\omega=0$ we obtain
\begin{equation}\label{eq:retarded G1}
\begin{split}
\go_{\alpha \beta}(\omega=0, \vec{k})= &-\left\langle \hat{c}_{\vec{k}, \alpha}[ \hat{H} - E_0]^{-1} \hat{c}_{\vec{k}, \beta}^{\dagger} \right\rangle_{\mathrm{GS}} \\
&+ \left\langle \hat{c}_{\vec{k}, \beta}^{\dagger} [ \hat{H} - E_0 ]^{-1} \hat{c}_{\vec{k}, \alpha} \right\rangle_{\mathrm{GS}}
\end{split}
\end{equation}
with $\hat{H}$ the many-body Hamiltonian of the system, and $E_0$ the energy of the ground state.
Although Eq.~\eqref{eq:retarded G1} results in a single number, for a specific choice of $\{\Vec{k}, \alpha, \beta\}$, the collection of all the possible $\go_{\alpha\beta}(\omega=0, \Vec{k})$ can be interpreted as a matrix $\Go (\omega=0, \vec{k})$ with indices $\alpha, \beta$, for each sector of fixed $\vec{k}$ -- where here and in the following we use the underscore to denote matrices.
This interpretation of the single-particle Green's function as a matrix allows to compute a spectrum of $\Go (\omega=0, \vec{k})$. Note that, $\Go(\omega=0, \Vec{k})$ is a hermitian matrix, yielding a real spectrum.  The eigenstates $v^{\xi}$ of $\Go (\omega=0, \vec{k})$ naturally define operators of the type
\begin{equation}\label{eq:eigenstates G1}
    \hat{a}^{\dagger}_{\vec{k}, \xi} = \sum_{\alpha} v^{\xi}_{\alpha}
    \hat{c}^{\dagger}_{\vec{k},\alpha},
\end{equation}
which we refer to as the eigenstates of $\Go (\omega=0, \vec{k})$.

We first recall the role of $\Go(\omega, \vec{k})$ for noninteracting systems. In this case, the Hamiltonian is written as
\begin{equation}
    \hat{H} = \sum_{\vec{k}, \alpha, \beta} \hat{c}^{\dagger}_{\vec{k},\alpha} h(\vec{k})_{\alpha \beta} \hat{c}_{\vec{k},\beta},
\end{equation}
and the single-particle Bloch Hamiltonian $h(\bm{k})$ is related to $\Go(\omega, \vec{k})$ through the relation $\Go(\omega, \bm{k})= [\mathrm{i}\omega - h(\bm{k})]^{-1}$. 
Based on these considerations, TQC can be thought of as a classification scheme for the spectrum of $\Go(\omega=0, \bm{k})$  instead of the band representations of $h(\bm{k})$~\cite{PhysRevB.78.195424_Qi_Hughes_Zhang}. (In the following, we refer to $\Go(\omega=0, \bm{k})$ as $\Go(\bm{k})$ or simply $\Go$, when we consider all the $\vec{k}$ sectors at once, as we always set the frequency to be equal to zero, unless otherwise stated.) The eigenvectors of $\Go(\bm{k})$, as defined in Eq.~\eqref{eq:eigenstates G1},  are thus the single-particle states which make up the single Slater determinant ground state, and the eigenvalues yield the inverse energies. Also, states obtained by applying the operators~\eqref{eq:eigenstates G1} on the vacuum are eigenstates of the Hamiltonian. In fact, the spectrum of $-[\Go]^{-1}$ can be interpreted as a band structure, and in this sense we speak about the ``band representation'' of $\Go$. 
Turning to the case with interactions, one possible avenue to extend TQC is to calculate $\Go(\omega=0, \vec{k})$ and invert it to obtain an effective Hamiltonian. Indeed, previous works have focused on applying the TQC framework to the single-particle Green's function in the interacting case, and the resulting effective Hamiltonian was termed \emph{topological Hamiltonian}~\cite{PhysRevB.83.085426,PhysRevX.2.031008,Wang_2013,Iraola,Lessnich_2021,PhysRevB.85.165126}, which also includes the self-energy. The choice of considering $\Go$ at zero frequency relies on it being sufficient to capture the topological properties of the system. 
This approach has proven successful in capturing topological properties of states that are adiabatically connected to noninteracting systems (single Slater determinants). The only significant difference to classifying band structures of Bloch Hamiltonians regards the notion of equivalence classes via band gaps: For two band representations of  $\Go$ to be equivalent, they have to be deformable into each other (while retaining symmetries and the locality of the operator) not only without any eigenvalue crossing infinity (corresponding to the normal noninteracting band gap), but also with no eigenvalue crossing zero (corresponding to poles in the self-energy when Mott gaps open)~\cite{Lessnich_2021,Wagner2022}. 

However, unlike the noninteracting case, an interacting state is generically not fully specified by $\Go$, and for intrinsically interacting topological states the aforementioned approach to identify topology fails. For interacting states, there are in general a number of non-trivial $n$-particle Green's functions $\gn$, where $n=1,\cdots,N$ for a system with particle number $N$, which cannot be reconstructed from the sole knowledge of $\Go$. While evaluating all the $N$ correlation functions is not feasible in practice, one can truncate the series, as is done in the Bogoliubov–Born–Green–Kirkwood–Yvon hierarchy~\cite{BBGKY}. For some classes of states, this truncation is exact: for instance, in the $n$-MALs states the $n$-particle Green's function completely describes the system, since correlations are constrained to involve at most $n$ electrons at a time, if the $n$-particle operators creating clusters of electrons at different unit cells do not overlap. Based on this statement, we will see in Sec.~\ref{Sec:GF classification} how to use $\Gt$ to diagnose the interacting or noninteracting nature and symmetry properties of an MAL ground state. In the present paper, we focus on the information contained in the two-particle Green's function $\Gt$, although in principle the generalization to higher-order Green's functions $\Gn$ should be straightforward.

\bigskip\tocless\subsection{Two-particle retarded Green's function}{sec:2p Gf}
We build the iTQC formalism based on the two-particle retarded Green's function describing the amplitude for the creation of two electrons at real time $t=0$, and the annihilation of two electrons at a later time $t$
\begin{equation}\label{eq:retarded 2 particle GF in imaginary time real space operators}
\begin{split}
    &\gt_{\bm{r}_1\bm{r}_2 , \bm{r}'_2\bm{r}'_1; \alpha \beta, \gamma \delta}(t)
    \\ &= -\Theta(t) \expval{ [
    \hat{c}_{\bm{r}_2, \beta}(t)
    \hat{c}_{\bm{r}_1, \alpha}(t),
    \hat{c}^{\dagger}_{\bm{r}'_1, \gamma}(0) \hat{c}^{\dagger}_{\bm{r}'_2, \delta}(0) ] }_{\mathrm{GS}}.
\end{split}
\end{equation}
Again, the expectation value $\langle \cdots \rangle_{\mathrm{GS}}$ is taken over the many-body ground state of the Hamiltonian.
The transformed version of Eq.~\eqref{eq:retarded 2 particle GF in imaginary time real space operators} as a function of imaginary (Matsubara) frequency $\omega$ reads
\begin{equation}\label{c}
\begin{split}
    &\gt_{\bm{r}_1\bm{r}_2 , \bm{r}'_2\bm{r}'_1; \alpha \beta, \gamma \delta}(\omega)
    \\&=- \expval{\hat{c}_{\bm{r}_2,\beta} \hat{c}_{\bm{r}_1,\alpha} [\mathrm{i} \omega + E_0 - \hat{H}]^{-1} \hat{c}^{\dagger}_{\bm{r}'_1,\gamma} \hat{c}^{\dagger}_{\bm{r}'_2,\delta}}_{\mathrm{GS}}\\ 
    & \quad + \expval{\hat{c}^{\dagger}_{\bm{r}_1',\gamma} \hat{c}^{\dagger}_{\bm{r}_2',\delta} [\mathrm{i} \omega - E_0 + \hat{H}]^{-1}  \hat{c}_{\bm{r}_2,\beta} \hat{c}_{\bm{r}_1,\alpha}}_{\mathrm{GS}},
\end{split}
\end{equation}
where $E_0$ indicates the ground state energy and $\hat{H}$ is the many-body Hamiltonian of the system.
It is useful to write Eq.~\eqref{c} in the Lehmann decomposition~\cite{BruusFlensberg}
\begin{equation}\label{eq:Lehman G2 real space}
\begin{split}
    &\gt_{\bm{r}_1\bm{r}_2 , \bm{r}'_2\bm{r}'_1; \alpha \beta, \gamma \delta}(\omega) \\&=
    - \sum_{m \in \mathscr{H}_{N+2}} \frac{\bra{\mathrm{GS}} \hat{c}_{\bm{r}_2, \beta} \hat{c}_{\bm{r}_1, \alpha} \ket{m}\bra{m} \hat{c}^{\dagger}_{\bm{r}'_1, \gamma} \hat{c}^{\dagger}_{\bm{r}'_2, \delta}\ket{\mathrm{GS}}}{\mathrm{i} \omega + E_0 - E_{m}} \\
    &+ \sum_{n \in \mathscr{H}_{N-2}} \frac{\bra{\mathrm{GS}} \hat{c}^{\dagger}_{\bm{r}'_1, \gamma} \hat{c}^{\dagger}_{\bm{r}'_2, \delta}  \ket{n}\bra{n} \hat{c}_{\bm{r}_2, \beta} \hat{c}_{\bm{r}_1, \alpha} \ket{\mathrm{GS}}}{\mathrm{i} \omega - E_0 + E_{n}} ,
\end{split}
\end{equation}
where $\mathscr{H}_{N\pm2}$ indicates the Hilbert space of $N\pm2$ particles, and we assumed that the many-body ground state lies in the $N$-particle sector of the Fock space. 
Expressions analogous to Eqs.~\eqref{eq:retarded 2 particle GF in imaginary time real space operators},~\eqref{c} and~\eqref{eq:Lehman G2 real space} apply for electronic creation and annihilation operators acting in momentum space.
The counterpart of~\eqref{c} with electronic operators acting in momentum space is
\begin{equation}\label{eq:retarded 2 particle GF in omega}
\begin{split}
    &\gt_{\bm{k}_1,\bm{k}_2, \alpha \beta, \gamma \delta}(\omega, \vec{q})
    \\&=- \expval{\hat{c}_{-\bm{k}_1+ \bm{q}, \beta} \hat{c}_{\bm{k}_1, \alpha} [\mathrm{i} \omega + E_0 - \hat{H}]^{-1} \hat{c}^{\dagger}_{\bm{k}_2, \gamma} \hat{c}^{\dagger}_{-\bm{k}_2 + \bm{q}, \delta}}_{\mathrm{GS}}\\
    & \quad + \expval{\hat{c}^{\dagger}_{\bm{k}_2, \gamma} \hat{c}^{\dagger}_{-\bm{k}_2 + \bm{q}, \delta} [\mathrm{i} \omega - E_0 + \hat{H}]^{-1}  \hat{c}_{-\bm{k}_1+ \bm{q}, \beta} \hat{c}_{\bm{k}_1, \alpha}}_{\mathrm{GS}},
\end{split}
\end{equation}
which is determined by three different momenta: the internal momenta $\vec{k}_1$, $\vec{k}_2$ and the total momentum exchanged $\vec{q}$.
We now restrict our attention to a smaller class of Green's functions, as compared to Eq.~\eqref{eq:retarded 2 particle GF in omega}, by introducing the constraint that the pair of operators that create (annihilate) electrons are local in space. The requirement of locality is obtained by demanding that they are not further than a certain distance $u_{\max}$ apart, meaning $\vec{r}_2=\vec{r}_1 + \vec{u}$ and $\vec{r}'_2=\vec{r}'_1 + \vec{v}$, for a range of $\abs{\vec{u}},\abs{\vec{v}} < u_{\max}$. This allows to include correlations between pairs of electrons on sufficiently short distances, which we expect to be the ones dominating the essential entanglement structure of a gapped ground state. Note that $\vec{u}, \vec{v}$ also depend on the indices $\alpha, \gamma$, and in principle also condition the allowed $\beta, \delta$ (App.~\ref{sec:App_Delta}). This constraint of locality is necessary to obtain a finite dimensional matrix from the set of $\gt$ expectation values, such that its dimensions are independent of system size -- at fixed $\vec{q}$. The more generic type of $\gt$, as the one in Eq.~\eqref{eq:retarded 2 particle GF in omega}, has the momentum labels $\bm{k}_1, \bm{k}_2$, apart from $\vec{q}$, over which it needs to be diagonalized, leading to a number of eigenvalues that does not increase linearly with system size. 

With the locality constraint in place and by taking into account translational invariance, one obtains
\begin{equation}\label{eq:G2q explicit}
    \begin{split}
   &\gt_{ \alpha \beta \vec{u}, \gamma \delta \vec{v}}(\omega,\bm{q})\\
    &= \frac{1}{N} \sum_{\bm{r}, \bm{r}'} e^{\mathrm{i}\bm{q}\cdot (\bm{r}-\bm{r}')} 
 \gt_{\bm{r},\bm{r}+\vec{u},\bm{r}', \bm{r}'+\vec{v},\alpha \beta, \gamma \delta}(\omega)
    \\
    &:= \frac{1}{N} \sum_{\bm{r}, \bm{r}'} e^{\mathrm{i}\bm{q}\cdot (\bm{r}-\bm{r}')} 
 \gt_{\bm{r},\bm{r}',\alpha \beta \vec{u}, \gamma \delta \vec{v}}(\omega) \\
 &=  \frac{1}{N} \sum_{\bm{k}_1, \bm{k}_2} e^{\mathrm{i}(-\vec{k}_1+\vec{q})\cdot \vec{u}} e^{-\mathrm{i}(-\vec{k}_2+\vec{q})\cdot \vec{v}}\gt_{\bm{k}_1,\bm{k}_2, \alpha \beta, \gamma \delta}(\omega,\bm{q}),
    \end{split}
\end{equation}
showing that our requirement of locality is in fact equivalent to tracing out the internal momenta $\vec{k}_1$ and $\vec{k}_2$.

In analogy to the discussion in Sec.~\ref{sec:G1}, the collection of all the expectation values~\eqref{eq:G2q explicit} can be recast into a matrix, indicated as $\Gt(\omega, \vec{q})$.
For each value of momentum $\vec{q}$, we define $\{(\alpha, \beta,\vec{u}),(\gamma, \delta,\vec{v})\}$ as compact indices  for $\Gt(\omega, \vec{q})$. This corresponds to considering the two electronic creation (annihilation) operators as a single operator, and the Green's function can be seen as a matrix with entries
\begin{equation}\label{eq:G2 with O operators}
\begin{split}
    \gt_{(\alpha\beta\vec{u}), (\gamma\delta\vec{v})}&(\omega=0, \Vec{q})  \\
    =&\expval{\hat{O}_{\vec{q},(\alpha\beta\vec{u})}[ \hat{H}-E_0]^{-1}\hat{O}^{\dagger}_{\vec{q},(\gamma\delta\vec{v})}}_{\mathrm{GS}}\\
    &+ \expval{\hat{O}^{\dagger}_{\vec{q},(\gamma\delta\vec{v})} [\hat{H}-E_0]^{-1}\hat{O}_{\vec{q},(\alpha\beta\vec{u})}}_{\mathrm{GS}},
\end{split}
\end{equation}
where we defined
\begin{equation}
    \hat{O}^{\dagger}_{\vec{q},(\gamma\delta\vec{v})} = \frac{1}{\sqrt{N}}\sum_{\vec{k}} e^{\mathrm{i}(- \vec{k}+\vec{q}) \cdot \vec{v}} \hat{c}^{\dagger}_{\vec{k},\gamma}\hat{c}^{\dagger}_{-\vec{k}+\vec{q},\delta}.
\end{equation}
We dub the matrix $\Gt(\omega=0, \vec{q})$ with entries defined in Eq.~\eqref{eq:G2 with O operators} the \textit{two-particle radius confined Green's function}, which is the central object for the classification of MAL states presented in this work.
In the following, we refer to $\Gt(\omega=0, \vec{q})$ as $\Gt(\vec{q})$ -- or simply $\Gt$ when all the $\vec{q}$ sectors are considered-- for compactness. We note that for bosonic correlation functions, there can be an order-of-limits ambiguity with respect to taking the limits $\omega\to0$ and $\vec{q}\to0$. In our computations we always perform the $\omega\to0$ limit first by setting $\omega=0$ from the start.
Since Eq.~\eqref{eq:G2 with O operators} depends on a single momentum $\bm{q}$, the number of its eigenvalues scales linearly with the system size. Hence, by diagonalizing the matrix $\Gt(\vec{q})$ at each $\vec{q}$, one obtains a set of eigenvalues that can be interpreted as a ``band structure'' of $\Gt$. 
Each eigenstate $v^{\xi}_{\vec{q}(\alpha\beta\vec{u})}$ of the matrix $\Gt$ can now naturally be used to define a two-particle operator
\begin{equation}\label{eq:eigenstate G2}
    \hat{O}^{\dagger}_{\vec{q}, \xi} = \frac{1}{\sqrt{N}} \sum_{(\alpha\beta\vec{u})} v^{\xi}_{\vec{q}(\alpha\beta\vec{u})} \sum_{\vec{k}}e^{\mathrm{i}(- \vec{k}+\vec{q}) \cdot \vec{u}}\hat{c}^{\dagger}_{\vec{k},\alpha}\hat{c}^{\dagger}_{-\vec{k}+\vec{q},\beta},
\end{equation}
which we will refer to as an eigenstate of $\Gt$.
After diagonalizing $\Gt$, we consider its inverse spectrum rather than the original set of eigenvalues, in analogy to the approach used to analyze $\Go$ and such that the resulting eigenvalues can be expressed in units of energy.

Note that, when the ground state is a noninteracting state (hence a single Slater determinant) and at $\omega=0$, for any eigenvalue $\lambda_{2}$ of $\Gt$ it holds that
\begin{equation}
    \frac{1}{\lambda_{2}} \geq \Delta_{\Gt}:= \min(\Delta_{N+2},  \ \Delta_{N-2}),
    \label{eq:G2_bound}
\end{equation}
where we defined
\begin{equation}
\begin{split}
\Delta_{N+2} &:= E_{\mathrm{min}}(N+2) - E_0 >0,  \\
\Delta_{N-2} &:= E_{\mathrm{min}}(N-2) - E_0 > 0,
\end{split}
\end{equation}
with $E_{\mathrm{min}}(N\pm2)$ the lowest energy eigenvalue of the Hamiltonian in the $N\pm2$ particle sector. The latter statement can be understood by considering the electronic operators in Eq.~\eqref{eq:eigenstate G2} in the diagonal basis of the noninteracting Hamiltonian $\hat{H}$. Each eigenstate of $\Gt$ can lead to a non-zero expectation value for one and only one of the two terms in Eq.~\eqref{eq:Lehman G2 real space}, while the other one has to give a zero contribution. When both the single particle operators appearing in the two-particle eigenstates correspond to non-occupied states in the noninteracting ground state, they contribute to the first term in Eq.~\eqref{eq:G2 with O operators}, while if they are both occupied they only contribute to the second term in Eq.~\eqref{eq:G2 with O operators}. For a two-particle operator with one occupied and one non-occupied operator, the total contribution is zero. Therefore, any eigenvalue of $\Gt$ that violates the bound~\eqref{eq:G2_bound} is an indication of an interacting ground state, since it would not be present in a purely Slater determinant state. These are the eigenvalues that we focus on in our classification scheme. 

The eigenvalues of $\Go$ and $\Gt$ taken at zero frequency do not have an immediate physical interpretation in terms of quasiparticles~\cite{Wang_2013}, but nevertheless some understanding in this direction can be developed. 
An eigenstate $\hat{O}^{\dagger}_{\vec{q}, \xi}$ of the two-particle Green's function naturally corresponds to two particle excited states of the form
\begin{equation}\label{eq:bound state G2}
    \ket{\Psi_{\text{ex}}} = \hat{O}^{\dagger}_{\vec{q}, \xi} \ket{\mathrm{GS}}, \quad \ket{\bar{\Psi}_{\text{ex}}} = \hat{O}_{\vec{q}, \xi} \ket{\mathrm{GS}},
\end{equation}
with $\hat{O}^{\dagger}_{\vec{q}, \xi}$ as defined in Eq.~\eqref{eq:eigenstate G2}. In the case of a generic state, Eq.~\eqref{eq:bound state G2} will not give exact eigenstates of the Hamiltonian, however they can serve as an ansatz for two-particle excitations. The $\Gti$ eigenvalue with eigenstate $\hat{O}^{\dagger}_{\vec{q}, \xi}$ gives an estimate of the energy difference between the particle-particle ($\ket{\Psi_{\text{ex}}}$) and hole-hole ($\ket{\bar{\Psi}_{\text{ex}}}$) excited states.
Therefore, states of the form Eq.~\eqref{eq:bound state G2} corresponding to eigenstates violating the bound~\eqref{eq:G2_bound} can be interpreted as ``bound states'' in the particle-particle spectra. States that are far above the bound~\eqref{eq:G2_bound} will correspond to the particle-particle continuum and are therefore less interesting. For the systems we study, the particle-particle continuum is generally well-separated from the bound states below the bound, allowing these eigenstates to be clearly identified. For systems where this separation of scales is not clear, our method will not be applicable.

Let us finally comment on the choice of Green's function. 
Whereas an AL is created by acting with one-particle electron operators, Eq.~\eqref{eq:decomp}, a 2-MAL is created by acting on the vacuum with a two-particle operator, Eq.~\eqref{eq:decomp_MAL}. The noninteracting topology is successfully diagnosed with the Green's function of Eq.~\eqref{eq:retarded G1} and this motivates the choice of Eq.~\eqref{eq:G2 with O operators}, i.\,e., the particle-particle response function, to diagnose the topology of interacting states, as it resembles the structure of the single-particle correlation function, with the replacement of single particle operators with two-particle operators. Indeed, as discussed in the later sections, this quantity shows a clear and easily identifiable signature of a MAL phase in terms of a single low-lying band.

However, this choice of two-particle correlation function is not the only possibility.
Instead of considering the Green's function of the form of Eq.~\eqref{eq:G2q explicit}, we could for instance evaluate the Green's function of the particle-hole type, i.\,e., 
\begin{equation}\label{eq:particle-hole green's function}
\begin{split}
    &\gt_{\mathrm{ph}, \vec{r}_1, \vec{r}_2, \vec{r}_1', \vec{r}_2', \alpha, \beta, \gamma, \delta}(\omega) \\
    &= \mathrm{FT}_{\omega}[- \Theta(t)\expval{ [\hat{c}^{\dagger}_{\bm{r}_2, \beta}(t)\hat{c}_{\bm{r}_1, \alpha}(t),\hat{c}^{\dagger}_{\bm{r}'_1, \gamma}(0) \hat{c}_{\bm{r}'_2, \delta}(0) ] }],
\end{split}
\end{equation} 
which probes particle-hole like excitations on the ground state, as opposed to particle-particle excitations.
This correlation function, rearranged into a matrix and after setting $\omega=0$, has a positive semidefinite spectrum, and it has a noninteracting bound analogous to the one of Eq.~\eqref{eq:G2_bound}. In this case, the bound is set by $1/\Delta_N$, where $\Delta_{N} = E_{\text{ex}}(N) - E_0$ is the energy gap in the $N$-particle sector, with $ E_{\text{ex}}(N)$ the energy of the first excited state in the $N$-particle sector. This is true if one discards the contribution coming from the overlap with the ground state in the Lehmann decomposition, which would result into a divergent eigenvalue.
Although the particle-hole Green's function may be useful to distinguish different MALs, in general it results into a more complicated band structure, as we show in Sec.~\ref{sec:flat H} and App.~\ref{Appendix:particle-hole G2}.
Hence, we choose the particle-particle correlation function $\gt$, which shows a clear signature in its spectrum through which MALs can be more straightforwardly detected.

\bigskip\tocless\subsection{Transformation properties}{sec:transf}
In this section, we shortly outline how to derive the transformation properties of $\Gt$. Consider an element $h \in G$ of the space group $G$ acting on the electronic operators by a unitary symmetry operator $U_{h}$.
A MAL cluster operator that respects the symmetries of the space group $G$ transforms linearly according to a set of real-space representations $A$ of the space group $G$
\begin{equation}
    U_h \hat{O}^{\dagger}_{\vec{r}, \xi} U_h^{-1} = \sum_{\xi'} A^{\vec{r}}_{\xi' \xi}(h) \hat{O}^{\dagger}_{\vec{r}', \xi'},
\end{equation}
with $\vec{r}'$ the position of the cluster operator after transformation. As a  MAL operator transformed to momentum space, e.\,g., as in Eq.~\eqref{eq:MAL operator in k space}, can only depend on the momentum $\vec{q}$, it must transform under a consistent representation $\rho^{\vec{q}}$ of the space group $G$
\begin{equation}
     U_h \hat{O}^{\dagger}_{\vec{q}, \xi} U_h^{-1} = \sum_{\xi'} \rho^{\vec{q}}_{\xi' \xi}(h) \hat{O}^{\dagger}_{\vec{q}', \xi'},
\end{equation}
with $\vec{q}'=R\vec{q}$, $R$ the action of $h$ in momentum space.
From this latter expression and Eq.~\eqref{eq:G2 with O operators}, we deduce that $\Gt$ transforms under the action of $h$ as
\begin{equation}\label{eq:transformation of G2}
\begin{split}
    &\Gt_{(\alpha, \beta, \vec{u}), (\gamma, \delta, \vec{v})}(\vec{q}) \\
    &= 
  \sum_{(\alpha, \beta,\vec{u})',(\gamma, \delta,\vec{v})'} [\rho^{\vec{q}}(h)_{(\alpha, \beta, \vec{u})' (\alpha, \beta, \vec{u})}]^*  \\
    & \qquad\qquad
    \rho^{\vec{q}}(h)_{(\gamma, \delta, \vec{v})' (\gamma, \delta, \vec{v})}   \Gt_{(\alpha, \beta, \vec{u})', (\gamma, \delta, \vec{v})'}(\vec{q}'),
\end{split}
\end{equation}
see App.~\ref{sec:App_G2_transf} for a proof.

\bigskip\tocless\section{Classification of Green's function band structures}{Sec:GF classification}
So far, we have introduced the concept of $n$-MALs, and particularly the one of MALs, and we have defined the two-particle radius confined Green's function for general many-body states.
In this section, we will bridge the two concepts and see how from the spectrum of $\Gt$ we can infer properties of ground states that can be written exactly as MALs, or that can be adiabatically connected to those. 

As we mentioned shortly at the end of Sec.~\ref{sec:G1}, an $n$-particle correlation function $\gn$ completely determines a state characterized by up to $n$-body correlations. This is exactly what is realized in \mbox{$n$-MALs}, as they are constructed as product states of non-overlapping $n$-particle operators and therefore confine correlations to engage at most $n$ electrons.

In Sec.~\ref{sec:flat H}, we derive the band representations of $\Gt$ for MAL states in the limit  of a spectrally flattened many-body Hamiltonian. In this simplified scenario, the potential of $\Gt$ in diagnosing MAL states properties becomes apparent. In Sec.~\ref{sec:Band repr G2}, we propose a framework to derive a full classification of the $\Gt$ band representations of the class of MAL states in the limit of spectrally flattened many-body Hamiltonians. Subsequently, in Sec.~\ref{sec:numerics}, we explore how this extends to more realistic systems, and we present  numerical results for several representative model Hamiltonians in 0D, 1D and 2D.

\bigskip\tocless\subsection{Spectrally flattened many-body Hamiltonian limit}{sec:flat H}
As a simplified example to understand the connections between two-particle radius confined Green's functions and MAL states, we compute the spectrum of $\Go$ and $\Gt$ in the case of an AL or an MAL ground state, using the spectrally flattened many-body Hamiltonian. The latter is a many-body extension of the concept of `flat band Hamiltonian', or `spectral flattening', used in the context of topological band theory~\cite{doi:10.1063/1.3149495}.
In the following, we refer to the spectrally flattened many-body Hamiltonian simply as flattened Hamiltonian, for compactness.

The flattened Hamiltonian is defined as
\begin{equation}\label{eq:flattened Hamiltonian}
    \hat{H} = \mathbb{1} - \Delta \ket{\mathrm{GS}} \bra{\mathrm{GS}},
\end{equation}
where $\ket{\mathrm{GS}}$ indicates the many-body ground state of the system, with energy $(1-\Delta)$. Any generic many-body state orthogonal to $\ket{\mathrm{GS}}$ has energy $1$, hence the system has a many-body gap $\Delta$ which separates the ground state from  all excited states.
We consider a lattice with a set of unit cells $\Lambda$, where the local Hilbert space of each unit cell is composed of three orbitals, each containing two single-particle states connected by TRS. We label these states by 1, 2, 3 and their TRS partners by $\bar{1}$, $\bar{2}$, $\bar{3}$. The two states belonging to each orbital are related by TRS as follows
\begin{equation}\label{eq:TRS of 1,2,3 orbitals}
   \mathcal{T} \hat{c}^{\dagger}_{\vec{r}, j}\mathcal{T}^{-1} = \hat{c}^{\dagger}_{\vec{r}, \bar{j}},\ \  \mathcal{T} \hat{c}^{\dagger}_{\vec{r}, \bar{j}}\mathcal{T}^{-1} = -\hat{c}^{\dagger}_{\vec{r},j}, \ \ j = 1, 2, 3.
\end{equation}
We successively consider the AL ground state
\begin{equation}\label{eq:AL gs flat limit}
        \ket{\mathrm{AL}} =
   \displaystyle  \prod_{\vec{r} \in \Lambda} \hat{c}^{\dagger
    }_{\vec{r},1} \hat{c}^{\dagger}_{\vec{r},\bar{1}} \ket{0},
\end{equation}
and the MAL ground state
\begin{equation}\label{eq:MAL gs flat limit}
\begin{split}
   \ket{\mathrm{MAL}} =\displaystyle \prod_{\vec{r} \in \Lambda} \hat{O}^{\dagger}_{\vec{r}}\ket{0}=\displaystyle\prod_{\vec{r} \in \Lambda} \frac{\big(\hat{c}^{\dagger
    }_{\vec{r},1} \hat{c}^{\dagger}_{\vec{r}, \bar{2}}-\hat{c}^{\dagger
    }_{\vec{r}, \bar{1}} \hat{c}^{\dagger}_{\vec{r},2} )}{\sqrt{2}}\ket{0},
\end{split}
\end{equation}
where we explicitly wrote an expression for the $M^{\xi}$ coefficients of Eq.~\eqref{eq:2-MAL wf}.
From Eqs.~\eqref{eq:AL gs flat limit} and ~\eqref{eq:MAL gs flat limit}, it follows that at a filling of two electrons per unit cell we require two single-particle states to construct an AL, whereas for an MAL operator we require four single-particle states connected pairwise by TRS. 

As a basis set of operators entering in $\Go$, we consider the creation and annihilation operators of the three orbitals in each unit cell $(\hat{c}^{\dagger}_{\vec{r}, j}, \ j =1, 2, 3,\bar{1}, \bar{2}, \bar{3}, \ \forall \vec{r} \in \Lambda)$. As a basis for $\Gt$, we consider all the possible products of two such single-particle operators taken at the same $\vec{r}$, i.\,e., we set $u_{\max}=0$. The case of $u_{\max}>0$ is discussed in App.~\ref{appendix:flat H}, where we show that the extension of the radius beyond $u_{\max}=0$ does not change the universal features in the spectrum of $\Gt$ relevant to us~\footnote{Here we set $\vec{u}=\vec{0}$ in the AL and MAL states proposed, with $\vec{u}$ as introduced Eq.~(\textcolor{red}{3}). It is for this reason that $u_{\text{max}}=0$ is sufficient to capture the relevant correlations of the MAL state in Eq.~(\textcolor{red}{34}).}.
By virtue of the bound in Eq.~\eqref{eq:G2_bound}, we separate the flattened Hamiltonian spectrum of $\Gt$ into two contributions: the \textit{continuum}, composed by eigenvalues whose inverse lies above or at the many-body gap between the $N$ and $N\pm2$ sectors of the Hilbert space $\Delta_{\Gt}$ ($\lambda_2^{-1} \geq \Delta_{\Gt}=\Delta$), and the \textit{interaction driven} spectrum of $\Gt$, with inverse eigenvalues lying strictly below this gap ($\lambda_2^{-1} < \Delta$). With the considerations that follow, we will reproduce the schematic forms of $\Go$ and $\Gt$ already outlined in Fig.~\ref{fig:IntroPlot}\textcolor{red}{e}-\textcolor{red}{g}.

\begin{figure}[t!]
    \centering
    \includegraphics[width=\columnwidth]{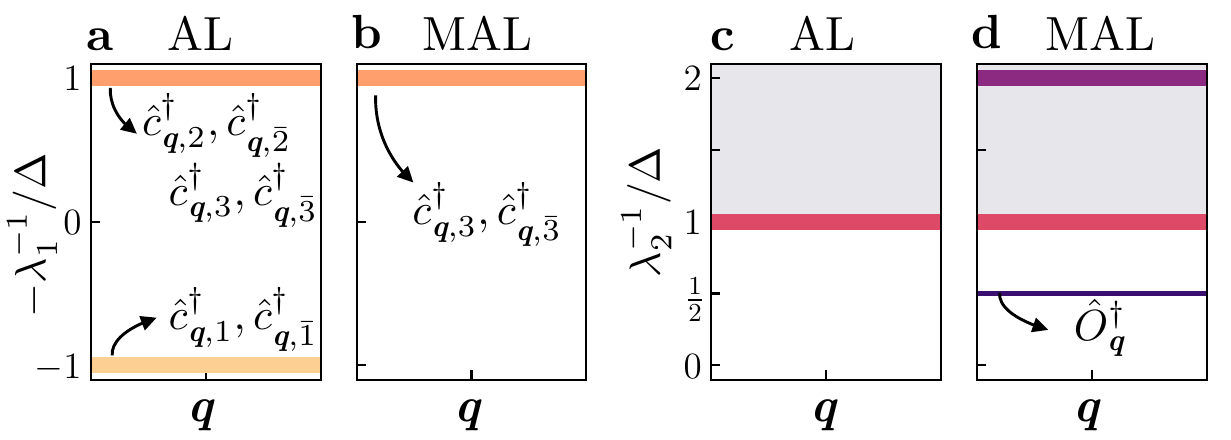}
    \caption{\textbf{Green's functions spectra with a spectrally flattened Hamiltonian.} \textbf{a}-\textbf{b} One- and \textbf{c}-\textbf{d} two-particle Green's functions inverse spectra for a AL or MAL ground state, evaluated in the flattened Hamiltonian limit. We plot $-\lambda^{-1}_1$, instead of $\lambda^{-1}_1$ to maintain the analogy with the topological Hamiltonian. Thick lines indicate eigenvalues with multiplicity higher than one, while the thin line in the MAL $\Gt$ inverted spectrum indicates a singly degenerate eigenvalue. In \textbf{b}, there is a band of eigenvalues corresponding to the empty orbitals $3, \bar{3}$ at $-\lambda_1^{-1}/\Delta = 1$, whereas, for comparison, in Fig.~\ref{fig:IntroPlot}\textcolor{red}{g} this line is missing since there are no empty orbitals remaining in the local Hilbert space.}
    \label{fig:FlatH_G}
\end{figure}

With the AL in Eq.~\eqref{eq:AL gs flat limit} as a ground state of the flattened Hamiltonian~\eqref{eq:flattened Hamiltonian}, $\Go$ and $\Gt$ are diagonal in our chosen basis. The eigenvalues of $\Go$ split into $\lambda_1=-1/\Delta$ and $\lambda_1=+1/\Delta$ contributions (see Fig.~\ref{fig:FlatH_G}\textcolor{red}{a}), respectively given by the eigenstates of $\Go$ corresponding to non-occupied ($\hat{c}^{\dagger}_{\vec{r},j}, \ j = 2, \bar{2}, 3, \bar{3}$) and occupied  ($\hat{c}^{\dagger}_{\vec{r},j}, \ j = 1, \bar{1}$) single-particle operators in the ground state. By listing all the eigenstates of $\Go$ with positive eigenvalue, the AL ground state can be exactly reconstructed.
On the other hand, $\Gt$ has a $\lambda_2 = 1/\Delta$ eigenvalue for any two-particle operator obtained from single-particle operators either both occupied or both non-occupied in the ground state, while it has $\lambda_2 = 0$ for any product mixing an occupied and a non-occupied single-particle operator (see Fig.~\ref{fig:FlatH_G}\textcolor{red}{c}). Hence, all eigenvalues of $\Gt$ for an AL belong to the continuum spectrum, meaning that it is not possible to identify the exact form of the AL ground state from the $\Gt$ spectrum alone.

Note that, the identification of $-\Goi$ with the single-particle Hamiltonian is only justified for a noninteracting Hamiltonian. This is very clear in the flat-band Hamiltonian scenario; since the band structure of $-\Goi$ bears no resemblance with the many-body Hamiltonian. Although the flattened Hamiltonian spectrum as well as the spectrum of $\Goi$ have two levels, their degeneracies are different.

For the MAL ground state of Eq.~\eqref{eq:MAL gs flat limit},
$\Go$ is again diagonal in our chosen basis, and it has eigenvalues $\lambda_1=-1/\Delta$ deriving from non-occupied single-particle operators, and $\lambda_1=0$ for any single-particle operator that appears in the ground state (see Fig.~\ref{fig:FlatH_G}\textcolor{red}{b}). The spectrum of $\Go$ is therefore not useful to determine the exact structure of the MAL ground state. For the MAL state, $\Gt$ has a diagonal block, with eigenvalues $\lambda_2 = 1/\Delta$ for eigenstates composed by two non-occupied operators, and $\lambda_2 = 1/(2\Delta)$ for operators that mix non-occupied single-particle operators and operators appearing in the MAL state.
There is a remaining non-diagonal sector of the $\Gt$ matrix spanned by the operators $V_{\vec{r}} = \{\hat{c}^{\dagger}_{\vec{r}, 1}\hat{c}^{\dagger}_{\vec{r}, \bar{2}},\ \hat{c}^{\dagger}_{\vec{r}, \bar{1}}\hat{c}^{\dagger}_{\vec{r}, 2}\}$ (already Pauli anti-symmetrized). In this subspace of two-particle operators, $\Gt$ assumes the form
\begin{equation}\label{eq:G2 flat H sector MAL}
   \Gt|_{V_{\vec{r}}} = \frac{1}{\Delta}\begin{pmatrix}
    +1 & -1\\
    -1 & +1
    \end{pmatrix}.
\end{equation}
This leads to a set of eigenvalues $\lambda_2 = 2/\Delta$ belonging to the interaction driven spectrum (see Fig.~\ref{fig:FlatH_G}\textcolor{red}{d}). The associated eigenstates are given by the two-particle operators $\hat{O}^{\dagger}_{\vec{r}}$ appearing in Eq.~\eqref{eq:MAL gs flat limit}. The block of $\Gt$ in Eq.~\eqref{eq:G2 flat H sector MAL} yields another eigenvalue $\lambda_2 =0$, with eigenstates given by the two-particle operator $(\hat{c}^{\dagger
    }_{\vec{r}, 1} \hat{c}^{\dagger}_{\vec{r}\bar{2}}+\hat{c}^{\dagger
    }_{\vec{r},\bar{1}} \hat{c}^{\dagger}_{\vec{r},2})/\sqrt{2}$.
Note that the set of eigenvalues discussed above are the same for operators in every decoupled unit cell $\vec{r}$, leading to a flat band in momentum space.

Importantly, the Fourier transform of the eigenstates of this interaction driven band are exactly the two-particle MAL operators composing the ground state. Hence, the eigenstates of the interaction driven band carry the same transformation properties under the symmetries of the system as the MAL operators of the ground state. We say that an MAL operator \textit{induces} an interaction driven band in the spectrum of $\Gt$, meaning that the presence of such an operator in the ground state leads to the existence of a band belonging to the interaction driven spectrum of $\Gt$, and the eigenstates of this band are the momentum-space transformed version of the MAL operators of the ground state.

Note that a band belonging to the interaction driven spectrum of $\Gt$ can only exist if the ground state is an entangled state, adiabatically disconnected from any single Slater determinant state.
This can be understood for instance by considering $\Gt$ in real space: when adding or removing a cluster $\hat{O}^{\dagger}_{\vec{r}}$ to the ground state of Eq.~\eqref{eq:MAL gs flat limit}, the resulting state is non-zero in both cases due to the entanglement between the electrons. As both terms in Eq.~\eqref{eq:G2 with O operators} contribute to the final expectation value, the bound in Eq.~\eqref{eq:G2_bound} can be violated.

As a conclusion of this section, we shortly consider the spectrum of $\Gt_{\mathrm{ph}}$, as introduced in Eq.~\eqref{eq:particle-hole green's function}, for the AL and MAL states of Eqs.~\eqref{eq:AL gs flat limit},~\eqref{eq:MAL gs flat limit}.
For the AL ground state, there is no eigenvalue lying in the interaction driven spectrum, which is once more identified as the part of the inverted spectrum falling below $\Delta$, therefore all the eigenvalues lie at either $\lambda_{\mathrm{ph}}^{-1}/\Delta = 1$ or they diverge $\lambda^{-1}_{\mathrm{ph}}/\Delta \rightarrow \infty$. In contrast to the case of $\Gt$, there is also a divergent eigenvalue $\lambda_{\mathrm{ph}}^{-1}/\Delta = 0$ due to the overlap between the MAL after the action of density-like operators, i.\,e., of the form $\hat{c}^{\dagger}_{i} \hat{c}_i$, with the ground state. 
For the MAL case, let us first note that the MAL state can be equivalently re-expressed as follows
\begin{equation}
\begin{split}
      \ket{\mathrm{MAL}}=
      \prod_{\vec{r}\in\Lambda}\begin{cases}  
       \frac{1}{\sqrt{2}}(\hat{c}^{\dagger}_{\vec{r},1} \hat{c}_{\vec{r},\bar{1}} - \hat{c}^{\dagger}_{\vec{r}, 2} \hat{c}_{\vec{r}, \bar{2}} ) \hat{c}^{\dagger}_{\vec{r}, \bar{1}}\hat{c}^{\dagger}_{\vec{r}, \bar{2}}\ket{0}\\[6pt]
       \frac{1}{\sqrt{2}}(\hat{c}^{\dagger}_{\vec{r}, 1} \hat{c}_{\vec{r}, 2} + \hat{c}^{\dagger}_{\vec{r}, \bar{1}} \hat{c}_{\vec{r}, \bar{2}}) \hat{c}^{\dagger}_{\vec{r}, 2} \hat{c}^{\dagger}_{\vec{r}, \bar{2}}\ket{0}\\[6pt]
    \frac{1}{\sqrt{2}}(\hat{c}^{\dagger}_{\vec{r}, \bar{2}} \hat{c}_{\vec{r}, \bar{1}} + \hat{c}^{\dagger}_{\vec{r}, 2} \hat{c}_{\vec{r}, 1} ) \hat{c}^{\dagger}_{\vec{r}, 1}\hat{c}^{\dagger}_{\vec{r}, \bar{1}}\ket{0}\\[6pt]
       \frac{1}{\sqrt{2}}(\hat{c}^{\dagger}_{\vec{r}, \bar{1}} \hat{c}_{\vec{r}, 1}-\hat{c}^{\dagger}_{\vec{r}, \bar{2}} \hat{c}_{\vec{r}, 2}) \hat{c}^{\dagger}_{\vec{r}, 2}\hat{c}^{\dagger}_{\vec{r}, 1}\ket{0}.
       \end{cases}
\end{split}
\end{equation}
For each of these lines, the operator in parenthesis contributes to the spectrum of $\gt_{\mathrm{ph}}$ with an eigenvalue $\lambda_{\mathrm{ph}}^{-1}/\Delta=1/2$. In addition, there are two eigenvalues stemming from density-like operators, which appear in the interaction driven part of the spectrum, one with $\lambda_{\mathrm{ph}}^{-1}/\Delta=1/2$ and another with $\lambda_{\mathrm{ph}}^{-1}/\Delta=0$, which is divergent due to a non-zero overlap with the ground state when acting with density terms on the MAL state.

Further details on the calculations outlined in this section are discussed in App.~\ref{appendix:flat H} and App.~\ref{Appendix:particle-hole G2}. 

In summary, from the results obtained in this section we infer that the $\Go$ spectrum is useful in determining the form of an AL ground state, but it is not enough to characterize an MAL. On the other hand, the spectrum of $\Gt$ allows to determine the form of an MAL. The $\Gt_{\mathrm{ph}}$ spectrum also carries information on the structure of an MAL, although it has an increased complexity.

\bigskip\tocless\subsection{Classification of $\Gt$ band structures}{sec:Band repr G2}
In this section, we outline how to classify MAL ground states in the limit of a flattened Hamiltonian, based on their $\Gt$ spectrum.

Different ALs can be distinguished by their $\Go$ spectrum, as discussed in Sec.~\ref{sec:G1}, and their classification was already exhausted in the context of TQC. To classify distinct TRS MALs, we solely consider the interaction driven part of the $\Gt$ spectrum, which is absent for ALs but appears for MALs with two-particle entanglement. 

We define the \textit{interaction driven band representation} of $\Gt$ as the collection of representations $\rho_{\vec{q}}$ of the little groups $G_{\vec{q}}$ in which the eigenstates of the interaction driven bands of $\Gt$ transform. It is only well defined if a gap separating the interaction driven and the continuum spectrum of $\Gt$ exists.

The $\Gt$ spectrum of any MAL state with a flattened Hamiltonian is obtained by combining the $\Gt$ spectra that would result from each two-particle operator in Eq.~\eqref{eq:2-MAL wf}, at a fixed unit cell $\vec{r}$, taken individually: every 2-AL operator induces a set of eigenvalues in the continuum spectrum, and every MAL operator disconnected from any AL operator induces an interaction driven band in the spectrum of $\Gt$.
Hence, in order to fully classify the interaction driven band representations of $\Gt$, it is enough to consider all the possible bands induced by states where there is a single MAL cluster operator acting in each unit cell.
All the remaining cases can be deduced from the latter by superimposing the bands induced by each MAL cluster operator appearing in the MAL state. We call the minimal set of band representations that span all the possible interaction driven band representations of $\Gt$ the \textit{elementary MAL-induced band representations} (EMAL).

In the following, we describe how to obtain the list of EMALs for every space group $G$.
We consider a crystalline insulator with space group $G$ and a set of orbitals placed at sites $\vec{x}_a$, $\vec{x}_b, \cdots$ ($a=1, \cdots, m_1$, $b = 1, \cdots, m_2$, $\cdots$) belonging to Wyckoff positions with multiplicities $m_1, m_2, \cdots$, and transforming in a direct sum of representations of the site-symmetry groups $G_{\vec{x}_a}, G_{\vec{x}_b}, \cdots$.
Interaction driven band representations obtained from orbitals that are placed at non-maximal Wyckoff positions can be adiabatically connected to those at maximal Wyckoff positions without the breaking of any symmetry, hence they do not contribute to generating distinct band representations (App.~\ref{subsec:adiabatic connectivity of MALs}), as it happens in TQC~\cite{Bradlyn2017}. Therefore, we only consider orbitals placed at maximal Wyckoff positions.

As a first step towards generating a band representation of $\Gt$, we need to find all the two-particle local cluster representations $A$ in which two-particle MAL cluster operators transform. These representations are constructed out of the single-particle representations of the orbitals, and have to be compatible with the space group $G$. This aspect is discussed in App.~\ref{App: MAL operators}.
This cluster representation induces a representation of $\Gt$ in the space group $G$
\begin{equation}\label{eq:rho_G}
    (A \uparrow G) = \rho.
\end{equation}
Then, the method to classify band representations proceeds as in TQC.
To find the band representation induced by $\rho$ at some specific value of the momentum  $\vec{q}$, one has to subduce the representation~\eqref{eq:rho_G} to the little group $G_{\vec{q}}$, leading to a two-particle momentum representation
\begin{equation}\label{eq:rho_q subduced}
    (\rho \downarrow G_{\vec{q}}) = \rho_{\vec{q}}.
\end{equation}
The collection of representations~\eqref{eq:rho_q subduced} at the maximal momenta of the Brillouin zone have to be continuously connected via maximal momentum lines, and this leads to certain compatibility relations. In principle, these compatibility relations need to be solved numerically using a graph theory algorithm~\cite{Bradlyn2017}. 

This way, all the possible band representations induced by MAL states compatible with the space group $G$ have been listed, provided that the list of $A$ representations is exhaustive. By comparing the band representation of a Green's function of interest to the such constructed band representations, in the limit of flattened Hamiltonian and by restricting the ground states to the class of MAL states, iTQC provides a new definition of topological states: 
\emph{If the band representation associated to the interaction-driven spectrum of a Green's function cannot be induced from ALs and \mbox{$2$-MALs}, the state is either (i) an SPT that cannot be induced from 0-dimensional blocks or (ii) it is a many-body state that is dominated by many-body correlations that involve more than two electrons.}

In Sec.~\ref{sec:1D}, we discuss the full classification of EMALs in 1D, which gives an explicit example of the procedure outlined above. In addition, simple 2D examples where the classification
can be done by hand are discussed in App.~\ref{app:2D_Examples}, and in future work we plan to compile the full tables for all 2D and 3D space groups.

\bigskip\tocless\subsection{Complete classification of ALs and MALs in 1D}{sec:1D}
\begin{figure}[t]
    \centering
    \includegraphics[width=1.\columnwidth]{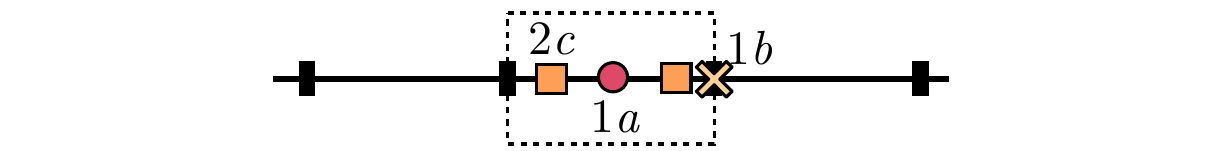}
    \caption{\textbf{One-dimensional lattice.} Wyckoff positions of a 1D lattice. The inversion center coincides with the Wyckoff position $1a$.}
    \label{fig:1D_unitcell_irreps}
\end{figure}
One-dimensional systems are already an interesting testing ground for our formalism. Within TQC, they only allow for trivial phases. These are either ALs or obstructed ALs, for which the atomic positions do not coincide with the Wyckoff positions of the AL Wannier functions. An example for the latter is the Su-Schrieffer-Heeger model for spinful electrons~\cite{SSH}.  With interactions, however, it is possible to find phases described by MALs. In Sec.~\ref{sec:diamond chain}, we provide a 1D model and demonstrate how iTQC can be used to identify different phases.

In 1D crystals there are three Wyckoff positions: $1a$, $1b$ and $2c$ (Fig.~\ref{fig:1D_unitcell_irreps}), and as ALs and MALs operators placed at $2c$ are adiabatically connected to the ones placed at $1a$ and $1b$, these cases can be discarded. In this section, we consider only inversion symmetry among the spatial crystalline symmetries, since different AL phases are not mirror indicated in 1D TRS crystals. The site symmetry groups of the sites at Wyckoff positions $1a$ and $1b$ are isomorphic to $C_i$.
Hence, the relevant point-group is the double group of $C_i$ (No. 2), and it contains the identity ($E$), inversion ($\mathcal{I}$), $2\pi$ rotation ($\bar{E}$), and the double inversion ($\bar{E}\mathcal{I}$) operations~\cite{BilbaoElcoro:ks5574}.
This group has four 1D irreps: the spinless ($\text{Tr}\rho(\bar{E})=+1$) representations $A_g$ and $A_u$, and the spinful ($\text{Tr}\rho(\bar{E})=-1$) representations $\bar{A}_g$ and $\bar{A}_u$, where the subscript $g$ ($u$) indicates that the representation is even (odd) under inversion (see Tab.~\ref{tab:character table 1}).

Hence, the site symmetry groups of the sites at Wyckoff positions $1a$ and $1b$ both admit two types of spinful orbitals, either with physical irrep even $(\bar{A}_g \bar{A}_g)$ or odd $(\bar{A}_u \bar{A}_u)$ under inversion, which are often simply referred to as $s$ and $p$ orbitals~\footnote{We denote by $\bar{A}_1\bar{A}_2$ the physical irrep that realizes TRS, constructed out of the two irreducible representations $\bar{A}_1, \, \bar{A}_2$ of a certain point-group $G_{\vec{x}}$. These physical representations are the corepresentations of the magnetic point-group constructed as $G_{\vec{x}} + \mathcal{T}G_{\vec{x}}$, which is a type-II magnetic group.}. The corresponding creation operators acting in momentum space are labeled by $\hat{c}^{\dagger}_{k, \alpha}$, with $\alpha = (W, \tau, \sigma)$. Here, $\tau \in \{\bar{A}_g \bar{A}_g, \bar{A}_u \bar{A}_u\}$ indicates the spinful orbital representation of the site-symmetry group $G_{\vec{x}}$ of a site $\vec{x}$ at Wyckoff positions $W \in \{ 1a, 1b\}$, and $\sigma\in\{\uparrow, \downarrow\}$ is the spin degree of freedom, which enumerates the two states connected by TRS in each orbital.

The action of inversion and TRS on the orbitals transforming in the two possible $\tau$ irreps is
\begin{equation}\label{eq:single orbital irrep}
   \tau_{\sigma' \sigma}(\mathcal{I}) = \pm \mathbb{1}_{\sigma' \sigma}, \quad \tau_{\sigma' \sigma}(\mathcal{T}) = \mathrm{i} \sigma^{(2)}
    _{\sigma' \sigma},
\end{equation}
where the plus sign holds for $\tau= \bar{A}_g \bar{A}_g$ and minus sign for $\tau = \bar{A}_u \bar{A}_u$. Note that inversion acts trivially on the spin degree of freedom.
The site symmetry group representation in momentum space induces the Wyckoff position-dependent representations at momentum $q$, which read
\begin{equation}\label{eq:site-symm induced rho_q 1D}
    \rho^q_{1a}(\mathcal{I}) = 1, \quad \rho^q_{1b}(\mathcal{I}) = e^{\mathrm{i}q},
\end{equation}
for inversion, while TRS acts by complex conjugation.

After having listed the allowed single-particle representations~$\tau$, we need to find the representations of two-particle operators consistent with the space group and constructed from single-particle state representations.
In general, this task can be rather cumbersome, but it is enough to narrow down the search to some restricted cases to obtain a full classification of EMALs in 1D: we only consider MAL operators constructed out of single particle operators that are placed at the same (maximal) Wyckoff position and same unit cell. Then, the local cluster representation $A$ is obtained by taking tensor products of pairs of the single-particle representations $\tau$, and the two-particle representation $\rho$ becomes the tensor product between $A$ and the momentum space representations induced by the site symmetry groups~\eqref{eq:site-symm induced rho_q 1D} (see App.~\ref{app:MAL representation from single-particle representations}). The relevant multiplication rules are $\bar{A}_g \otimes \bar{A}_g =\bar{A}_u \otimes \bar{A}_u = A_g$, and $\bar{A}_g \otimes \bar{A}_u = A_u$. Explicitly, the possible $A$ representations are given by
\begin{equation}\label{eq:list of MAL irreps}
\begin{split}
    &(\bar{A}_g \bar{A}_g) \otimes (\bar{A}_g \bar{A}_g)  = A_g \oplus  A_g \oplus  A_g \oplus  A_g,\\
    & (\bar{A}_u \bar{A}_u)\otimes (\bar{A}_u \bar{A}_u) = A_g \oplus  A_g \oplus  A_g \oplus  A_g,\\
     & (\bar{A}_g \bar{A}_g)\otimes (\bar{A}_u \bar{A}_u)= A_u \oplus  A_u \oplus  A_u \oplus  A_u,
\end{split}
\end{equation}
for both the Wyckoff positions $1a$ and $1b$.
Each tensor product in~\eqref{eq:list of MAL irreps} is four-dimensional and splits into a 1D spin-0 and a 3D spin-1 sector. The constraints of TRS and uniqueness of the ground state single out the spin-0 sector, as it is the only non-degenerate representation appearing in the presence of $SU(2)$ symmetry, leaving only a single term, out of four, in each line of Eq.~\eqref{eq:list of MAL irreps}. For the spin-1 sector, three 1D representations should be filled to fulfill $SU(2)$ symmetry, leading to an AL state. This leaves us with two possible types of $A$ representation, $A_g$ or $A_u$, irrespective of whether we consider orbitals at $1a$ or $1b$.

To obtain the full two-particle representation $\rho$, we consider the tensor product between local cluster spin-singlet irreps $A$ and one of the momentum space representation $\rho^q_W$ of Eq.~\eqref{eq:site-symm induced rho_q 1D}, depending on whether the orbitals are placed at the $1a$ or $1b$ Wyckoff position sites. This leads to the two-particle representations
\begin{equation}\label{eq:two particle q MAL representations 1D}
    \rho_{W, A_g} = A_g \otimes \rho^q_{W}, \quad   \rho_{W, A_u} = A_u \otimes \rho^q_{W},
\end{equation}
with $W \in \{1a, 1b\}$.

The two-particle representations are labeled by the $A$ representation from which they originate and the position of the two orbitals, which is unique as the two single orbital locations coincide.

To give an explicit example, we write an MAL operator transforming in the $\rho_{1a, A_u}$ representation, subduced to a specific momentum $q$, as
\begin{equation}\label{eq:example 1D MAL}
\begin{split}
    \hat{O}^{\dagger}_{q, 1a, A_u} = 
    \frac{1}{\sqrt{2N}} & \sum_k 
    \big(\hat{c}^{\dagger}_{k, 1a, \bar{A}_g \bar{A}_g, \uparrow}\hat{c}^{\dagger}_{-k+q, 1a, \bar{A}_u \bar{A}_u, \downarrow} \\
    &-\hat{c}^{\dagger}_{k, 1a,\bar{A}_g \bar{A}_g, \downarrow}\hat{c}^{\dagger}_{-k+q, 1a, \bar{A}_u \bar{A}_u, \uparrow}\big).
\end{split}
\end{equation}
with $N$ the number of unit cells in the 1D lattice. 

Based on the set of 1D two-particle representations of Eq.~\eqref{eq:two particle q MAL representations 1D}, the list of EMALs for the 1D space group with inversion ($\bar{1}$) can be inferred.
Table~\ref{tab:List of AL and MAL in 1D} lists the irrep of the $\Gt$ interaction driven bands at the maximal momenta $\Gamma$ ($k=0$) and $X$ ($k=\pi$) of the Brillouin zone, obtained by subducing the representation $\rho$ of each MAL operator to the little groups $G_{\Gamma}$ and $G_{X}$.
Note that some of the EMALs can be induced by a 2-AL operator (e.\,g., $\hat{O}^{\dagger}_{1a, A_g}$), while others cannot be obtained by any 2-AL operator (e.\,g., $\hat{O}^{\dagger}_{1a, A_u}$). This follows from the fact that some of the MALs can be adiabatically connected to ALs without the breaking of any relevant symmetry of the crystal point-group, and by tuning off interactions in the system, while there are intrinsically interacting MALs disconnected from any noninteracting state.

\begin{table*}[t]
    \centering\setlength\tabcolsep{0pt}
    \begin{tabular*}{\linewidth}{@{\extracolsep{\fill}}l c cc cc}
    \hline
    \hline
    
    &  & \multicolumn{2}{c}{$\Gt$} & \multicolumn{2}{c}{$\Gt_{\mathrm{ph}}$}  \\\cmidrule(lr){3-4}\cmidrule(lr){5-6}
     $\tau \otimes \tau'  $ & $\rho$ & $\Gamma$ & $X$ & $\Gamma$ & $X$  \\ 
     \hline
     \hline
     &&&&& \\
     & \\
       $(1a, \bar{A}_g \bar{A}_g) \otimes (1a, \bar{A}_g \bar{A}_g)$,  $(1a, \bar{A}_u \bar{A}_u) \otimes(1a, \bar{A}_u \bar{A}_u)$ & $1a$, $A_g$  & $\Gamma^+_1$ & $X^+_1$ & $\displaystyle\bigoplus_{i=1}^5\Gamma^+_1$ & $\displaystyle\bigoplus_{i=1}^5 X^+_1$ \\
       \hline
       &&&&& \\
       $(1b, \bar{A}_g \bar{A}_g) \otimes(1b, \bar{A}_g \bar{A}_g)$,
        $(1b, \bar{A}_u \bar{A}_u) \otimes(1b, \bar{A}_u \bar{A}_u)$  & $1b$, $A_g$ & $\Gamma^+_1$ & $X^-_1$ & $\displaystyle\bigoplus_{i=1}^5 \Gamma^+_1$ & $ \displaystyle\bigoplus_{i=1}^5 X^-_1$  \\
    \hline  
    &&&&& \\
        $(1a, \bar{A}_g \bar{A}_g) \otimes(1a, \bar{A}_u \bar{A}_u)$ &  $1a$, $A_u$ & $\Gamma^-_1$ & $X^-_1$ & $ \displaystyle\bigoplus_{i=1}^3 \Gamma^+_1 \displaystyle\bigoplus_{i=1}^2 \Gamma^-_1$ & $ \displaystyle\bigoplus_{i=1}^3 X^+_1 \displaystyle\bigoplus_{i=1}^2 X^-_1$\\
        \hline
        &&&&& \\
        $(1b, \bar{A}_g \bar{A}_g) \otimes(1b, \bar{A}_u \bar{A}_u)$ &  $1b$, $A_u$ & $\Gamma^-_1$ & $X^+_1$ &  $\displaystyle\bigoplus_{i=1}^3 \Gamma^+_1 \displaystyle\bigoplus_{i=1}^2 \Gamma^-_1$ & $ \displaystyle\bigoplus_{i=1}^3 X^-_1 \displaystyle\bigoplus_{i=1}^2 X^+_1$\\
        \hline
    \end{tabular*}
    \caption{\textbf{MAL band representation in 1D.} The first  column lists all the labels of the two single particle representations $\tau, \tau'$ whose tensor products give rise to an allowed two-particle representation $\rho$.  The resulting $\rho$'s labels are listed in the second column. The columns for the $\Gt$ representations contain the representations of the lowest lying band in $\Gt$ obtained by subducing $\rho$ to the maximal momenta $\Gamma$ and $X$. In the last two columns, the irreps of the lowest lying bands of the particle-hole Green's function $\Gt_{\mathrm{ph}}$ are listed, see Sec.~\ref{sec:flat H} and App.~\ref{Appendix:particle-hole G2}, and the contribution of the trivial zero eigenvalue is removed.}
    \label{tab:List of AL and MAL in 1D}
\end{table*}

In 1D, the MAL classification exhausts all the possibilities in terms of inversion eigenvalues at maximal momenta, meaning that any single interaction driven band representation of $\Gt$ is equivalent to one of the EMALs listed in Tab.~\ref{tab:List of AL and MAL in 1D}. For a state constructed as a product of more than one MAL operator per unit cell, there will be multiple interaction driven bands in the spectrum of $\Gt$, each one described by one of the band representations of Tab.~\ref{tab:List of AL and MAL in 1D}. However, states not adiabatically connected to MALs or ALs will not be captured by this classification, and may for instance result in a number of interaction driven bands which is not compatible with the number of predicted interaction driven bands at the same filling.

In 2D, the classification is richer, and MALs do not realize all the possible combinations of representations at maximal points, in principle allowing for topologically non-trivial band representations in $\Gt$, or, once more, for bands that are not captured by the MAL ansatz.

\bigskip\tocless\section{Numerical results}{sec:numerics}

So far, we have only made statements on the classification of MALs in the limit of a flattened Hamiltonian.
With more realistic Hamiltonians, our conclusions obtained in the idealized scenario of Sec.~\ref{sec:flat H} no longer hold exactly. However, our numerical calculations show that as long as the ground state is a gapped state adiabatically connected to an MAL and dominated by two-body correlations, the spectrum of $\Gt$ still conveys information on the properties of the ground state. This indicates that the spectrum of $\Gt$ is a powerful tool that can be useful beyond the perturbative regime of small interaction strength. 

Away from the ideal limit, the interaction driven bands predicted in the flattened Hamiltonian limit show a momentum dependence, while still retaining their symmetry properties and a gap separation from the bands characterized by larger eigenvalues. The lowest bands in the spectrum of $\Gti$ may not necessarily lie below the bound~\eqref{eq:G2_bound}, but still retain a gap separation from the remaining bands. Conversely, the presence of these bands in the spectrum of $\Gti$ hints at a ground state that is either closely approximated by an MAL state or adiabatically connected to an MAL state and still dominated by two-body correlations, rather than correlations involving a higher number of particles.

In our numerical calculations, we find that this signature of MAL ground state is robust at large values of the interaction strength, as long as the many-body Hamiltonian remains fully gapped and the ground state is dominated by two-body correlations rather than higher order correlations. Therefore we speak of \emph{MAL phases} to indicate a phase of a system where the ground state can be adiabatically connected to an MAL but not to any AL, without any gap closings or the breaking of symmetries of the system. On the other hand, we  say a system is in an AL phase, when its ground state can be adiabatically connected to an AL, without gap closings or symmetry breaking.

In this section, we will present a series of numerical calculations for different model Hamiltonians where the considerations of Sec.~\ref{Sec:GF classification} still hold for a range of parameters. These examples showcase how the classification of the interaction driven band representations of $\Gt$ can be used to infer properties of certain interacting ground states, beyond the limit of flattened Hamiltonian or MAL state.

\begin{figure}[t]
    \centering
    \includegraphics[width=\columnwidth]{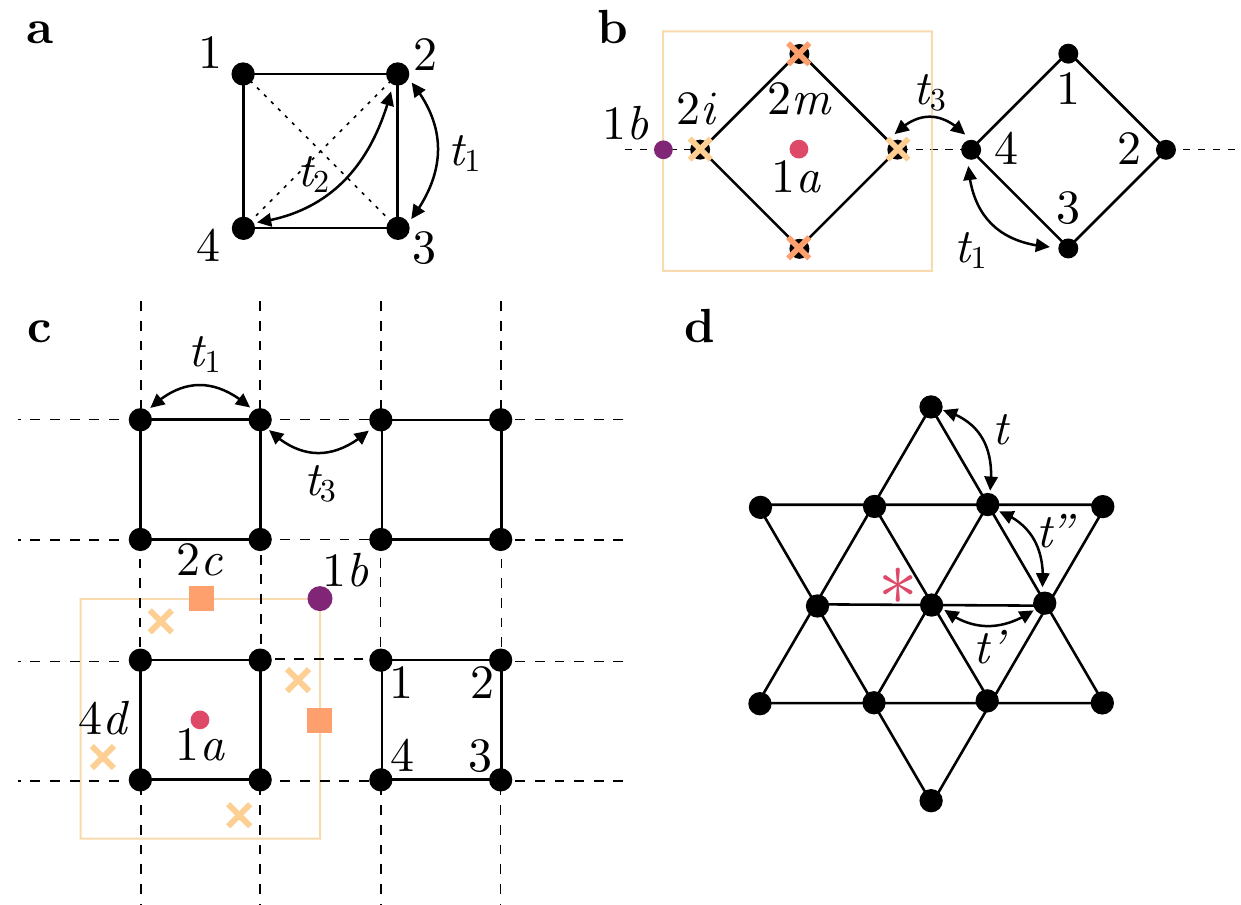}
    \caption{\textbf{Schematics of the models considered.} Schematics of \textbf{a} the Hubbard square (Sec.~\ref{sec:hubbard square}), \textbf{b} the Hubbard diamond chain (Sec.~\ref{sec:diamond chain}),  \textbf{c} the checkerboard lattice of Hubbard squares (Sec.~\ref{sec:checkerboard hubbard}) and \textbf{d} the star of David cluster (Sec.~\ref{sec:star of David}). In \textbf{b} and \textbf{c} some Wyckoff positions are explicitly marked for the two lattices, and the unit cell is highlighted by the yellow rectangle. The tunneling amplitudes discussed in the various models are highlighted in each schematic.}
    \label{fig:models_schematics}
\end{figure}

We will first discuss the Hubbard square (Sec.~\ref{sec:hubbard square}), which realizes a 0D MAL state in the limit of vanishing interactions. We then study two examples where the Hubbard square is used as a building block to construct first a 1D lattice, the Hubbard diamond chain (Sec.~\ref{sec:diamond chain}), and then a 2D checkerboard lattice (Sec.~\ref{sec:checkerboard hubbard}). As a last model, we present another example of 0D Hubbard cluster, the Hubbard star of David (Sec.~\ref{sec:star of David}), which may find application in the context of the layered material $\mathrm{1T-TaS_2}$~\cite{PhysRevX.7.041054}. The four schematics of the models considered throughout this section are shown in Fig.~\ref{fig:models_schematics}.

For each model presented in the following sections, we use symmetry groups that contain a minimal set of symmetries that allow to distinguish the various phases appearing in the phase diagrams, rather than the full symmetry groups. This choice is motivated by the aim of keeping the notation simple, since considering the full symmetry group would lead to a more complicated analysis while the results would remain substantially unchanged.

There are several numerical techniques which enable one to calculate the two-particle Green's function $\Gt$, such as exact diagonalization (ED) and QMC. In this section we will focus on systems amenable to these two techniques. In principle, these numerical techniques are also useful in the computation of $n$-particle Green's functions with generic $n$.
A more detailed discussion on the numerical calculations whose results are shown in the following sections is presented in App.~\ref{app:Numerics}.

Importantly, the QMC results presented in Sec.~\ref{sec:checkerboard hubbard} exemplify how different SPT phases can be distinguished by means of $\Gt$ for system sizes beyond the reach of methods such as ED and DMRG. While ED and DMRG provide access to the full many-body ground state, in our formulation the knowledge of the ground state is not required, nor the calculation of correlation functions involving an extensive number of electronic operators. This has to be contrasted with conventional techniques developed to diagnose SPT phases, such as the evaluation of string order parameters~\cite{PhysRevB.86.125441} and partial symmetry operations applied to subsystems of size comparable with the total size of the system~\cite{PhysRevLett.118.216402,PhysRevB.98.035151,PhysRevB.95.205139}. Both of these approaches to obtain SPT invariants involve an extensive number of fermionic and swap operators, respectively, which can be obtained efficiently when the ground state is known and in 1D systems~\cite{PhysRevB.86.125441}, but become computationally inaccessible in more than 1D and as the size grows. In addition, canonical pgSPT and cSPT invariants are quantized~\cite{PhysRevB.98.035151,PhysRevLett.118.216402,PhysRevB.95.205139}, and therefore it is reasonable to expect that their value heavily suffers from finite-size effects. Our method is still applicable for modest system sizes, as it only requires the presence of a gap in the many-body and in the $\Gt$ spectrum, while it does not rely on a quantized response.

\bigskip\tocless\subsection{Hubbard square}{sec:hubbard square}
As a 0D numerical example, we compute the $\Gt$ spectrum in ED for the Hubbard square, a four-site interacting fermionic model studied in Refs.~\cite{HubbardSquare1,HubbardSquare2,FMIPhysRevLett.105.166402,Hubbard_mol}.
A spinful electron is placed at each site of the square~\footnote{For each site of the Hubbard square, the only symmetry leaving the site invariant is the diagonal mirror operation passing through the site. Hence, the site symmetry group of each of the sites at the corner is the double group of $C_s$. A spinful orbital transforming in the spinful trivial physical representation ($^1\bar{E} \, ^2\bar{E}$) of the double group of $C_s$ is placed at each site of the square (see Tab.~\ref{tab:character table Cs}). This corresponds to a spin-1/2 with quantization axis along the out-of-plane direction.}, and
the Hamiltonian is defined as
\begin{align}\label{eq:Hamiltonian Hubbard square}
   \hat{H}_{\mathrm{HS}} &= -t_1 \sum_{i=1}^4\sum_{\sigma} (\hat{c}^{\dagger}_{i, \sigma} \hat{c}_{i+1, \sigma} + \text{h.c.}) \\ \nonumber
   &+ t_2 \sum_{i=1}^4\sum_{\sigma} \hat{c}^{\dagger}_{i \sigma} \hat{c}_{i+2, \sigma}+ U \sum_{i=1}^4 (\hat{n}_{i, \uparrow}-\frac{1}{2})(\hat{n}_{i, \downarrow}-\frac{1}{2}),
\end{align}
with $\hat{c}^{\dagger}_{i\sigma}$ ($\hat{c}_{i\sigma}$) creating an electron at site $i=1,\dots, 4$, with the spin $\sigma\in{\uparrow, \downarrow}$ labeling the two single-particle states of each orbital. 
It is convenient to transform the single-particle states in eigenstates of the four-fold rotation operation ($C_4$)
\begin{equation}
    \hat{c}^{\dagger}_{\ell,\sigma} = \frac{1}{2} \sum_{j=1}^4 e^{\mathrm{i}\ell j} \hat{c}^{\dagger}_{j,\sigma}, \ \ \hat{c}_{\ell,\sigma} = \frac{1}{2} \sum_{j=1}^4 e^{-\mathrm{i}\ell j} \hat{c}_{j,\sigma},
\end{equation}
with eigenvalues $\ell\in\{0, \frac{\pi}{2}, \pi, -\frac{\pi}{2}\}$.
The Hamiltonian becomes
\begin{align}
    \hat{H}_{\mathrm{HS}} =& \sum_{\ell, \sigma} \left(\varepsilon(\ell)- \mu - \frac{U}{2}\right) \hat{c}^{\dagger}_{\ell, \sigma} \hat{c}_{\ell, \sigma}\\ \nonumber
    & + \frac{U}{N} \sum_{\ell, \ell', \ell''} \hat{c}^{\dagger}_{\ell, \uparrow} \hat{c}^{\dagger}_{\ell', \downarrow} \hat{c}_{\ell''+\ell, \downarrow} \hat{c}_{-\ell''+\ell', \uparrow},
\end{align}
with $\varepsilon(\ell) = -2 t_1 \text{cos}(\ell) + t_2 \text{cos}(2\ell)$, and $N=4$ the number of sites.
The $C_4$ symmetric electronic single particle states transform as
\begin{equation}\label{eq:single part C4 transformation}
    C_{4} \hat{c}^{\dagger}_{\ell, \sigma}   C_{4}^{-1} = \sum_{\sigma'} e^{\mathrm{i}(\ell + \frac{\pi}{4} \sigma^{(3)}_{\sigma\sigma'})} \hat{c}^{\dagger}_{\ell, \sigma'}.
\end{equation}
The minimal symmetry group to distinguish the phases of the Hubbard square is the double group $C_{4}^D$ (see Tab.~\ref{tab:c4 double character table}), and the eight $\hat{c}^{\dagger}_{k, \sigma}$ operators transform as a set of double-valued physical representations of $C^{D}_{4}$ (see App.~\ref{App: example MAL construction: Hubbard square}).
Focusing on the case of half-filling, the Hubbard square is characterized by two distinct phases: trivial ($t_2>t_1$) and non-trivial ($t_2 <t_1$). 
\begin{figure}[t!]\centering\includegraphics[width=1\columnwidth]{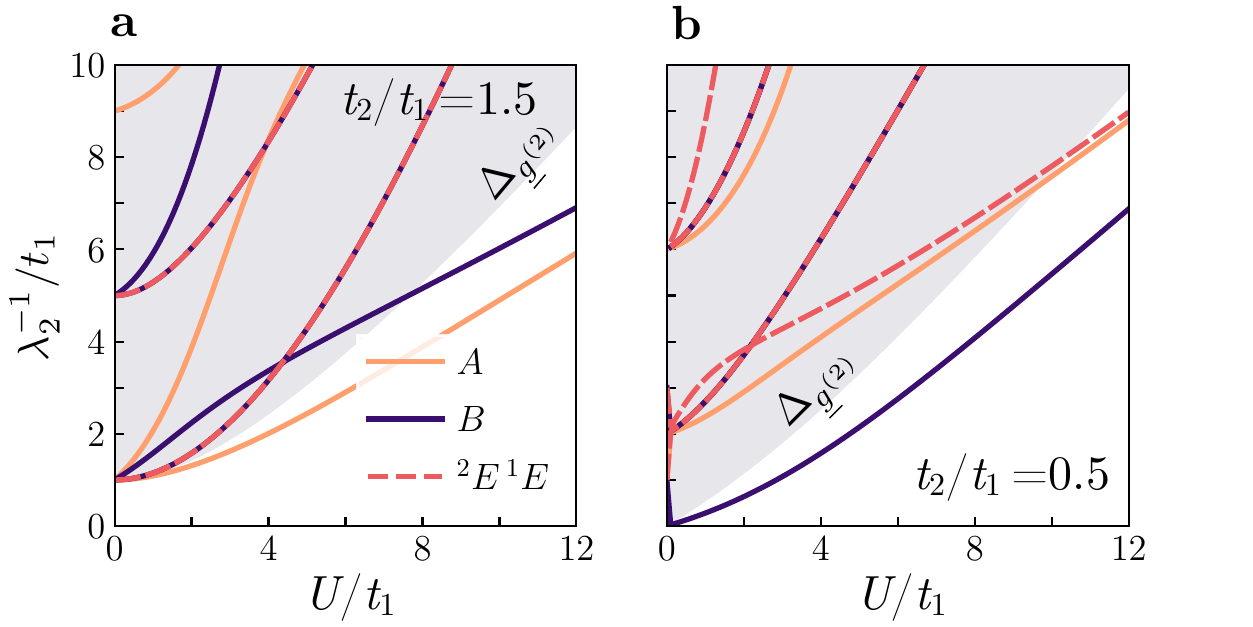}
    \caption{\textbf{Hubbard square.} Inverse eigenvalues of the 2-particle Green's function for \textbf{a} the trivial phase ($t_2>t_1$) and \textbf{b} the non-trivial phase ($t_2<t_1$) of the Hubbard square~\eqref{eq:Hamiltonian Hubbard square}. The boundary of the region shaded in yellow marks the two-particle gap $\Delta_{\Gt}$ at each value of the Hubbard interaction strength $U$. The eigenvalues of $\Gt$ are colored according to the legend in \textbf{a}, with the color-code distinguishing the $C^D_4$ representation in which their eigenstates transform, either $A$, $B$ or the corepresentation $^1E \, ^2 E$, see Tab.~\ref{tab:c4 double character table}. Solid (dashed) lines indicate singly (doubly) degenerate eigenvalues.
    The state in \textbf{b} below $\Delta_{\Gt}$ transforming in the $B$ irrep of $C_4$ indicates the non-trivial MAL character of the ground state. All quantities are expressed in units of $t_1$.}
    \label{fig:FMIsquare}
\end{figure}

In the limit of $U/t_1\rightarrow0$, the ground state of the trivial phase is a gapped state of the single Slater determinant form
\begin{equation}
    \ket{\mathrm{T}} = \hat{c}^{\dagger}_{\frac{\pi}{2}, \uparrow} \hat{c}^{\dagger}_{\frac{\pi}{2}, \downarrow} \hat{c}^{\dagger}_{-\frac{\pi}{2}, \uparrow} \hat{c}^{\dagger}_{-\frac{\pi}{2}, \downarrow} \ket{0}.
\end{equation}
This wave function transforms under the trivial representation of the point-group ($A$), meaning $C_4 \ket{\mathrm{GS}}=+\ket{\mathrm{GS}}$. In the limit $U/t_2\to\infty$ the wave function takes an alternative form with the same $C_4$ eigenvalue, 
\begin{equation}
|\mathrm{T'}\rangle=\hat{O}_{13}^{\dagger} \hat{O}_{24}^{\dagger}|0\rangle,
\end{equation}
where $\hat{O}_{i j}^{\dagger} \equiv[c_{i, \uparrow}^{\dagger} c_{j, \downarrow}^{\dagger}+c_{j, \uparrow}^{\dagger} c_{i, \downarrow}^{\dagger}] / \sqrt{2}$. The states $|\mathrm{T}\rangle$ and $|\mathrm{T'}\rangle$ can be continuously connected as they are characterized by the same $C_4$ symmetry eigenvalues.

For the non trivial phase, in the limit of $U/t_1\rightarrow0$, the square has the unique ground state
\begin{equation}\label{eq:GS Hubbard square non trivial}
    \ket{\mathrm{NT}} = \frac{1}{\sqrt{2}}\hat{c}^{\dagger}_{0, \uparrow} \hat{c}^{\dagger}_{0, \downarrow} (\hat{c}^{\dagger}_{\frac{\pi}{2}, \uparrow} \hat{c}^{\dagger}_{\frac{\pi}{2}, \downarrow}+\hat{c}^{\dagger}_{-\frac{\pi}{2}, \uparrow} \hat{c}^{\dagger}_{-\frac{\pi}{2}, \downarrow}) \ket{0},
\end{equation}
and in the limit $U/t_1\to\infty$
\begin{equation}
    \left|\mathrm{NT'}\right\rangle=\frac{1}{\sqrt{3}}\left[\hat{O}_{12}^{\dagger} \hat{O}_{34}^{\dagger}-\hat{O}_{14}^{\dagger} \hat{O}_{23}^{\dagger}\right]\ket{0},
\end{equation}
where both states have $C_4\ket{\mathrm{GS}} = - \ket{\mathrm{GS}}$, and transform in the $B$ representation of the point-group (App.~\ref{App: example MAL construction: Hubbard square}). The state in Eq.~\eqref{eq:GS Hubbard square non trivial} realizes an example of MAL, with a $C_4$ eigenvalue that is non-trivial, in the sense that it cannot be reproduced by any TRS AL. For this state, there is a single MAL cluster operator that leads to a contribution in the interaction driven spectrum of $\gt$, the one enclosed in parenthesis in Eq.~\eqref{eq:GS Hubbard square non trivial}. Once more, the states $|\mathrm{NT}\rangle$ and $|\mathrm{NT'}\rangle$ can be continuously connected as they have the same eigenvalue under $C_4$ operation.

Figure~\ref{fig:FMIsquare} shows the ED inverse $\Gt$ spectra obtained in the two phases, trivial (Fig.~\ref{fig:FMIsquare}\textcolor{red}{a}) and non-trivial (Fig.~\ref{fig:FMIsquare}\textcolor{red}{b}), for a range of interaction strength $U/t_1$.
From the spectra of $\Gt$, we find that in each of the two phases there is a single eigenvalue below the two-particle gap $\Delta_{\Gt}$, for $U/t_1$ not too large, whose symmetry eigenvalue under the action of $C_4$ rotation reflects the symmetry of the ground state. As $U$ becomes increasingly larger, the relevant correlation function describing the Hubbard square state becomes the four-particle correlation function. Hence, additional eigenvalues begin to appear in the lower part of the inverted spectrum of $\Gt$.

Larger ``Hubbard molecules" can exhibit even more complex phase diagrams of band insulating and fragile Mott insulating phases~\cite{Mobius}. In Sec.~\ref{sec:star of David}, we present an example of larger cluster, the Hubbard star of David, where we observe a rich phase diagram that includes an MAL phase.

\bigskip\tocless\subsection{Hubbard diamond chain}{sec:diamond chain}
As a 1D example, we consider the Hubbard diamond chain defined in Ref.~\cite{Iraola}. This is constructed by connecting several Hubbard squares through their corners, see Fig.~\ref{fig:models_schematics}\textcolor{red}{b}. 
In this case, the relevant space group is $Pmmm$ (No. 47), and the lattice sites are placed at the $2m$ and $2i$ Wyckoff positions. The site-symmetry groups of sites in Wyckoff position $2m$ and $2i$ are isomorphic to $m2m$ and $2mm$, respectively, and the orbitals localized at these four sites can be continuously connected to orbitals placed at the maximal Wyckoff position $1a$ of the lattice.
Here, we only consider the point-group $C_{2v}$ as it is sufficient in diagnosing the MAL phase (while in Ref.~\cite{Iraola} the analysis is carried out for the full symmetry group $D_{2h}$). As for the Hubbard square, each site has spinful electrons.
The Hamiltonian is
\begin{equation}\label{eq:Hamiltonian DC}
\begin{aligned}
    \hat{H}= &\, U \sum_{j, \alpha} \hat{n}_{j, \alpha, \uparrow} \hat{n}_{j, \alpha, \downarrow}+\sum_{j, \sigma} \sum_{\alpha, \beta} \hat{c}_{j, \alpha, \sigma}^{\dagger} \mathbb{T}_{\alpha \beta} \hat{c}_{j, \beta, \sigma} \\
&-\sum_{\sigma, j}\left(t_{3} \hat{c}_{j, 1, \sigma}^{\dagger} \hat{c}_{j+1, 3, \sigma}+\text {h.c.}\right) - \mu \sum_{j, \alpha, \sigma} \hat{n}_{j, \alpha, \sigma},
\end{aligned}
\end{equation}
with
\begin{equation}
\mathbb{T}=-\left[\begin{array}{cccc}
0 & t_{1} & t_{2} & t_{1} \\
t_{1} & 0 & t_{1} & t_{2} \\
t_{2} & t_{1} & 0 & t_{1} \\
t_{1} & t_{2} & t_{1} & 0
\end{array}\right],
\end{equation}
where $\hat{c}_{j, \alpha, \sigma}^{\dagger}\left(\hat{c}_{j, \alpha, \sigma}\right)$ creates (annihilates) an electron of spin $\sigma$ at site $\alpha \in\{1,2,3,4\}$ of the cell labeled by $j=1, \ldots, N$, with $N$ the number of unit cells. In Eq.~\eqref{eq:Hamiltonian DC}, $t_1$ and $t_2$ are hopping amplitudes within a single square, $t_3$ the one between different squares, and $\mu$ is the chemical potential.
\begin{figure*}[t!]
    \centering
    \includegraphics[width=\textwidth, page=1]{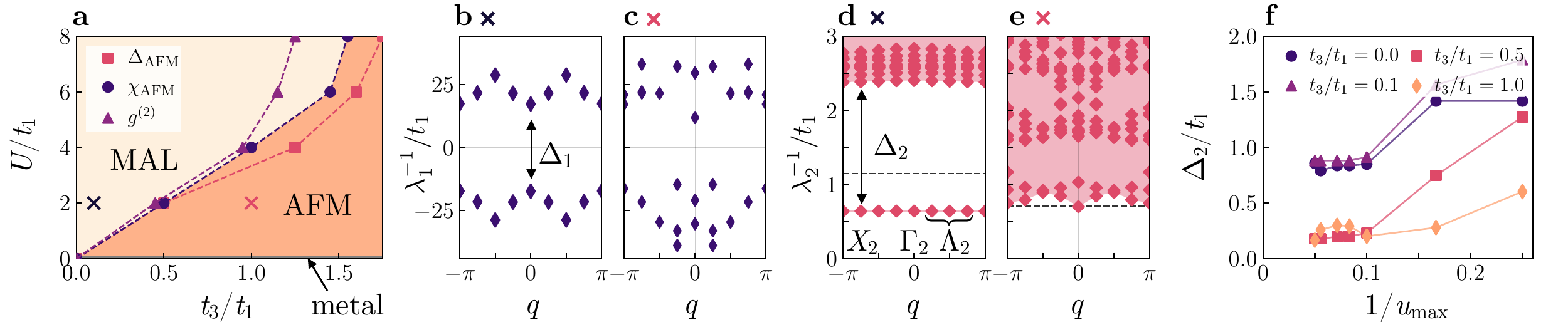}
    \caption{{\bf Hubbard diamond chain.} \textbf{a} Phase diagram of the diamond chain model in the $(U / t_1, t_3 / t_1)$ plane. In the upper part, the system is in an MAL phase, while in the lower it is in an AFM phase. The phase transition is marked using three approaches:
        (i) $\Delta_{\text{AFM}}$: by the vanishing gap between the ground state and the $S = 1$, $k = \pi$ excited state computed within $L \to \infty$ extrapolation of the QMC data, (ii) $\chi_{\text{AFM}}$: inflection point in susceptibility towards the respective Néel order computed within QMC on $L = 10$ unit cells (iii) $\Gt$: by the vanishing gap between the lowest-lying band and the rest of the inverted spectrum of $\Gt$, computed within QMC on the $L = 8$ chain (see App.~\ref{sec:App_QMC}). The two crosses mark the points in the phase diagram corresponding to the plots \textbf{b--e}. {\bf b--c} Inverted band structure of $\Go$ of the diamond chain obtained using QMC simulations, computed at $U / t_1 = 2$ and $t_3 / t_1 = 0.1,\,1.0$, for \textbf{b} and \textbf{c} respectively. The spectrum is doubly degenerate at each value of $q$ due to the spin degree of freedom. {\bf d--e} Inverted band structure of $\Gt$, for $U / t_1 = 2$, $u_{\max}=6$, $L=8$ and $t_3 / t_1 = 0.1,\,1.0$, for \textbf{c} and \textbf{d}, respectively. The dashed line indicates the two-particle gap $\Delta_{\Gt}$ estimated from the spectral gap of $\Goi$, meaning $\Delta_{\Gt} \approx2 \Delta_1$. \textbf{f} Spectral gap $\Delta_2$ between the lowest-lying and the remaining eigenvalues of $\Gti$, as a function of $1/u_{\max}$, computed at fixed $U / t_1 = 2$, and $t_3 / t_1 = 0.0,\,0.1,\,0.5$ and $1.0$.
    }
    \label{fig:hubbard_diamond_chain}
\end{figure*}

The noninteracting model at half-filling has three phases~\cite{Iraola}:
If the hopping $t_2$ dominates, then the individual squares are in the trivial phase of the Hubbard square. This is an insulating AL state. If the hopping $t_3$ dominates, then the model can be adiabatically connected to the interacting SSH chain~\cite{SSH}, with the addition of two weakly coupled sites.
This insulating state is an obstructed AL. These two insulating states will be unaffected by a small $U$ since they are protected by a gap. In fact, it is known that there is no phase transition as a function of $U$ for both the SSH chain~\cite{Lessnich_2021} and the trivial phase of the Hubbard square~\cite{FMIPhysRevLett.105.166402}, it therefore is plausible that there is no phase transition as a function of $U$ in these phases of the Hubbard diamond chain.
On the other hand, if $t_1$ dominates, then the individual squares can be described by the non-trivial phase of the Hubbard square. For $U=0$ this phase is metallic, but for any finite $U$ a gap in the many-body spectrum opens up, analogously with the case of the isolated Hubbard square. Therefore, the system realizes an MAL phase at finite $U$ and dominant $t_1$.

We perform ED and QMC  calculations on the system at $t_2=0$, while varying $t_3/t_1$ and $U/t_1$.
In Fig.~\ref{fig:hubbard_diamond_chain}\textcolor{red}{a}, the phase diagram of the system as a function of these two parameters is shown: at $U=0$ the system is gapless (therefore we label it as metal), at large $U/t_1$ and small $t_3/t_1$ the ground state is described by an MAL state, and at small $U/t_1$ and large $t_3/t_1$ the system realizes an antiferromagnetic (AFM) phase. Figures~\ref{fig:hubbard_diamond_chain}\textcolor{red}{b}--\textcolor{red}{e} show the $\Go$ and $\Gt$ inverse spectra for the case of $u_{\max}=6$, with the unit cell defined as a single diamond (see Fig.~\ref{fig:models_schematics}\textcolor{red}{b}).
In the MAL regime, corresponding to Fig.~\ref{fig:hubbard_diamond_chain}\textcolor{red}{d}, there is a single low-lying band in the inverted spectra of $\Gt$, whose eigenstates have mirror $M_y$ eigenvalue equals to $-1$ at both momenta $q=0,\, \pi$. This signals that the ground state is adiabatically connected to an MAL placed at the $1a$ Wyckoff position of the unit cell, transforming in the $A_{2}$ irrep of the point-group $C^D_{2v}$.
On the other hand, in the large $t_3$ regime of Fig.~\ref{fig:hubbard_diamond_chain}\textcolor{red}{e}, there is no band in the spectrum of $\Gti$ separated from the continuum.

\bigskip\tocless\subsection{Checkerboard lattice of Hubbard squares}{sec:checkerboard hubbard}
As a 2D example, we consider the model proposed in Ref.~\cite{FMIPhysRevLett.105.166402}. This consists of a checkerboard lattice where Hubbard squares, with on-site interaction $U$ and nearest-neighbor hopping $t_1$, are coupled to the neighboring unit cell by a hopping parameter $t_3$, see Fig.~\ref{fig:models_schematics}\textcolor{red}{c}. 
The Hamiltonian for the model reads
\begin{equation}
    \hat{H} = \sum_{\vec{r}} \hat{H}_{\mathrm{HS}, \vec{r}} - \sum_{\substack{\expval{ \vec{r}, i, \vec{r}', j} \\  \vec{r} \neq  \vec{r}'}}\sum_{\sigma} t_3(\hat{c}^{\dagger}_{ \vec{r}, i, \sigma}\hat{c}_{ \vec{r}', j, \sigma} + \text {h.c.}),
\end{equation}
with $ \vec{r},  \vec{r}'$ labeling different unit cells, $i, j \in \{1,\cdots,4\}$ indicating the sites within each unit cell, and $H_{\mathrm{HS},  \vec{r}}$ is the Hamiltonian in Eq.~\eqref{eq:Hamiltonian Hubbard square} for the square in the unit cell labeled by  $\vec{r}$. Here, we neglect the diagonal hopping $t_2$ in Eq.~\eqref{eq:Hamiltonian Hubbard square} as our focus is on the non-trivial phase of the Hubbard square.

\begin{figure*}
    \centering
    \includegraphics[width=\textwidth]{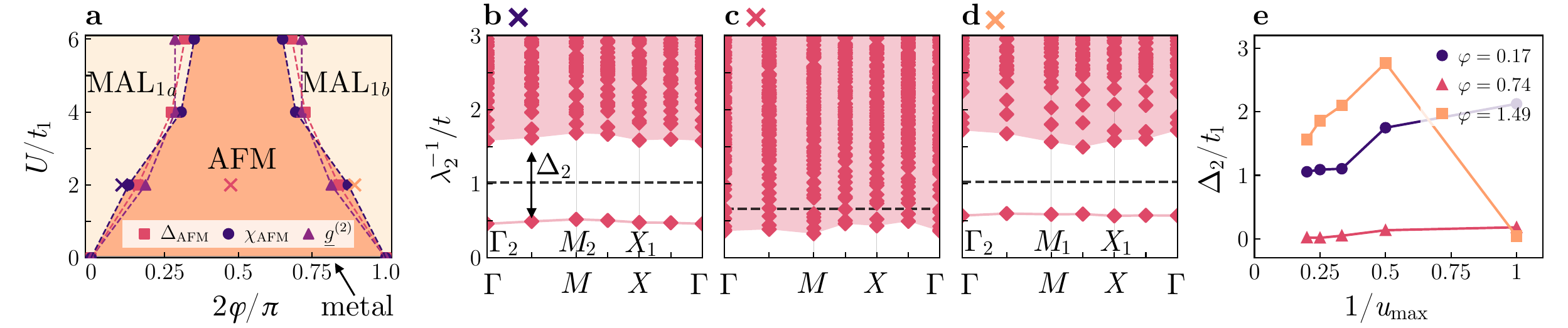}
    \caption{\textbf{Checkerboard lattice of Hubbard squares.} \textbf{a} Phase diagram as a function of $\varphi/(\pi/2)$ and $U/t$ ($t_1= t \cos\varphi, \ t_3 = t \sin \varphi$). The phase boundaries are evaluated using (i) $\Delta_{\text{AFM}}$: the vanishing gap between the ground state and the $S=1$ and $\vec{k}=(\pi, \pi)$ excited state, extrapolated for $L\rightarrow\infty$ from the QMC data, (ii) $\chi_{\text{AFM}}$: the inflection point in susceptibility towards the Néel order, obtained within QMC on the $L=5$ lattice  (corresponding to $L\times L$ unit cells) and (iii) $\Gt$: the gap closing in the inverse spectra of $\Gt$, obtained in QMC for a system with $L = 3$ (see App.~\ref{app:Numerics}). The three crosses mark the points in the phase diagram corresponding to the \textbf{b--d} plots. \textbf{b}--\textbf{d} Inverted spectrum of $\Gt$ in the three regimes \textbf{b} $t_1\gg t_3$ ($\varphi=0.08$), \textbf{c} $t_1 \approx t_3$ ($\varphi=0.74$) and \textbf{d} $t_1 \ll t_3$ ($\varphi=1.49$), for $U/t=2$ and $\varphi=\arctan(t_1/t_3)$. The spectra in \textbf{b}--\textbf{d} are evaluated with QMC, for a system of size $L=4$. \textbf{e}~Convergence of the spectral gap in the inverted spectrum of $\Gt$ as a function of inverse cluster size $1/u_{\max}$, for systems with $L=4$ and $\varphi=0.00,\,0.17,\,1.41,\,1.57$.}
    \label{fig:panel_checkerboard}
\end{figure*}

In the limit $t_1 \gg t_3$ and $U\rightarrow0$, the ground state is described by a product of MAL operators, each transforming in the $B$ representation of the point-group $C^D_{4}$ and placed at the Wyckoff position $1a$ of the lattice. 
In the opposite regime of dominating $t_3$ ($t_1 \ll t_3$ and $U\rightarrow0$) the ground state is also described by a product of MAL operators transforming in the $B$ representation of $C_4^D$, which are, however, placed at the $1b$ Wyckoff position of the lattice. Therefore, at finite $U$, the two regimes of dominating $t_1$ or dominating $t_3$ realize two distinct MAL phases. For the intermediate region $t_1\approx t_3$, the ground state is an AFM at finite $U$ and is gapless (metal) at $U=0$.
The two MAL phases can be distinguished by the band representations that the distinct MALs induce in the spectrum of $\Gt$.
Figures~\ref{fig:panel_checkerboard}\textcolor{red}{b}--\textcolor{red}{d} show the inverted spectrum of $\Gt$ in the three phases. For the two MAL phases, the irreps of the lowest-lying bands at maximal momenta in the Brillouin zone are marked, and they correspond to the band representation induced by an MAL cluster transforming in the $B$ representation of $C_4^D$ placed respectively at the $1a$ and $1b$ Wyckoff position. In App.~\ref{subsec:App 2D examples E1 E2 at 1B}, we derive explicitly the full band representation induced by MAL operators placed at the $1a$ Wyckoff position. The case in which the orbitals are placed at the $1b$ Wyckoff position follows analogously.

\bigskip\tocless\subsection{Hubbard star of David}{sec:star of David}
As a last example of application of our method, we consider a single cluster of atomic sites forming a star of David shape.
This model is motivated by the monolayer material $\mathrm{T}-\mathrm{Ta S}_2$, a cluster Mott insulator where the originally triangular lattice of each layer undergoes a charge density wave instability that leads to the formation of a star of David pattern~\cite{FazekasTosatti1979, doi:10.1073/pnas.1706769114}. While detailed theoretical models have been developed to discuss this system~\cite{PhysRevX.7.041054}, here we present a simplified single-orbital model to analyze an individual cluster.

We consider the star of David cluster shown in Fig.~\ref{fig:models_schematics}\textcolor{red}{d}, with a single trivial spinful orbital placed at each one of the thirteen sites. To distinguish between AL and MAL phases, a possible minimal symmetry group is the double group of $C_{6}$ (see Tab.~\ref{tab:character table C6}).
We consider a tight binding model with nearest neighbor hoppings and on-site Hubbard interaction. In addition, we introduce a local chemical potential $\mu_*$ acting on the central site only, which is distinguished by an asterisk in Fig.~\ref{fig:models_schematics}\textcolor{red}{d}.
The model Hamiltonian for this star of David cluster reads
\begin{gather}\label{eq:Hamiltonian David Star}
        \hat{H} =
        \sum_{i=1}^{13}\sum_{\sigma}t( \hat{c}^{\dagger}_{i,\sigma} \hat{c}_{i+1, \sigma}+\text {h.c.})\\\nonumber
        +\sum_{i=1}^{6}\sum_{\sigma} \big[t'(\hat{c}^{\dagger}_{2i,\sigma} \hat{c}_{1,\sigma} + \text {h.c.}) + t''( \hat{c}^{\dagger}_{2i,\sigma} \hat{c}_{2i+2, \sigma}+\text {h.c.})\big]\\\nonumber
        + U \sum_{i=1}^{13} \hat{n}_{i, \uparrow} \hat{n}_{i, \downarrow}+\mu \sum_{i=1 }^{13}\sum_{\sigma}  \hat{n}_{i, \sigma} + \mu_* \sum_{\sigma} \hat{n}_{1, \sigma},
\end{gather}
where we identified the sites $14\equiv 2$, the parameters $t$, $t'$ and $t''$ describe the hopping amplitudes, $U$ the strength of the Hubbard interaction, $\mu$ the global chemical potential, and $\mu_*$ is the local chemical potential.

\begin{figure*}[t!]
    \centering
    \includegraphics[width=\textwidth]{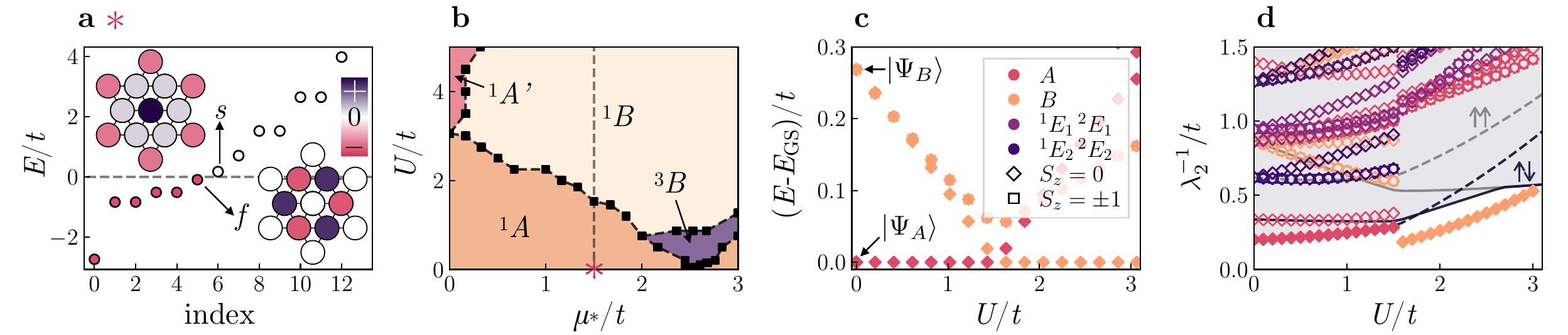}
    \caption{\textbf{Star of David.} \textbf{a} Single-particle spectrum of the star of David at $U/t=0$, $\mu_*/t=1.5$. Each eigenvalue is doubly degenerate due to the spin degree of freedom.
    The two levels closest to zero energy correspond to the $s$ and $f$ states. Their wave function weights are shown in the insets. 
    \textbf{b} Phase diagram obtained within ED of the star of David as a function of $\mu_*/t$ and $U/t$ in the Hilbert space of $N=12$ electrons, at fixed $t'/t=1$ and $t''/t = 0.4$. Phases are labeled by $^{2S+1}R$, where $R$ indicates the $C_{6}$ representation in which the ground state transforms and $S$ its total spin. (The representation labeled by $A$ corresponds to $A_1$ for $C_{6v}$, while  $A'$ corresponds to $A_2$) The pink asterisk marks the parameters of \textbf{a}.
    \textbf{c} Low-energy spectrum at $N = 12$ as a function of $U$, plotted with respect to the lowest energy eigenvalue (i.\,e., the ground state energy).
    \textbf{d} Inverted $\Gt$ spectrum obtained in ED with a cutoff of $m_{\max}=300$ in each symmetry sector as a function of $U$ (App.~\ref{app:DavidStar_ED}). The lowest eigenvalues are colored according to the physical representation of $C_{6}$ in which their eigenstates transform. The continuous (dashed) lines mark the $N-2$ ($N+2$) many-body gap, $\Delta_{N\pm2} = E_{\min, N\pm 2} - E_{\mathrm{GS}, N}$ with $N=12$. The gray (black) lines refer to excitations with $S_z = \pm1$ ($S_z=0$).
    In \textbf{c}-\textbf{d}, the parameters are $\mu_*/t=1.5$, $t'/t=1$ and $t''/t = 0.4$, corresponding to the dashed line in \textbf{b}. The legend in \textbf{c} applies to both \textbf{c} and \textbf{d}. In \textbf{d}, only the lowest eigenvalues are filled with color, for visibility.}
    \label{fig:star_of_david}
\end{figure*}

We focus on the case of electron filling $N = 12$, which allows for a TRS ground state, and we fix $t'/t=1$, $t''/t = 0.4$. The regime of twelve electrons per star of David cluster may be reached experimentally upon sample doping.
The phase diagram evaluated in ED as a function of $U$ and $\mu_*$ is shown in Fig.~\ref{fig:star_of_david}\textcolor{red}{b}.

In the single-particle sector of the Hilbert space there are thirteen energy levels, each one two-fold degenerate due to the spin degree of freedom.
At $U=0$, $\mu_*=0$ and $N=12$, these energy levels are filled up to some states, which we label by $f$, whose symmetry eigenvalue under six-fold rotation ($C_6$) is $-1$, see Fig~\ref{fig:star_of_david}\textcolor{red}{a}. We define the creation operators for these states as $\hat{c}^{\dagger}_{f, \sigma}$, with \mbox{$\sigma = \{\uparrow, \downarrow\}$}. In this limit, the many-body ground state is a gapped AL, with the first twelve levels completely filled
\begin{equation}\label{eq:AL GS in David Star}
    \ket{\Psi_A} = \hat{c}^{\dagger}_{f, \uparrow}\hat{c}^{\dagger}_{f, \downarrow} \ket{\mathrm{\Phi}},
\end{equation}
where $\ket{\Phi}$ indicates the many-body state where all the single-particle states with energy below the one of the $f$ states are completely filled.
At the Fermi energy $-\mu$, there are two pairs of spin-degenerate single-particle states with $C_6$ eigenvalue $+1$ and $-1$ respectively.
For finite values of $\mu_*>0$, the spin-degenerate single-particle states at the Fermi energy characterized by $C_6$ eigenvalue $+1$, which we call $s$ states, become lower in energy, and eventually reach the $f$ energy value, for $\mu^c_*/t\sim 2.5$. We indicate creation operators that create electrons in the $s$ state by $\hat{c}^{\dagger}_{s, \sigma}$, \mbox{$\sigma =\{\uparrow, \downarrow\}$}.
The inset of Fig.~\ref{fig:star_of_david}\textcolor{red}{a} shows the single-particle noninteracting spectrum at the value $\mu_*/t=1.5$.

As $\mu_*$ and $U$ increase, the relevant low-energy sector of the Hilbert space at finite but small $U$ involves the states where all the levels with energy below the one of $f$ are filled, while the remaining two electrons occupy some of the $s$ and $f$ states.
In the limit $U\rightarrow0$ and $\mu_*<\mu_*^c$, the ground state remains the gapped AL of Eq.~\eqref{eq:AL GS in David Star}. For small but finite $U$, the ground state is adiabatically connected to the AL in Eq.~\eqref{eq:AL GS in David Star}, and it transforms in the $A$ representation of $C_{6}$.
For larger values of $U$, the antiferromagnetic exchange favors the singlet configuration mixing the $s$ and $f$ states, leading to a level crossing in the many-body spectrum (see Fig.~\ref{fig:star_of_david}\textcolor{red}{c}). After the crossing, the new many-body ground state is a gapped singly degenerate state that transforms in the $B$ representation of the point-group, and has total spin zero.
This ground state can be adiabatically continued to the state
\begin{equation}\label{eq:excited state David Star U to 0}
    \ket{\Psi_B} = \frac{1}{\sqrt{2}}(\hat{c}^{\dagger}_{f, \uparrow}\hat{c}^{\dagger}_{s, \downarrow} - \hat{c}^{\dagger}_{f, \downarrow}\hat{c}^{\dagger}_{s, \uparrow} ) \ket{\Phi},
\end{equation}
which appears as an excited state in the many-body spectrum at $U\rightarrow0$.
The state in Eq.~\eqref{eq:excited state David Star U to 0} is an MAL state, and from this we deduce that the ground state in the $B$ phase can be adiabiatically connected to an MAL state.

We compute the spectrum of $\Gt$ for the star of David cluster in ED and by truncating the Hamiltonian (see App.~\ref{app:DavidStar_ED} for further details), keeping $\mu_*/t=1.5$ fixed while varying $U$.
Figure~\ref{fig:star_of_david}\textcolor{red}{d} shows the resulting inverted spectrum of $\Gt$. At values of $U$ lower that the critical interaction strength at which there is the transition between the $A$ phase and the $B$ phase, the lowest inverse eigenvalue of $\Gt$ transforms in the $A$ representation, and lies at the edge of the continuum of the spectrum. In the $B$ phase, there is a gap separation between the continuum of the inverse spectrum of $\Gt$ and a lowest-lying eigenvalue transforming in the $B$ representation. Therefore, for the star of David cluster one can distinguish between the AL phase, labeled by $A$, and the MAL phase, labeled by $B$, by looking at the spectrum of $\Gt$.

As seen in the cases of the Hubbard diamond chain (Sec.~\ref{sec:diamond chain}) and the checkerboard lattice of Hubbard squares (Sec.~\ref{sec:checkerboard hubbard}), weakly coupling several clusters that realize an MAL state results in an MAL phase that extends over the full lattice. This in principle will also apply to the case of the star of David clusters, which can then be diagnosed through the spectrum of $\Gt$ computed for the whole lattice.

\bigskip\tocless\section{Conclusion}{sec:conclusion}
Topological quantum chemistry is a framework for classifying materials based on the single-particle band structure developed with noninteracting electrons in mind. In particular, TQC lists all possible AL states for a given space group and compares a given first principles band structure to the respective list. If the band structure can not be constructed from an AL, it is said to be topological. While it is possible to apply the tools of TQC to interacting materials by using the (inverse of the) one-particle Green's function as an effective Hamiltonian, such an approach is blind to much of the richness of interacting states. 

Electronic interactions are important in many materials of interest and may drive a system into a state that is not adiabatically connected to a noninteracting state. It is for this setting that the framework of iTQC provides new insights through the symmetry properties of the many-body Green's function. Consequently, while TQC naturally works `from first principles', in other words the analytic result of classifying AL-induced band structures is compared to bands obtained from density functional theory, the starting point of iTQC is inherently an interacting electron system. A natural testing ground for iTQC, which we discussed in this paper, are thus toy models of interacting electrons on a given lattice. Still, such models can in principle be obtained from first principles through, e.\,g., via constrained RPA~\cite{cRPA,PhysRevB.74.125106} or the linear response approach~\cite{LinearResponsePhysRevB.71.035105, https://doi.org/10.48550/arxiv.2201.04213}.

The basic building block of iTQC's trial states is the $n$-MAL operator, which creates a cluster of $n$-electrons with non-trivial transformation properties under spatial symmetries. They are 0D-block cSPTs, including classes of cSPTs that are disconnected from any noninteracting state. Existing frameworks to identify the nature of entangled and featureless many-body states typically rely on nonlocal observables like string order parameters~\cite{PhysRevB.86.125441} or entanglemtent spectra~\cite{Haldane2008} that are hard to access in traditional condensed matter experiments. A strength of the Green's function based iTQC formalism is that it is based on substantially less exotic correlation functions that could be measurable.
In ED, there is no computational gain from using our method. Yet, in QMC and in quantum simulations the type of correlators we consider are generically easier to compute than SPT topological invariants. Therefore for most physically relevant cases, the recipe proposed in the present paper is comparatively computationally cheaper.
Specifically, it is possible to infer information about the particle-particle Green's function $\Gt$ by relating it to the superconducting susceptibility
\begin{align}
&\chi_{\alpha\beta\gamma\delta}\left(\vec{q}, \tau\right)=\\& \sum_{\vec{k}, \vec{k}^{\prime}}\left\langle T_{\tau} \hat{c}_{\vec{k}-\vec{q}, \alpha}^{\dagger}(\tau) \hat{c}_{-\vec{k}, \beta}^{\dagger}(\tau) \hat{c}_{\vec{k}^{\prime}+\vec{q}, \gamma}(0) \hat{c}_{-\vec{k}^{\prime}, \delta }(0)\right\rangle\nonumber,
\end{align}
where $T_{\tau}$ is the time-ordering operator~\footnote{The superconducting susceptibility is thus related to the time-ordered Green's function, as opposed to the retarded Green's function considered in this work.}. 
However, in experiments it may be easier to measure the particle-hole Green's function $\Gt_{\mathrm{ph}}$. For example, in neutron scattering one measures the spin-spin correlation function and certain electromagnetic properties are related to the current-current correlator, which can be computed from $\Gt_{\mathrm{ph}}$~\cite{Schwinger}. Furthermore, Raman scattering cross sections depend on $\Gt_{\mathrm{ph}}$~\cite{Raman}. A measurement of the anticorrelation of electrons in momentum space can show the ``correlation hole" originating from the Pauli principle and repulsive Coulomb interactions 
\cite{PhysRevB.73.041404,PhysRevLett.98.257604}. Alternatively, one can observe the 2-photon 2-electron spectra with intense pulses of light~\cite{PhysRevLett.103.063002}.

While we have focused in this paper entirely on the case of $n=1$ and $n=2$, in other words ALs and 2-MALs, the generalization to higher \mbox{$n$-MALs} is in principle straight forward in the iTQC formalism and is a possible extension of our current work. To do so, one would consider the correlation functions $\langle \hat O^\dagger(t) ,\hat{O}(0)\rangle$, where $\hat O^\dagger$ is a $n$-particle operator, and there is a single bosonic or fermionic time, for $n$-even or $n$-odd respectively. Still, the computational complexity of the Green's functions increases with $n$ and therefore ---in the absence of an efficient numerical scheme to evaluate many-body Green's functions--- this approach is likely to be useful only for small $n$. Focusing on small $n$ is equivalent to a truncation in the entanglement entropy of the states we study, therefore it is likely that tensor network states are a natural language in which to think about this problem. 
Using tensor networks, one can obtain the $\Gt$ band structures for a larger class of model systems, including topological ones, to diagnose their properties. 

In this work, we focused on the MAL-induced bandstructures of the two-particle Green's function, however, fragile topological phases also have signatures appearing in the Cooper pair spectrum \cite{JonahSC} and should therefore be amenable to our approach. We leave the full extension of our scheme to fragile phases to future work.

Another important future goal is the application of iTQC to real materials. Given the finite size of the clusters we consider, any crystal structure that already has a natural substructure would be well suited for a realization of the topological analysis we are presenting. A possible candidate is for example Y-kapellasite~\cite{Ykapellasite} whose underlying lattice structure is
that of hexagonal clusters of Cu 3d$^9$ atoms arranged in a triangular lattice. 

\bigskip\tocless\section{Acknowledgements}{sec:Acknowledgements}

We thank the authors of Ref.~\cite{MBTQC} for sharing an advance copy of their preprint. We would like to thank B.~Andrei Bernevig, Jonah Arbeitman-Herzog, Juven C. Wang, Ken Shiozaki, Xi Dai, Frank Pollmann, Axel F\"unfhaus and Juan Luis Mañes for useful discussions. GW acknowledges NCCR MARVEL funding from the Swiss National Science Foundation. MOS and TN acknowledge funding from the Swiss National Science Foundation (Project 200021E\_198011) as part of the QUAST 
FOR 5249-449872909 (Project P3). RV and MGV
acknowledge support from the  
 Deutsche Forschungsgemeinschaft (DFG, German Research Foundation) through QUAST FOR 5249-449872909 (Project P4) and MGV thanks European Research Council (ERC) grant agreement no. 101020833. MGV and MI acknowledge Spanish Ministerio de
Ciencia e Innovacion (grant PID2019-109905GBC21). AT is supported by the Swedish Research Council (VR) through grants number
2019-04736 and 2020-00214. Exact diagonalization was performed using the \texttt{quspin} package~\cite{Weinberg_2017}. The QMC and ED simulations were supported
by the RSF grant (project No. 21-12-00237) and used the computing resources of the federal collective usage center ``Complex for simulation and data processing for mega-science facilities'' at NRC ``Kurchatov Institute''.

\clearpage
\renewcommand{\bibsection}{\subsection*{References}}
\bibliography{references}

\clearpage
\newpage

\renewcommand{\arraystretch}{1.2}

\onecolumngrid

	\setcounter{equation}{0}
	\setcounter{figure}{0}
	\setcounter{table}{0}
	\setcounter{page}{1}

	\renewcommand{\theequation}{S\arabic{equation}}
	\renewcommand{\thefigure}{S\arabic{figure}}
	\renewcommand{\thetable}{S\arabic{table}}

\begin{appendix}

\begin{center}
    \textbf{\large Interacting topological quantum chemistry of Mott atomic limits}\\ 
    \medskip
   \normalsize Martina O. Soldini$^1$, Nikita Astrakhantsev$^1$, Mikel Iraola$^{2, 3}$, Apoorv Tiwari$^{4}$, Mark~H.\,Fischer$^1$,\\
    Roser Valent\'{i}$^5$, Maia~G.\,Vergniory$^{2, 6}$, Glenn Wagner$^1$, Titus Neupert$^1$,\\
    \medskip
    \textit{$^1$University of Zurich, Winterthurerstrasse 190, 8057 Zurich, Switzerland}\\
    \textit{$^2$Donostia International Physics Center, 20018 Donostia-San Sebastian, Spain}\\
    \textit{$^3$Department of Physics, University of the Basque Country UPV/EHU, 48080 Bilbao, Spain}\\
    \textit{$^4$Department of Physics, KTH Royal Institute of Technology, \\Roslagstullsbacken 21, 114 21 Stockholm, Sweden.}\\
    \textit{$^5$Institut für Theoretische Physik, Goethe-Universität Frankfurt, 60438 Frankfurt am Main, Germany}\\
    \textit{$^6$Max Planck Institute for Chemical Physics of Solids, 01187 Dresden, Germany}
\end{center}

\tableofcontents

\numberwithin{equation}{section}

\section{MAL operators}\label{App: MAL operators}
\subsection{Transformation properties of single- and two-particle operators}\label{App: transformation MAL operators}
Here, we first recall the transformation properties of single-particle operators and we then discuss the transformation properties of MAL operators at a general level. In Sec.~\ref{app:MAL representation from single-particle representations}, we show how to write the two-particle representations in terms of single-particle representations.

Let us consider a lattice with space group $G$. We denote by $\hat{c}_{\vec r,\alpha}^\dagger$ the operator that creates an electron in the exponentially localized Wannier function with quantum numbers described by $\alpha=(W, a, A, i)$ and in the unit cell labeled by the lattice vector $\vec r$. This indicates an orbital transforming under the representation $A$ of the site-symmetry group of the site $\vec{x}_{a}$ ($i=1,\cdots, \text{dim}A$), for an orbital located at the site $\vec x_a$ belonging to a Wyckoff position $W$ with multiplicity $m$ ($a = 1, \cdots, m$).
For the sake of simplicity, we will use the shorthand notation $\alpha=(a,i)$.
The creation and annihilation operators $\hat{c}_{\vec r,\alpha}^\dagger$ and $\hat{c}_{\vec r,\alpha}$ transform under the representation $A$ of the site-symmetry group $G_{\vec{x}_a}$, and under the action of an element of the space group $h=\{R | \vec v\} \in G$ the transformation reads~\cite{Bradlyn2017, Lessnich_2021}
\begin{equation}\label{eq:single particle operator transformation in real space Appendix}
    \begin{gathered}
   U_h \hat{c}^{\dagger}_{\vec{r},\alpha} U^{\dagger}_h = \sum_{\vec{r}',\alpha'}  A^{\vec{r}' \vec{r}}_{\alpha' \alpha}(h) \hat{c}^{\dagger}_{\vec{r}',\alpha'}
   \\
   U_h \hat{c}_{\vec{r}_, \alpha} U^{\dagger}_h = \sum_{\vec{r}',\alpha'}  \big[A^{\vec{r}'\bm{r}}_{\alpha' \alpha}(h)\big]^* \hat{c}_{\vec{r}', \alpha'},
    \end{gathered}
\end{equation}
where the representation $A$ of $G_{\vec{x}_a}$ is defined in real space. If the site $\vec{x}_a$ is mapped to $\vec{x}_{\bar{a}}$ (modulo lattice translations) under the action of $h$, the matrix elements of $A$ assume the form ($\alpha'=(a', i')$)
\begin{equation} \label{eq:tranformation A real space}
    A^{\vec r' \vec r}_{\alpha'\alpha}
    = \delta_{\vec r', R\vec r + \vec{v} + \vec t_{\bar{a} a}} 
      \delta_{a',\bar{a}} \
      \rho_{i'i}(g),
\end{equation}
where $\vec{t}_{\bar{a} a} = R \vec x_a - \vec x_{\bar{a}}$ is a vector of the lattice, and $\rho$ is the representation in which the orbitals at site $\vec x_1$ transform under its site-symmetry group $G_{\vec{x}_a}$. The element $g$ and $\bar{a}$ are uniquely determined by the coset decomposition $h g_{a} = \left\{E \mid \vec{t}_{\bar{a} a}\right\} g_{\bar{a}} g$~\cite{Bradlyn2017}.

Transforming the creation operators defined above to reciprocal space leads to
\begin{equation}
\hat{c}^{\dagger}_{\vec k, \alpha}
=\frac{1}{\sqrt{N}}\sum_{\vec r} e^{\mathrm{i}\vec k \cdot \vec r} \hat{c}^\dagger_{\vec{r},\alpha},
\end{equation}
where, once more, we define $\alpha=(a, i)$, and now the operator creates an electron in the Bloch state with momentum $\vec{k}$ constructed from the orbital $i$ on sites labeled by $a$.
Under the symmetry $h=\{R|\vec v\} \in G$, the reciprocal space operator transforms as~\cite{Lessnich_2021, Bradlyn2017}
\begin{equation}\label{eq:single particle operator transformation Appendix}
    \begin{gathered}
   U_h \hat{c}^{\dagger}_{\vec{k},\alpha} U^{\dagger}_h = \sum_{\alpha'}  \rho^{\vec{k}}_{\alpha' \alpha}(h) \ \hat{c}^{\dagger}_{R\bm{k},\alpha'} = e^{\mathrm{i}(R\vec{k}) \cdot  \vec{t}_{\bar{a}a}} \sum_{i'}  \rho^{\bm{k}}_{i' i}(g) \ \hat{c}^{\dagger}_{R\bm{k}, \bar{a}, i'} \\
   U_h \hat{c}_{\bm{k}, \alpha} U^{\dagger}_h = \sum_{\alpha'}  [\rho^{\bm{k}}_{\alpha' \alpha}(h)]^* \ \hat{c}_{R\vec{k},\alpha'} = e^{-\mathrm{i}(R\vec{k}) \cdot  \vec{t}_{\bar{a}a}} \sum_{i'} [\rho^{\bm{k}}_{i' i}(g)]^* \ \hat{c}^{\dagger}_{R\bm{k}, \bar{a}, i'},
    \end{gathered}
\end{equation}
where $\rho$ is the band representation in which Bloch states labeled by $\alpha$ and $\vec{k}$ transform, $R\vec{k}$ is defined modulo reciprocal lattice vectors, and $\bar{a}$ and $g$ are determined by the coset decomposition.

\vspace{10pt}

Let us now consider the transformations of MAL operators. A two-particle operator transforms linearly with a set of representations $A$ of the space group $G$, which can be recast into a real-spaced value representation $\rho$ acting on the two-particle operator
\begin{equation}\label{eq:transf Or real space}
\begin{split}
    U_h\hat{O}_{\vec{r}, \xi}U_h^{\dagger} =& \sum_{\alpha, \beta, \vec{u}} M^{\xi}_{\alpha \beta \vec{u}} U_h\hat{c}^{\dagger}_{\vec{r},\alpha} \hat{c}^{\dagger}_{\vec{r}+\vec{u},\beta}U_h^{\dagger} \\
    =& \sum_{\alpha, \beta, \vec{u}}\sum_{\vec{u}',\alpha', \beta'} M^{\xi}_{\alpha \beta \vec{u}} \big[  A^{\bm{r}}_{\vec{u}'\Vec{u},\alpha'\alpha,\beta' \beta}(h)\big] \hat{c}^{\dagger}_{\vec{r}',\alpha'} \hat{c}^{\dagger}_{(\vec{r}'+\vec{u}'),\beta'} \\
    =& \sum_{\xi',\vec r'} \rho^{\vec{r}',\vec{r}}_{\xi', \xi} \hat{O}^{\dagger}_{\vec{r}', \xi'}.
\end{split}
\end{equation}

As an example, let us consider the case of a 1D lattice with inversion symmetry ($\mathcal{I}$) and spinless orbitals placed at the Wyckoff position $1a$. 
These single-particle states transform either as the representation $A_g$ or $A_u$ of the site-symmetry group $\Bar{1}$ (Tab.~\ref{tab:character table 1}) of the site at Wyckoff position $1a$.
For the two-particle operator $\hat{O}^{\dagger}_{r, u} = \hat{c}^{\dagger}_{r, 1a, A_g}\hat{c}^{\dagger}_{r+u,1a, A_u}$, the transformation reads
\begin{equation}\label{eq:real space transfm of Or in 1D}
    \mathcal{I} \hat{O}^{\dagger}_{r,u} \mathcal{I}^{-1} =  \hat{c}^{\dagger}_{-r,1a, A_g}(-\hat{c}^{\dagger}_{-r-u, 1a, A_u}) = \sum_{r',v} A^{r',r}_{v,u} \hat{O}^{\dagger}_{r', v}  = -\hat{O}^{\dagger}_{-r, -u},
\end{equation}
with $A^r_{v, u}=-\delta_{v, -u}$.

\begin{table}[t!]
    \centering
    \begin{tabular}{c@{\hspace{10pt}}c@{\hspace{10pt}}|@{\hspace{5pt}}c@{\hspace{10pt}}c@{\hspace{10pt}}c@{\hspace{10pt}}c}
    \hline
    \hline
    
     (a) & (b) & $E$ & $\mathcal{I}$ & $\bar{E}$ & $\bar{\mathcal{I}}$ \\ 
     
     \hline
       $A_g$  & $\Gamma_1^+$  & $1$ & $1$ & $1$ & $1$  \\
        $A_u$ &  $\Gamma_1^-$   & $1$ & $-1$ & $1$ & $-1$ \\
        $\bar{A}_{u}$ & $\bar{\Gamma}_2$  & $1$ & $-1$ & $-1$ &  $1$  \\
         $\bar{A}_{g}$ &  $\bar{\Gamma}_3$  & $1$ &  $-1$ & $-1$ &  $1$ \\
         \hline
         \hline
    \end{tabular}
    \caption{Character table of the double point-group of $C_i$~\cite{BilbaoElcoro:ks5574}.}
    \label{tab:character table 1}
\end{table}

A two-particle operator obtained by transforming an operator $\hat{O}^{\dagger}_{\vec{r}, \xi}$ to momentum space, $\hat{O}^{\dagger}_{\vec{q}, \xi}$, must transform under a representation that at most depends on the momentum $\vec{q}$, but is independent of any other momentum. In fact, the operation of Fourier transforming a two-particle real space operator 
\begin{equation}
\hat{O}^{\dagger}_{\vec{q},\xi}=\frac{1}{\sqrt{N}} \sum_{\vec{r}} e^{\mathrm{i}\vec{q}\cdot \vec{r}} \hat{O}^{\dagger}_{\vec{r},\xi}
\end{equation}
can only introduce a $\vec{q}$ dependence in the otherwise momentum independent transformation rule of Eq.~\eqref{eq:transf Or real space}.
Hence, the operator $\hat{O}^{\dagger}_{\vec{q}, \xi}$ has to transform in a consistent representation $\rho$ of the space group $G$
\begin{equation}\label{eq:rho q of two-particle operator}
    U_h \hat{O}^{\dagger}_{\vec{q}, \xi} U_h^{\dagger} = \sum_{\vec{r}, \vec{r}', \vec{q}', \xi'} e^{\mathrm{i}\vec{q}\cdot \vec{r}} \ \rho^{\vec{r}' \vec{r}}_{\xi', \xi}  \ e^{-\mathrm{i} \vec{q}' \cdot \vec{r}'}  \ \hat{O}^{\dagger}_{\vec{q}', \xi'} =\sum_{\xi', \vec{q}'} \rho^{\vec{q}'\vec{q}}_{\xi' \xi} (h) \hat{O}^{\dagger}_{\vec{q}', \xi'},
\end{equation}
with $\rho$ that can be inferred by Fourier transforming the representation $A$.
Considering once more the example of Eq.~\eqref{eq:real space transfm of Or in 1D}, the momentum-space operator $\hat{O}^{\dagger}_{q, u} = \sum_{r} e^{\mathrm{i}qr} \hat{O}^{\dagger}_{r, u}/\sqrt{N}$ transforms as 
\begin{equation}
\begin{split}
    \mathcal{I} \hat{O}^{\dagger}_{q, u}  \mathcal{I}^{-1} =& \frac{1}{\sqrt{N}}\sum_{r} e^{\mathrm{i}qr}  \mathcal{I}\hat{O}^{\dagger}_{r, u} \mathcal{I}^{-1} = \frac{1}{\sqrt{N}}\sum_{r} e^{\mathrm{i}qr} \sum_{r',v} A^{r'r}_{v u}  \hat{O}^{\dagger}_{r', v}    \\
    =&\sum_{q', v} \big[\frac{1}{N}\sum_{r, r'} e^{\mathrm{i}qr} A^{r'r}_{v u} e^{-\mathrm{i}q'r'}\big] \hat{O}^{\dagger}_{q', v} = \sum_{q', v} \rho^{q'q}_{vu}(\mathcal{I}) O^{\dagger}_{q', v} =- O^{\dagger}_{-q, -u},
\end{split}
\end{equation}
where we obtained the final result by inserting the explicit form of the representation found in Eq.~\eqref{eq:real space transfm of Or in 1D}.

\subsection{MAL representations from single-particle representations}\label{app:MAL representation from single-particle representations}
We now turn to the question of how to construct a representation of the space group $G$ for an MAL operator expressed in momentum space out of single-particle representations.

The constraints that MAL states have to fulfill are: (i) they have to be non-degenerate (i.\,e., there should be no symmetry-enforced degeneracy between different MAL states), (ii) MAL operators generating the MAL state and acting on different unit cells should not overlap. The latter constraint ensures that the resulting states are not a long-range quantum superposition of electrons, but rather (classical) states where correlations are truly constrained to be within clusters of two-electrons. Operators that overlap at different unit cells create in general resonant-valence-bond states, where singlets are spread across the lattice and the state has long-range entanglement. 
This represents a distinct class of states as compared to the MAL states studied in this work, which will be the subject of future works. 
In addition to the contraints (i) and (ii), we focus on MAL operators that respect TRS, as with this condition AL states are constrained to transform trivially under the action of the symmetries of the group, while MAL states are not.

The general transformation of a two-particle operator under $h=\{R | \vec v\} \in G$, with the convention $\alpha= (W_1, a, \rho_1, i)$, $\beta = (W_2, b, \rho_2, j)$, can be written as
\begin{equation}\label{eq:appA2 transf Or real space}
    \begin{split}
    h \hat{O}^{\dagger}_{\Vec{r}, \xi} h^{-1} &= \sum_{\alpha, \beta, \vec{u}} M^{\xi}_{\alpha, \beta, \vec{u}} \big[h \hat{c}^{\dagger}_{\Vec{r}, \alpha} h^{-1}\big] \big[h \hat{c}^{\dagger}_{\Vec{r}+\vec{u}, \beta}h^{-1}\big] \\
    &=  \sum_{\alpha, \beta, \vec{u}} M^{\xi}_{\alpha, \beta, \vec{u}} \sum_{\vec{r}', \vec{u}', \alpha', \beta'} A^{(W1),\vec{r}'\vec{r}}_{\alpha'\alpha}(h) A^{(W2),\vec{r}'+\vec{u}',\vec{r}+\vec{u}}_{\beta'\beta}(h)\hat{c}^{\dagger}_{\vec{r}', \alpha'}\hat{c}^{\dagger}_{\vec{r}'+\vec{u}', \beta'}.
    \end{split}
\end{equation}
By comparing Eq.~\eqref{eq:appA2 transf Or real space} to Eq.~\eqref{eq:transf Or real space}, we identify the expression of $A^{\vec r'\vec r}_{\vec u' \vec u,\alpha'\alpha,\beta'\beta}$ in terms of the representations of single-particle states:
\begin{equation}\label{eq:two particle representation real space in terms of single particle}
\begin{split}
    A^{\vec r'\vec r}_{\vec u' \vec u,\alpha'\alpha,\beta'\beta}(h)
    &= 
    A^{(W1),\vec{r}'\vec{r}}_{\alpha'\alpha}(h) A^{(W2),\vec{r}'+\vec{u}',\vec{r}+\vec{u}}_{\beta'\beta}(h)\\
    &=
    \delta_{\vec r', R\vec r + \vec v + R\vec x_a - \vec x_{\tilde{a}}} \ \delta_{\vec{u}', \bar{\vec{u}}} \ \delta_{a',\tilde{a}} \ \rho^{(W1)}_{i'i}(g_1) \
     \ \delta_{b',\tilde{b}} \  \rho^{(W2)}_{j'j}(g_2),
\end{split}
\end{equation}
where we defined the transformed vector of $\vec{u}$ to be
\begin{equation}
\begin{split}
    \bar{\vec{u}}&= (R\vec r + \vec v + R\vec{u} + R\vec x_b - \vec x_{\tilde{b}}) - (R\vec r + \vec v + R\vec x_a - \vec x_{\tilde{a}}) \\
    &= R\vec{u} + R\vec x_b - \vec x_{\tilde{b}} - R\vec x_a + \vec x_{\tilde{a}} \\
    & =  V(R)\vec{u} + \vec{t}_{\bar{b} b} - \vec{t}_{\bar{a} a}
\end{split}
\end{equation}
and the vector $\vec{t}_{\bar{b} b} - \vec{t}_{\bar{a} a}$ is still a vector of the lattice, as it is obtained as the difference of lattice vectors.
In Eq.~\eqref{eq:two particle representation real space in terms of single particle}, $\rho^{(W1)}$ and $\rho^{(W2)}$ correspond to the representations in which the orbitals transform, $g_1$ ($g_2$) is the element of the site-symmetry group of a site of the Wyckoff position $W_1$ ($W_2$) related to $h$ via $h g_a = \{E|\vec{t}_{\bar{a}a}\}g_{\bar{a}}g_1$ ($h g_b = \{E|\vec{t}_{\bar{b}b}\}g_{\bar{b}}g_2$), and 
$V(R)$ is the vector representation of the rotational part of $h$ in which the vector $\vec{u}$ transforms (see App.~\ref{App: transformation of vector u}).

In momentum space, we obtain
\begin{equation}
\begin{split}
    \rho^{\vec{q}' \vec{q}}_{\alpha'\alpha, \beta'\beta, \vec{u}'\vec{u}} &= \sum_{\vec{r}, \vec{r}'} e^{\mathrm{i}(\vec{q}\cdot\vec{r}-\vec{q}'\cdot\vec{r}')} \delta_{\vec{r}', R\vec{r}+\vec{v}} \delta_{\vec{u}', \bar{\vec{u}}}  \delta_{a'\bar{a}}\delta_{b'\bar{b}} \big[\rho_{1, i' \bar{i}}(g_1) \rho_{2,j' \bar{j}}(g_2)\big] \\
    &=e^{\mathrm{i} \vec{q}' \cdot \vec{t}_{\bar{a}a}}  \ \delta_{\vec{q}',\bar{\vec{q}}} \delta_{\vec{u}', \bar{\vec{u}}}  \delta_{a' \bar{a}}\delta_{b' \bar{b}} \big[\rho_{1, i' \bar{i}}(g_1) \rho_{2,j' \bar{j}}(g_2)\big],
\end{split}
\end{equation}
which we can then write as acting on the $M^{\xi}$ coefficient, as in Eq.~\eqref{eq:rho q of two-particle operator}.

\vspace{10pt}

In practice, we consider a simpler avenue to obtain (some of the) two-particle representations $\rho$ of an MAL operator out of single particle representations, such that their MAL state fulfills the constraints (i)-(ii). To that end, we follow two steps: 

(1) Construct a `local cluster representation' as a tensor product of single-particle state representations of orbitals that belong to the same unit cell and the same maximal Wyckoff position of multiplicity $1$. Each orbital has to be compatible with the site symmetry group of their site, and the resulting representation has to be 1D to fulfill the constraint (i). 

(2) Take the tensor product of the local cluster representation with the representation induced in momentum space by the site symmetry group of the Wyckoff positions. 

This way, we find all the allowed two-particle representations $\rho$ where the two electrons of each cluster are located at the same Wyckoff position, with displacement $\vec{u}=\vec{0}$. 
With this choice, the representation in Eq.~\eqref{eq:transf Or real space} can be separated into the real space contribution and the orbital contributions
\begin{equation}\label{eq:real space transf O with simplified assumptions}
\begin{split}
    h \hat{O}^{\dagger}_{\vec{r}, \xi} h^{-1} &= \sum_{i, j} M^{\xi}_{i, j} \sum_{i', j', \vec{r}'}A^{\vec{r}'\vec{r}}_{i'i,j'j }(h) c^{\dagger}_{\vec{r}', W, A, i'} c^{\dagger}_{\vec{r}', W, B, j'} \\ 
    &= \sum_{i, j} M^{\xi}_{i, j}\sum_{\vec{r}', i', j'} [\delta_{\vec{r}', R(\vec{r}+\vec{x}) + \vec{v}-\vec{x}}  \rho^A_{i'i}(g) \rho^B_{j'j}](g) c^{\dagger}_{\vec{r}', W, A, i'} c^{\dagger}_{\vec{r}', W, B, j'}\\
    &= \sum_{i', j', \vec{r}'} \delta_{\vec{r}', R(\vec{r}+\vec{x}) + \vec{v}-\vec{x}}      \Big[ \sum_{i, j} \rho^A_{i'i}(g)\rho^B_{j'j}(g) M^{\xi}_{i, j} \Big] c^{\dagger}_{\vec{r}', W, A, i'} c^{\dagger}_{\vec{r}', W, B, j'}\\
    &= \sum_{\xi'} \rho(h)_{\xi', \xi} \hat{O}^{\dagger}_{R(\vec{r}+\vec{x}) + \vec{v}-\vec{x}, \xi_1},
\end{split}
\end{equation}

where $\vec{x}$ is the position of the site at Wyckoff position $W$ and $\rho^{A,B}$ are representations of the site symmetry group of this site.
The coefficients $M^{\xi}$ transform in the tensor product representation $\rho^A \otimes \rho^B$, and should be anti-symmetric under the exchange of the two particle operators in each term.
In momentum space, Eq.~\eqref{eq:real space transf O with simplified assumptions} leads to
\begin{equation}
    \rho^{\vec{q}'\vec{q}}_{\xi', \xi} 
    = \frac{1}{N} \sum_{\vec r, \vec r'} e^{\mathrm{i} (\vec{r}\vec{q} - \vec{r}'\vec{q}')} \delta_{\vec{r}',R(\vec{r}+\vec{x}) + \vec{v}-\vec{x}} \ \rho_{\xi', \xi}  
    = e^{-\mathrm{i}\vec{t}_{\vec x\vec x}\cdot \vec{q}'} \delta_{\vec{q}', R\Vec{q}} \  \rho_{\xi', \xi},
\end{equation}
where $\Vec{t}_{\vec x \vec x} = R\vec x - \vec x$  and $M^{\xi'} = (\rho^A \otimes \rho^B) M^{\xi}$. Therefore, it is enough to consider all the possible tensor products of the form $\rho^A \otimes \rho^B$ to determine the corresponding set of $\{M^{\xi}\}$ coefficients that give rise to MAL operators consistent with the space group representations.
Although this procedure is a simplification that does not capture all the possible cases, it is enough to obtain an exhaustive classification of interaction induced band representations in 1D, and to reproduce the numerical results presented in this paper (modulo an adiabatic deformation of the sites to a maximal Wyckoff position). In future works, we plan to provide a full general classification.

Let us continue with the 1D example introduced in App.~\ref{App: transformation MAL operators}. We consider the site symmetry group of the site corresponding to Wyckoff positions $1a$ or $1b$, both isomorphic to the double group of $C_i$ (see character table in Tab.~\ref{tab:character table 1}).

We now reintroduce the spin degree of freedom, and the two spinful single-particle physical irreps of the site symmetry group are $\bar{A}_{g} \bar{A}_{g}$ and $\bar{A}_{u} \bar{A}_{u}$, whose basis is given by
\begin{equation}\label{eq:single particle TR irreps}
    \bar{A}_{g} \bar{A}_{g} \ : \quad (\hat{c}^{\dagger}_{\bar{A}_{g}\bar{A}_{g}, \uparrow}, \ \hat{c}^{\dagger}_{\bar{A}_{g}\bar{A}_{g}, \downarrow}), \qquad 
    \bar{A}_{u}  \bar{A}_{u} \ : \quad (\hat{c}^{\dagger}_{\bar{A}_{u}\bar{A}_{u}, \uparrow}, \hat{c}^{\dagger}_{\bar{A}_{u}\bar{A}_{u}, \downarrow}),
\end{equation}
respectively.
For concreteness, we consider two spinful orbitals placed at $1a$, each one transforming in one of the two spinful representations.
To construct the representation of a local cluster operator obtained by combining single-particle representations, we consider two of the orbital representations of Eq.~\eqref{eq:single particle TR irreps}.

The two-particle coefficients $M^{\xi}$ transform according to
\begin{equation}
    (\bar{A}_{g} \bar{A}_{g}) \otimes (\bar{A}_{u} \bar{A}_{u}) = A_u \oplus A_u \oplus A_u \oplus A_u.
\end{equation}
A single one of the $A_u$ representations appearing on the right-hand-side realizes a spin-singlet representation, with a spin-0 character, while the three remaining ones realize three spin-1 individual representations. We only retain the former as it fulfills TRS, and its basis state can be expressed as
\begin{equation}
   \hat{O}^{\dagger}_{A_u} = \sum_{\sigma, \sigma'} M^{A_u}_{\sigma \sigma'} \hat{c}^{\dagger}_{\bar{A}_{g}\bar{A}_{g}, \sigma} \hat{c}^{\dagger}_{\bar{A}_{u} \bar{A}_{u}, \sigma'} =\frac{1}{\sqrt{2}} (\hat{c}^{\dagger}_{\bar{A}_{g} \bar{A}_{g}, \uparrow} \hat{c}^{\dagger}_{\bar{A}_{u} \bar{A}_{u}, \downarrow} - \hat{c}^{\dagger}_{\bar{A}_{g} \bar{A}_{g}, \downarrow} \hat{c}^{\dagger}_{\bar{A}_{u} \bar{A}_{u}, \uparrow}).
\end{equation}

Finally, we take the tensor product of the local cluster representation with the representation induced by the site symmetry group of $1a$, thereby including the momentum dependence
\begin{equation}
    \rho^q_{1a}(\mathcal{I}) = 1, \qquad \rho_{q, 1a, A_u} = A_u \otimes \rho^{q}_{1a},
\end{equation}
where $\rho^q_{1a}$ is the matrix element $\rho^{q' q}$ of the representation $\rho_{1a}^{q' q}(\mathcal{I}) = \delta_{q', \mathcal{I}q} = \delta_{q', -q}$.
The basis state for the two-particle representation in momentum space then reads
\begin{equation}
    O^{\dagger}_{q, A_u} = \frac{1}{\sqrt{N}} \sum_r e^{\mathrm{i} q r} \frac{1}{\sqrt{2}}(\hat{c}^{\dagger}_{r, 1a, \bar{A}_{g} \bar{A}_{g}, \uparrow} \hat{c}^{\dagger}_{r, 1a, \bar{A}_{u} \bar{A}_{u}, \downarrow} - \hat{c}^{\dagger}_{r, 1a, \bar{A}_{g} \bar{A}_{g}, \downarrow} \hat{c}^{\dagger}_{r, 1a,  \bar{A}_{u} \bar{A}_{u}, \uparrow}),
\end{equation}
which is an MAL operator transforming in the $\rho_{q, 1a, A_u}$ representation of the double group of $C_i$.

\subsection{Transformation properties of MALs spread over different unit cells} \label{App: transformation of vector u}
In this section, we focus on the transformation of the vector $\vec{u}$, as defined in Eq.~\eqref{eq:2-MAL wf}. This allows to extend the treatment of the previous sections, App.~\ref{App: transformation MAL operators} and~\ref{app:MAL representation from single-particle representations}, to MAL operators that are not localized in a single unit cell.

\vspace{10pt}

Let us consider the coset decomposition of the space group $G$ with respect to its translation subgroup $T$:
\begin{equation} \label{eq:coset decompostion wrt translation subgroup}
    G = \bar g_1 T + \bar g_2 T + \bar g_3 T + \cdots g_N T,
\end{equation}

where $\bar g_1, \bar g_2, \cdots, \bar g_N$ are the coset representatives of the decomposition, and $\bar g_1$ is usually taken to be the identity.
The set of coset representatives is always finite. Indeed, it can be chosen to form a point-group $\bar G$, in the case of a symmorphic space group $G$.
Note that, every symmetry $g_i=\{R | \vec{v}\} \in G$ has in correspondence a coset representative $\bar g_i = \{R | \vec{0}\} \in \bar G$, i.\,e.:

Based on this relation, a vector $\vec{u}_i$ transforms under the action of a symmetry operation $g \in G$ as follows:

\begin{equation} \label{eq:vector representation for u}
    g \vec{u}_i = V_{ji}(\bar g) \vec{u}_j,
\end{equation}

where $V$ denotes the vector representation of $\bar G$.
The set of vectors related to $\vec{u}$ by the symmetries of $\bar G$ is called the \textit{star of} $\vec{u}$: $\mathrm{star}(\vec{u}) = \{ V(\bar g) \vec u \ | \ \forall \bar g \in \bar G \}$ 

In the case of non-symmorphic space groups, the set of coset representatives can not be chosen to coincide with a point-group, as some of them will always have a translational part. 
However, it can be selected to be isomorphic to a point-group $\bar G$ whose elements are the rotational parts of the coset representatives.
Hence, the relation between symmetries of the space group $G$ and elements of the point-group $\bar G$ described above for symmorphic space groups still holds for non-symmorphic space groups.
As a consequence, in non-symmorphic lattices vectors $\{\vec{u}_n\}$ also transform as the vector representation in Eq.~\eqref{eq:vector representation for u}, with the only difference that $\bar g$ might not be a coset representative in the decomposition Eq.~\eqref{eq:coset decompostion wrt translation subgroup}, but a symmetry of the point-group $\bar G$ isomorphic to the set of coset representatives.
Note that vectors $\{\vec{u}_{n}\}$ and $p$-orbitals transform in the same way, i.\,e., as the vector representation $V$.

\subsubsection{Example: symmorphic lattice} \label{subsec:example symmorphic lattice}

\begin{figure}[t!]
    \centering
    \includegraphics[width=0.3\textwidth]{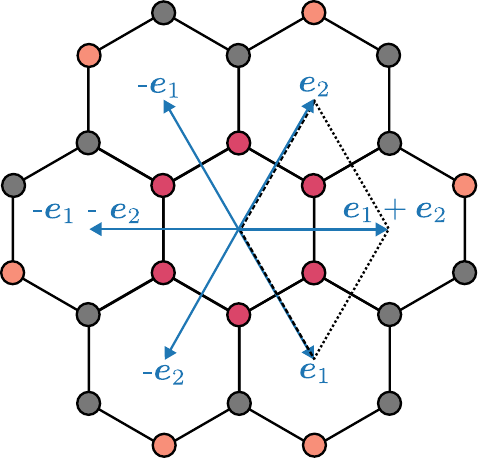}
    \caption{Hexagonal lattice considered as an example to illustrate the transformation of a vector $\vec{u}$ in symmorphic lattices. The unit cell is indicated by dashed lines and blue vectors form the star of $\vec{u} = \vec{e}_1 + \vec{e}_2$. Red circles denote the orbit of the site at $1/3 \vec{e}_1 + 2/3\vec{e}_2$, while orange circles correspond to the orbit of $1/3 \vec{e}_1 + 2/3\vec{e}_2 + \vec{u}$.}
    \label{fig:example symmorphic lattice}
\end{figure}

Let consider as an example of a symmorphic space group the hexagonal lattice of symmetry $G=P6$ (see Fig.~\ref{fig:example symmorphic lattice}), with spinless $s$-like Wannier functions at Wyckoff positions $2b$.
The coset decomposition in Eq.~\eqref{eq:coset decompostion wrt translation subgroup} can be written in the following way for the space group $P6$:

\begin{equation}
    P6 = T + C_6 T + C_3 T + C_2 T + C_3^{-1} T + C_6^{-1} T,
\end{equation}

where $T$ is the translation subgroup of $P6$.
The coset representatives in Eq.~\eqref{eq:vector representation for u} form the point-group $\bar G = C_6$.
According to the transformation in Eq.~\eqref{eq:vector representation for u}, the star of a (non-zero) vector $\vec{u}$ is $\mathrm{star}(\vec{u}) = \{ V(\bar g) \vec u |\ \forall \bar g \in \bar G \}$, where

\begin{equation} \label{eq:V matrix for C_6}
    V(C_6) = 
    \begin{pmatrix}
    1 & -1 \\
    1 & 0 \\
    \end{pmatrix}.
\end{equation}

The matrices $V$ for the rest of symmetries in $\bar G$ can be obtained from Eq.~\eqref{eq:V matrix for C_6}.
For example, the star of the vector $\vec u = \vec{e}_1 + \vec{e}_2$ is:

\begin{equation} \label{eq:star e1 + e2}
    \mathrm{star}(\vec{e}_1 + \vec{e}_2) = 
    \{\vec{e}_1 + \vec{e}_2, -\vec{e}_1, -\vec{e}_1-\vec{e}_2, -\vec{e}_2, \vec{e}_1 \}.
\end{equation}

We denote $A$ the site at $\frac{1}{3}\vec{e}_1 + \frac{2}{3} \vec{e}_2$ of the Wyckoff position $2b$ and $B$ the site at $\frac{2}{3} \vec{e}_1 + \frac{1}{3} \vec{e}_2$.
Each orbital in the lattice can be labeled by the triplet $(\vec{r}, \vec{x}_a, \vec{u})$, where $\vec{r}$ is a vector of the lattice, $\vec{x}_a$ is the position of the site $a$ ($a=A,B$) within the unit cell and $\vec{u}$ is also a vector of the lattice. 
The transformation of a state under a symmetry $g=\{R|\vec{v}\} \in P6$ is described by the following change of labels:

\begin{equation} \label{eq:transformation of r, x_a and u}
    g: (\vec{r}, \vec{x}_a, \vec{u}) 
    \rightarrow 
    (R\vec{r}+\vec{v}+\vec{t}_{a'a}, R\vec{x}_a ).
\end{equation}

Let us consider the orbitals at $\frac{1}{3} \vec{e}_1 + \frac{2}{3} \vec{e}_2$ and $\frac{4}{3} \vec{e}_1 + \frac{5}{3} \vec{e}_2$ as candidates to form a Mott cluster state.
In particular, the orbital at $\frac{1}{3} \vec{e}_1 + \frac{2}{3} \vec{e}_2$ is labeled as $(\vec{0}, \vec{x}_A, \vec{0})$, while the state at $\frac{4}{3} \vec{e}_1 + \frac{5}{3} \vec{e}_2$ is $(\vec{0}, \vec{x}_A, \vec{e}_1 + \vec{e}_2)$.
The Mott cluster would involve the product of creation operators $c_{(\vec{0}, \vec{x}_A, \vec{0})}^\dagger c_{(\vec{0}, \vec{x}_A, \vec{e}_1 + \vec{e}_2)}^\dagger$, where $\vec u = \vec{e}_1 + \vec{e}_2$.
Tab.~\ref{tab:example symmorphic lattice transformation u} shows the transformation of the labels of these states under the symmetries in $\bar G$, which is compatible with the description in Eq.~\eqref{eq:transformation of r, x_a and u}. 
Note that the transformation of the label $\vec u$ coincides with the star of $\vec u = \vec{e}_1 + \vec{e}_2$ in Eq.~\eqref{eq:star e1 + e2}.

\begin{table}
    \centering
    \begin{tabular}{c c c}
    \hline
    \hline
        symmetry & $(\vec{e}_{0}, \vec{x}_A, \vec{0})$ & $(\vec{0}, \vec{x}_A, \vec{e}_1 + \vec{e}_2)$ \\
              \hline
        $C_6$ & $(-\vec{e}_{1}, \vec{x}_B, \vec{0})$ & $(\vec{e}_1, \vec{x}_B, \vec{e}_2)$ \\
        $C_3$ & $(-\vec{e}_{1}-\vec e_{2}, \vec{x}_A, \vec{0})$ & $(-\vec{e}_1-\vec{e}_2, \vec{x}_A, -\vec{e}_1)$ \\
        $C_2$ & $(-\vec{e}_1-\vec{e}_2, \vec{x}_B, \vec{0})$ & $(-\vec{e}_1-\vec{e}_2, \vec{x}_B, -\vec{e}_1 -\vec{e}_2)$ \\
        $C_3^{-1}$ & $(-\vec{e}_2, \vec{x}_A, \vec{0})$ & $(-\vec{e}_2, \vec{x}_A, -\vec{e}_2)$ \\
        $C_6^{-1}$ & $(\vec{0}, \vec{x}_B, \vec{0})$ & $(\vec{0}, \vec{x}_B, \vec{e}_1)$ \\
        \hline
        \hline
    \end{tabular}
    \caption{Transformation of the labels $(\vec r, \vec x_a, \vec u)$ corresponding to the orbitals at $(1/3) \vec{e}_1 + (2/3) \vec{e}_2$ and $4/3 \vec{e}_1 + (5/3) \vec{e}_2$ under the point-group $\bar G$. Note that the orbit of $\vec{u}=\vec{e}_1 + \vec{e}_2$ coincides with the star in Eq.~\eqref{eq:star e1 + e2}.}
    \label{tab:example symmorphic lattice transformation u}
\end{table}

\subsubsection{Example: non-symmorphic lattice}

\begin{figure}[t!]
    \centering
    \includegraphics[width=0.45\textwidth]{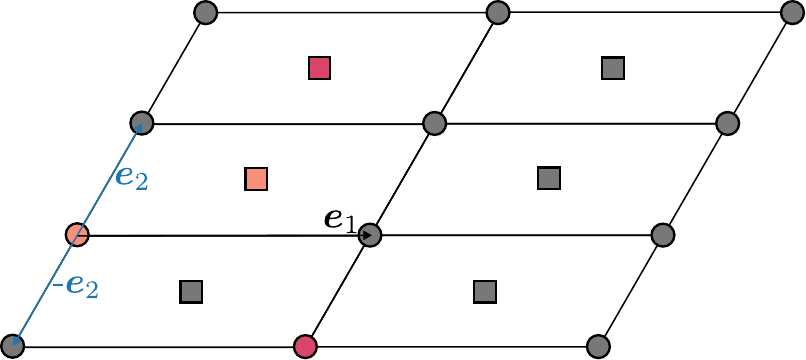}
    \caption{Monoclinic lattice (projection along monoclinic axis) considered as an example to illustrate the transformation of a vector $\vec{u}$ in nonsymmorphic lattices. Blue vectors form the star of $\vec{u} = \vec{e}_2$. Orange circles denote the orbit of the site at $z\vec{e}_3$, while orange circles correspond to the orbit of $ -z\vec{e}_3 + \vec{u}$.}
    \label{fig:example nonsymmorphic lattice}
\end{figure}

We will now give an example for the case of a non-symmorphic space group.
Let us consider the space group $P2_1$ corresponding to the 3D monoclinic lattice, with spinless $s$-like Wannier functions in Wyckoff position $2a$ (see  Fig.~\ref{fig:example nonsymmorphic lattice}).
The space group $P2_1$ can be decomposed in terms of its translation subgroup $T$ in the following way:

\begin{equation} \label{eq:decomposition P2_1}
    P2_1 = T + \{ C_{2x}\ | \ \vec{e}_1 + \vec{e}_2 \} T,
\end{equation}

where $\vec{e}_i$ with $i=1,2,3$ are the primitive lattice vectors.
Note that the coset representatives $\{E \ | \ \vec{0}\}$ and $\{ C_{2x} \ | \vec{e}_1 + \vec{e}_2 \}$ do not form a closed group, but their rotational parts $E$ and $C_{2x}$ do form the point-group $\bar G = C_2$.
Therefore, we can relate every element $h =\{ R|\vec{v} \} \in G$ with an element $R \in \bar G$.
Based on this relation, the star of the vector $\vec{u}$ is $\mathrm{star}(\vec{u}) = \{ \vec{u}, V(C_{2x}\vec{u}) \}$, where

\begin{equation}
    V(C_{2x}) =
    \begin{pmatrix}
    0 & -1 \\
    1 & 0
    \end{pmatrix}
\end{equation}

is the matrix of $C_{2x}$ in the vector representation.

The Wyckoff position $2a$ contains two sites $\vec{x}_A = z \vec{e}_3$ and $\vec{x}_B = 1/2 \vec{e}_1 + 1/2\vec{e}_2 - z \vec{e}_3$.
Like in Sec.~\ref{subsec:example symmorphic lattice}, each orbital can bi labeled by a triplet of vectors $(\vec{r}, \vec{x}_a, \vec{u})$ whose transformation is written in Eq.~\eqref{eq:transformation of r, x_a and u}.
In particular, we consider the orbitals at $z \vec{e}_3$ and $1/2 \vec{e}_1 + 3/2\vec{e}_2 - z \vec{e}_3$ as candidates to form a Mott cluster state via products like $c_{(\vec{0}, \vec{x}_A, \vec{0})}^\dagger c_{(\vec{0}, \vec{x}_B, \vec{u})}^\dagger$, where $\vec{u} = \vec{e}_2$.
The star of this vector is star$(\vec{e}_2) = \{ \vec{e}_2, -\vec{e}_2 \}$, as it is indicated in Fig.
Furthermore, under $\bar G$ these states turn into $(\vec{0}, \vec{x}_B, \vec{0})$ and $(\vec{e}_1, \vec{x}_A, -\vec{e}_2)$, according to Eq.~\eqref{eq:transformation of r, x_a and u}.

\subsection{Example of MAL operator construction: Hubbard square}\label{App: example MAL construction: Hubbard square}
Let us consider as an explicit example the system considered in the main text: the Hubbard square (Sec.~\ref{sec:hubbard square}).
More precisely, we want to derive the explicit form of the two-particle operator that realizes the non-trivial ground state in the case $t_1>t_2$.

As discussed in Sec.~\ref{sec:hubbard square}, the Hubbard square has four orbitals placed at the corners transforming in the trivial representation of the site-symmetry group that can be adiabatically connected to a set of orbitals placed at the center of the square transforming in some irreps of the double group of $C_4$. This corresponds to transforming the creation operators in the basis of the $C_4$ operator eigenstates, which are labeled by the $C_4$ orbital eigenvalues $\ell=0, \pm \pi/2, \pi$.
The double group of $C_4$ has four 1D spinful irreps~\cite{BilbaoElcoro:ks5574}, and they are pairwise degenerate due to TRS. We do not include spin-orbit coupling, and therefore the states at $\ell = \pm \pi/2$ are also degenerate by virtue of TRS.

\begin{figure}[t]
    \centering
    \includegraphics[width=\textwidth]{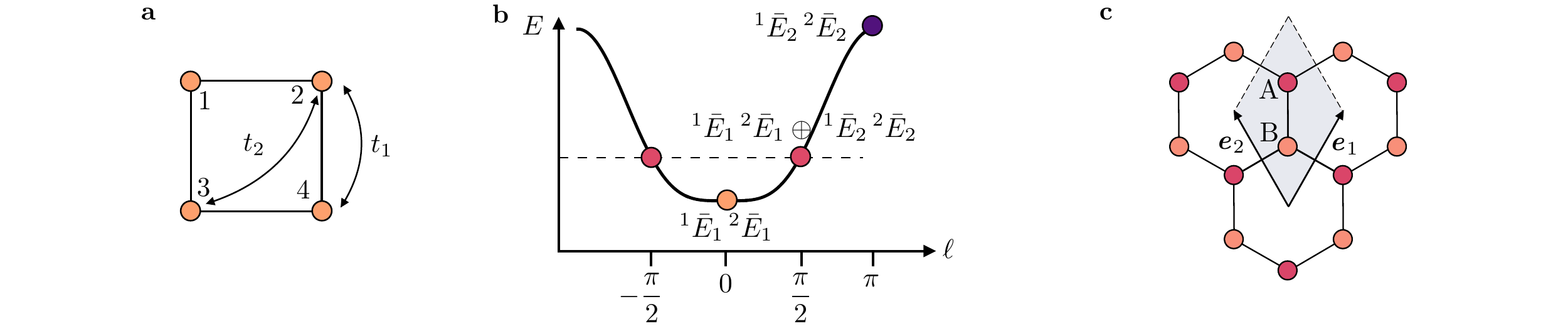}
    \caption{\textbf{a} Schematics of the Hubbard square and \textbf{b} spectrum for $t_1>t_2$ and no interaction ($U=0$) as a function of the $C_4$ orbital eigenvalue of the eigenstates ($\ell$). The Fermi level at half-filling is marked by a dashed line. \textbf{c} Lattice scheme for the honeycomb lattice. The unit cell is highlighted in gray, and sites belonging to the same sublattices are filled by the same color.}
    \label{fig:hubbardsquare_noninteracting_Repr}
\end{figure}

The noninteracting spectrum at half-filling and for $t_1>t_2$
has four-degenerate states lying at the Fermi level, with $C_4$ eigenvalues $\pm i$, corresponding to $\ell=\pm \pi/2$ (Fig.~\ref{fig:hubbardsquare_noninteracting_Repr}).
The operators creating electrons with $\ell=\pm \pi/2$ transform in the following irreps of $C^D_4$
\begin{equation}\label{eq:basis for barE_1+barE_2}
\hat{c}^{\dagger}_{\frac{\pi}{2}, \uparrow} \in \ ^2\bar{E}_2, \quad \hat{c}^{\dagger}_{-\frac{\pi}{2}, \downarrow} \in \ ^1\bar{E}_2, \quad \hat{c}^{\dagger}_{-\frac{\pi}{2}, \uparrow}\in \ ^2\bar{E}_1, \quad \hat{c}^{\dagger}_{\frac{\pi}{2}, \downarrow} \in \ ^1\bar{E}_1,
\end{equation}
the symmetry eigenvalue of each operator under $C_4$ was given in Eq.~\eqref{eq:single part C4 transformation} and the character table of $C^D_4$ is in Tab.~\ref{tab:c4 double character table}.
The physical irreps stemming from~\eqref{eq:basis for barE_1+barE_2} are then $^1 \bar{E}_1 \, ^2 \bar{E}_1$ and $^1 \bar{E}_2\, ^2\bar{E}_2$.

In the many-body ground state, the representation lying at $\ell=0$ is completely filled, and therefore transforming in the trivial representation of $C_{4}$, $A$. For this reason, we can retain the $\ell = \pm \pi/2$ states only in the low-energy and small $U$ description, while discarding the completely filled ($\ell=0$) and completely empty ($\ell=\pi$) states, which do not affect the symmetry properties of the ground state.

We now compute the tensor product between these low-energy half-filled orbitals, and consider only the resulting antisymmetrized (a.s.) set of representations
\begin{equation}\label{eq:product repr hubbard square}
   (^1\bar{E}_1 \, ^2\bar{E}_1 \, \oplus \,  ^1\bar{E}_2 \, ^2\bar{E}_2)\otimes ( ^1\bar{E}_1 \, ^2\bar{E}_1 \, \oplus \, ^1\bar{E}_2 \,  ^2\bar{E}_2)|_{\text{a.s.}}= \ ^2\Bar{E} \, \oplus \, A \, \oplus \, ^1\Bar{E} \, \oplus \, B \, \oplus \, ^1\Bar{E} \, \oplus \, B \, \oplus \, ^2\Bar{E} \, \oplus \, ^2\Bar{E} \, \oplus \, A \, \oplus \, ^1\Bar{E}.
\end{equation}
In terms of the starting operators of Eq.~\eqref{eq:basis for barE_1+barE_2}, the basis operators for the two representations $B$ appearing in Eq.~\eqref{eq:product repr hubbard square} are
\begin{equation}
    \hat{O}^{\dagger}_{B, 1} = \frac{1}{\sqrt{2}} (\hat{c}^{\dagger}_{\frac{\pi}{2}, \uparrow}\hat{c}^{\dagger}_{\frac{\pi}{2}, \downarrow}-\hat{c}^{\dagger}_{-\frac{\pi}{2}, \uparrow}\hat{c}^{\dagger}_{-\frac{\pi}{2}, \downarrow}), \qquad 
    \hat{O}^{\dagger}_{B, 2} = \frac{1}{\sqrt{2}} (\hat{c}^{\dagger}_{\frac{\pi}{2}, \uparrow}\hat{c}^{\dagger}_{\frac{\pi}{2}, \downarrow}+\hat{c}^{\dagger}_{-\frac{\pi}{2}, \uparrow}\hat{c}^{\dagger}_{-\frac{\pi}{2}, \downarrow}).
\end{equation}
The second operator $\hat{O}^{\dagger}_{B, 2}$ breaks TRS, and the first operator $\hat{O}^{\dagger}_{B, 1}$ is the two-particle operator appearing in the non-trivial ground state of the Hubbard square.

\begin{table}
    \centering
    \begin{tabular}{c @{\hspace{10pt}}c@{\hspace{10pt}}|@{\hspace{10pt}}c @{\hspace{10pt}}c @{\hspace{10pt}}c @{\hspace{10pt}}c @{\hspace{10pt}}c @{\hspace{10pt}}c @{\hspace{10pt}}c @{\hspace{10pt}}c}
    \hline
    \hline
     
       (a) & (b) & $E$ &  $C_2$ & $C_4$ & $C^{-1}_4$ & $\Bar{E}$ & $\Bar{C}_2$ & $\Bar{C}_4$ & $\bar{C}^{-1}_4$  \\
        
        \hline
        $A$ & $\Gamma_1$ & 1 & 1 & 1 & 1 & 1 & 1 & 1 & 1 \\
        $B$ & $\Gamma_2$ & 1 & 1 & $-1$ & $-1$ & 1 & 1 & $-1$ & $-1$ \\
        $^2E$ & $\Gamma_3$ & 1 & $-1$ & $\mathrm{i}$ & $-\mathrm{i}$ & 1 & $-1$ & $\mathrm{i}$ & $-\mathrm{i}$ \\
        $^1E$ & $\Gamma_4$ & 1 & $-1$ & $-\mathrm{i}$ & $\mathrm{i}$ & 1 & $-1$ & $-\mathrm{i}$ & $\mathrm{i}$ \\
        $^2\bar{E}_2$ & $\bar{\Gamma}_5$ & 1 & $-\mathrm{i}$ & $\omega^3$ & $-\omega$ & $-1$ & $\mathrm{i}$ & $-\omega^3$ & $\omega$\\
        $^2\bar{E}_1$ & $\bar{\Gamma}_6$ & 1 & $-\mathrm{i}$ & $-\omega^3$ & $\omega$ & $-1$ & $\mathrm{i}$ & $\omega^3$ & $-\omega$\\
        $^1\bar{E}_2$ & $\bar{\Gamma}_7$ & 1 & $\mathrm{i}$ & $-\omega$ & $\omega^3$ & $-1$ & $-\mathrm{i}$ & $\omega$ & $-\omega^3$\\
        $^1\bar{E}_1$ & $\bar{\Gamma}_8$ & 1 & $\mathrm{i}$ & $\omega$ & $-\omega^3$ & $-1$ & $-\mathrm{i}$ & $-\omega$ & $\omega^3$\\
    \hline
    \hline
    \end{tabular}
    \caption{Character table of $C^D_4$~\cite{BilbaoElcoro:ks5574}. Here we defined $\omega \equiv e^{\mathrm{i}\pi/4}$.}
    \label{tab:c4 double character table}
\end{table}

\subsection{Adiabatic connection of MALs}\label{subsec:adiabatic connectivity of MALs}
To prove that two MALs are equivalent we can rely on the argument presented in the context of TQC for the adiabatic connectivity of ALs. 
Let us now discuss a few explicit examples in the context of 1D lattices, where there are three Wyckoff positions: $1a$, $1b$ and $2c$ (Fig.~\ref{fig:1D_unitcell_irreps} of the double group of the site symmetry group $\Bar{1}$).
In the following, we adopt the convention that $a^{\dagger}_{r, \bar{A}_{g/u}\bar{A}_{g/u}, \sigma}$, $b^{\dagger}_{r, \bar{A}_{g/u}\bar{A}_{g/u}, \sigma}$ and $c^{\dagger}_{r,\alpha, \sigma}$ ($\alpha=1,2$) create electronic operators at Wyckoff position $1a$, $1b$ (in either the $\bar{A}_g\Bar{A}_g$ or the $\bar{A}_u\Bar{A}_u$ physical irreps) and the two $2c$ positions labeled by $\alpha$, respectively.
MALs at $1a$ and $2c$ can be adiabatically connected by the unitary transformation
\begin{equation}
    \hat{c}^\dagger_{r, \alpha, \sigma}
    =\frac{1}{\sqrt{2}}\left(\hat{a}^\dagger_{r,\bar{A}_g\bar{A}_g,\sigma}
    -(-1)^\alpha\hat{a}^\dagger_{r,\bar{A}_u\bar{A}_u,\sigma}\right), \qquad \alpha=1, 2 .
\end{equation}
On the other hand, MALs placed at $1b$ can be connected to MALs at $2c$
using the unitary transformation
\begin{equation}
    \hat{b}^\dagger_{r, \bar{A}_{g/u}\bar{A}_{g/u}, \sigma}
    =\frac{1}{\sqrt{2}}\left(\hat{c}^\dagger_{r, 1,\sigma}
    -(-)^{f(g/u)}\hat{c}^\dagger_{r-1,2,\sigma}\right),  \ \ f(g)=+1,\ f(u)=-1 .
\end{equation}

As an additional remark, some of the MALs can be connected to ALs without the breaking of any symmetry. For instance, consider the two operators
\begin{equation}
    \hat{O}^{\dagger}_{r, A_g, 1a} =  \frac{1}{\sqrt{2}} (\hat{a}^{\dagger}_{r, \bar{A}_g\bar{A}_g, \uparrow} \hat{\tilde{a}}^{\dagger}_{r, \bar{A}_g\bar{A}_g, \downarrow} - \hat{a}^{\dagger}_{r, \bar{A}_g\bar{A}_g, \downarrow} \hat{\tilde{a}}^{\dagger}_{r, \bar{A}_g\bar{A}_g, \uparrow}), \qquad \hat{\tilde{O}}_{r, A_g, 1a} = \hat{a}^{\dagger}_{r, \bar{A}_g\bar{A}_g, \uparrow}  \hat{a}^{\dagger}_{r, \bar{A}_g\bar{A}_g, \downarrow} .
\end{equation}
One can define the adiabatic unitary transformation
\begin{equation}
\hat{a}^{\dagger}_{r, \bar{A}_g\bar{A}_g, \downarrow}(t) = \frac{(1-t)\hat{a}^{\dagger}_{r, \bar{A}_g\bar{A}_g, \downarrow} - t \hat{\tilde{a}}^{\dagger}_{r, \bar{A}_g\bar{A}_g, \downarrow}}{\sqrt{t^2 + (1-t)^2}} , \quad \hat{\tilde{a}}^{\dagger}_{r, \bar{A}_g\bar{A}_g, \downarrow}(t) = \frac{(1-t) \hat{\tilde{a}}^{\dagger}_{r, \bar{A}_g\bar{A}_g, \downarrow} + t \hat{a}^{\dagger}_{r, \bar{A}_g\bar{A}_g, \downarrow}}{\sqrt{t^2 + (1-t)^2}}  \qquad t \in [0, 1]
\end{equation}
which maps $\hat{O}^{\dagger}_{r, A_g, 1a}$ into a linear combinations of ALs
\begin{equation}
    \hat{O}^{\dagger}_r(t)=\frac{1}{\sqrt{2 (t^2 + (1-t)^2)}} \big( \hat{a}^{\dagger}_{r, A_g, \uparrow} ((1-t)\hat{a}^{\dagger}_{r,\bar{A}_g\bar{A}_g, \downarrow} - t \hat{\tilde{a}}^{\dagger}_{r,\bar{A}_g\bar{A}_g, \downarrow}) - (((1-t) \hat{\tilde{a}}^{\dagger}_{r,\bar{A}_g\bar{A}_g, \downarrow} + t \hat{a}^{\dagger}_{r,\bar{A}_g\bar{A}_g, \downarrow}) \hat{\tilde{a}}^{\dagger}_{r,\bar{A}_g\bar{A}_g, \uparrow} \big).
\end{equation}
From this, there is an additional transformation that brings us from $\hat{O}^{\dagger}(t=1)$ to $\hat{\tilde{O}}^{\dagger}_{r, A_g, 1a}$, without breaking symmetries, for example
\begin{equation}
    \hat{O}^{\dagger}_{r,\mathrm{AL}}(t) = (1-t)\hat{O}^{\dagger}_r(1) + t \hat{\tilde{O}}^{\dagger}_{r, A_g, 1a} \qquad t \in [0, 1].
\end{equation}
On the other hand, some of the MAL operators are not adiabatically connected to any AL operator, and there is no symmetry preserving transformation that allows to map one to the other.

\section{Two-particle retarded Green's functions}

\subsection{Derivation of transformation of the two-particle Green's function}
\label{sec:App_G2_transf}
In this section, we derive the transformation law quoted in the main text in Eq.~\eqref{eq:transformation of G2}. To see how Eq.~\eqref{eq:retarded 2 particle GF in omega} transforms, we follow Ref.~\cite{Lessnich_2021}, and adopt a similar notation (see Eq.~(9) in~\cite{Lessnich_2021}). Consider a unitary symmetry operator $U_{h}$ with $h \in G$, where $G$ is the space group.
For a single ground state, it holds that $U_{h}\ket{\mathrm{GS}}=e^{i\phi}\ket{\mathrm{GS}}$, with $e^{i\phi} \in \mathbb{C}$. 
With Eq.~\eqref{eq:single particle operator transformation Appendix}, the transformation of the expectation value entering in the sum of Eq.~\eqref{eq:G2 with O operators}, under the action of $h \in G$ is
\begin{equation}\label{eq:G2q tranformation rule app}
\begin{split}
     g^{(2)}_{(\alpha, \beta, \vec{u}), (\gamma, \delta, \vec{v})}(t, \vec{q}) &= -\Theta(t) \expval{ [
    \hat{O}_{\vec{q}, (\alpha, \beta, \vec{u})}(t), \hat{O}^{\dagger}_{\vec{q}, (\gamma, \delta, \vec{v})}(0) ] }_{\mathrm{GS}} \\
     &=-\Theta(t)\expval{ [e^{\mathrm{i} t \hat{H}}
   \hat{O}_{\vec{q}, (\alpha, \beta, \vec{u})} e^{-\mathrm{i} t \hat{H}},
   \hat{O}^{\dagger}_{\vec{q}, (\gamma, \delta, \vec{v})} ] }_{\mathrm{GS}} \\
    &=-\Theta(t) \expval{ [e^{ \mathrm{i} t U^{\dagger}_h\hat{H}U_h} \hat{O}_{\vec{q}, (\alpha, \beta, \vec{u})} e^{-\mathrm{i} t U^{\dagger}_h\hat{H}U_h},\hat{O}^{\dagger}_{\vec{q}, (\gamma, \delta, \vec{v})} ] }_{\mathrm{GS}}\\
    &= -\Theta(t) \expval{ [U^{\dagger}_h e^{\mathrm{i} t \hat{H}}U_h
    \hat{O}_{\vec{q}, (\alpha, \beta, \vec{u})}  U^{\dagger}_h e^{-\mathrm{i} t \hat{H}}U_h, \hat{O}^{\dagger}_{\vec{q}, (\gamma, \delta, \vec{v})} ] }_{\mathrm{GS}}\\
    & = \sum_{\substack{(\alpha, \beta, \vec{u})',\\ (\gamma, \delta, \vec{v})'}}\rho^{\bm{q}}(h)^*_{(\alpha,\beta, \vec{u}), (\alpha,\beta, \vec{u})'}
    \rho^{\bm{q}}(h)_{ (\gamma, \delta, \vec{v})',  (\gamma, \delta, \vec{v})} \
    \gt_{ (\alpha,\beta, \vec{u})'(\gamma, \delta, \vec{v})'}(t, \vec{q}') 
\end{split}
\end{equation}
with $\vec{q}'=R\vec{q}$ and $\rho^{\vec{q}}$ as the one in Eq.~\eqref{eq:rho q of two-particle operator}.
From Eq.~\eqref{eq:G2q tranformation rule app}, it also follows that the eigenstates of $\Gt$, say $v_{(\gamma, \delta, \vec{v})}(t, \vec{v})$ transforms according to the representation $\rho^{\vec{q}}_G$ introduced in Eq.~\eqref{eq:rho q of two-particle operator}
\begin{equation}
\begin{split}
    \lambda =& \sum_{\substack{(\alpha, \beta, \vec{u}),\\ (\gamma, \delta, \vec{v})}} v^*_{(\alpha, \beta, \vec{u})}(t, \vec{q}) \gt_{(\alpha, \beta, \vec{u}), (\gamma, \delta, \vec{v})}(t, \vec{q})  v_{(\gamma, \delta, \vec{v})}(t, \vec{q})\\
    =& \sum_{\substack{(\alpha, \beta, \vec{u})',\\ (\gamma, \delta, \vec{v})'}}\sum_{\substack{(\alpha, \beta, \vec{u}),\\ (\gamma, \delta, \vec{v})}} v^*_{(\alpha, \beta, \vec{u})}(t, \vec{q}) [\rho^{\vec{q}}_{(\alpha, \beta, \vec{u})' (\alpha, \beta, \vec{u})}(h)]^* \gt_{(\alpha, \beta, \vec{u})', (\gamma, \delta, \vec{v})'}(t, \vec{q}') \rho^{\vec{q}}_{(\gamma, \delta, \vec{v})' (\gamma, \delta, \vec{v})}(h)  v_{(\gamma, \delta, \vec{v})}(t, \vec{q}) \\
    =& \sum_{\substack{(\alpha, \beta, \vec{u})',\\ (\gamma, \delta, \vec{v})'}}\Big[ \sum_{(\alpha, \beta, \vec{u})} \rho^{\vec{q}}_{(\alpha, \beta, \vec{u})' (\alpha, \beta, \vec{u})}(h) v_{(\alpha, \beta, \vec{u})}(t, \vec{q})\Big]^{*} \gt_{(\alpha, \beta, \vec{u})', (\gamma, \delta, \vec{v})'}(t, \vec{q}') \Big[\sum_{(\gamma, \delta, \vec{v})} \rho^{\vec{q}}_{(\gamma, \delta, \vec{v})' (\gamma, \delta, \vec{v})}(h)  v_{(\gamma, \delta, \vec{v})}(t, \vec{q}) \Big].
\end{split}
\end{equation}

\subsection{Real-space extent of the two-particle operators}\label{sec:App_Delta}
The real-space two-particle retarded Green's function for general position arguments is
\begin{equation}
\begin{split}
    g^{(2)}_{\bm{r}_1\bm{r}_2 , \bm{r}'_2\bm{r}'_1; \alpha \beta, \gamma \delta}(t)
     &= -\Theta(t) \expval{ [
    \hat{c}_{\bm{r}_2, \beta}(t)
    \hat{c}_{\bm{r}_1, \alpha}(t),
    \hat{c}^{\dagger}_{\bm{r}'_1, \gamma}(0) \hat{c}^{\dagger}_{\bm{r}'_2, \delta}(0) ] }_{\mathrm{GS}}.
\end{split}
\end{equation}
The Fourier transformed Green's function is
\begin{equation}
\begin{split}
      g^{(2)}_{\bm{k}_1, \bm{k}_2, \vec{k}'_1, \vec{k}'_2, \alpha \beta, \gamma \delta}(t)&= \frac{1}{N^2} \sum_{\bm{r}_1, \bm{r}_2, \bm{r}_1', \bm{r}_2'} e^{\mathrm{i} (\vec{k}_1 \cdot \vec{r}_1 + \vec{k}_2 \cdot \vec{r}_2 - \vec{k}_1' \cdot \vec{r}_1' - \vec{k}_2' \cdot \vec{r}_2')} g^{(2)}_{\bm{r}_1\bm{r}_2 , \bm{r}'_2\bm{r}'_1; \alpha \beta, \gamma \delta}(t).
\end{split}
\end{equation}
Inserting momentum conservation, $\delta_{\vec{k}_1+\vec{k}_2, \vec{k}'_1+\vec{k}'_2}$, one gets
\begin{equation}
\begin{split}
    g^{(2)}_{\bm{k}_1, \bm{k}_2, \alpha \beta, \gamma \delta}(t, \bm{q})&= \frac{1}{N^2} \sum_{\bm{r}_1, \bm{r}_2, \bm{r}_1', \bm{r}_2'} e^{\mathrm{i} (\vec{k}_1 \cdot \vec{r}_1 + (- \vec{k}_1+ \vec{q}) \cdot \vec{r}_2 - \vec{k}_2 \cdot \vec{r}_1' - (- \vec{k}_2+ \vec{q}) \cdot \vec{r}_2')} g^{(2)}_{\bm{r}_1\bm{r}_2 , \bm{r}'_2\bm{r}'_1; \alpha \beta, \gamma \delta}(t)\\
    &=\frac{1}{N^2} \sum_{\bm{r}_1, \bm{r}_2, \vec{u}, \vec{v}} e^{\mathrm{i} (\vec{k}_1 \cdot \vec{r}_1 + (- \vec{k}_1+ \vec{q}) \cdot (\vec{r}_1 + \vec{u}) - \vec{k}_2 \cdot \vec{r}_2 - (- \vec{k}_2+ \vec{q}) \cdot (\vec{r}_2 + \vec{v})} g^{(2)}_{\bm{r}_1,\bm{r}_1+\vec{u} , \bm{r}_2,\bm{r}_2+\vec{v}; \alpha \beta, \gamma \delta}\\
    &= \frac{1}{N} \sum_{\vec{u}, \vec{u'}} e^{\mathrm{i} (- \vec{k}_1+ \vec{q}) \cdot \vec{u} -\mathrm{i}  (- \vec{k}_2+ \vec{q}) \cdot\vec{v}} \Big\{ \frac{1}{N} \sum_{\bm{r}_1, \bm{r}_2} e^{\mathrm{i} \vec{q}\cdot(\vec{r}_1 - \vec{r}_2)} g^{(2)}_{\bm{r}_1,\bm{r}_2; \alpha \beta \vec{u}, \gamma \delta  \vec{v}}(t) \Big\}.
\end{split}
\end{equation}
In the last line, we single out the transformed Green's function that only depends on the pair momentum $\vec{q}$, corresponding to the term in curly brackets,
\begin{equation}
    g^{(2)}_{\alpha \beta \vec{u}, \gamma \delta  \vec{v}}(t, \vec{q}) := \frac{1}{N} \sum_{\bm{r}_1, \bm{r}_2} e^{\mathrm{i} \vec{q}\cdot(\vec{r}_1 - \vec{r}_2)} g^{(2)}_{\bm{r}_1,\bm{r}_2; \alpha \beta \vec{u}, \gamma \delta  \vec{v}}(t).
\end{equation}
This Green's function is obtained by only transforming the center of mass position of the created or annihilated pairs, while keeping the relative position as a parameter.
The outlined construction allows to include longer-range correlations perturbatively: the maximal value assumed by the correlation cluster, which is linked to the maximal length of $\vec{u}, \vec{v}$, corresponds to a radius within which correlations are probed. When constructing the basis for $\Gt(q)$, one should first fix the maximal value of $\vec{u}$, and include symmetry equivalent sites for every choice of $\vec{r}_1, \alpha$. Hence, the sets of $(\beta,\vec{u}), (\delta,\vec{v})$ depend on the value of $\alpha, \gamma$ respectively.

Let us consider an explicit example of basis construction, for the case of the honeycomb lattice.
This task requires listing all the possible pairs of creation operators that assume the indexes $\{\alpha, \beta,\vec{u}\}$, once $\vec{r}_1$ is fixed.
Let us consider the unit cell at position $\vec{r}$ enclosed in dashed lines in Figure~\ref{fig:hubbardsquare_noninteracting_Repr}\textcolor{red}{c}: the unit cell is composed by two sublattice sites, labeled by A and B, and we consider a single spinful $p_z$ orbital placed at each site.

The set of basis operator pairs for the zero-radius size of cluster are enumerated in Tab.~\ref{tab:hex_basis_0th}. Increasing the radius of the correlation to the first next smallest cluster, adds the list of operators of Tab.~\ref{tab:hex_basis_1st} to the basis set.

\begin{table}[t!]
    \centering
    \begin{tabular}{c @{\hspace{5pt}} c  @{\hspace{5pt}}  c @{\hspace{5pt}} c}
    \hline
    \hline
          $(\vec{r}$, A, $\uparrow)$ & $(\vec{r},$ A$,\downarrow)$ & $(\vec{r}$, B, $\uparrow)$ & $(\vec{r}$, B, $\downarrow)$ \\
    \hline
         $\hat{c}^{\dagger}_{\vec{r}, A \uparrow}\hat{c}^{\dagger}_{\vec{r}, A \downarrow}$
         & $\hat{c}^{\dagger}_{\vec{r}, A \downarrow}\hat{c}^{\dagger}_{\vec{r}, A \uparrow}$ 
         & $\hat{c}^{\dagger}_{\vec{r}, B \uparrow}\hat{c}^{\dagger}_{\vec{r}, B \downarrow}$
         & $\hat{c}^{\dagger}_{\vec{r}, B \downarrow}\hat{c}^{\dagger}_{\vec{r}, B \uparrow}$\\[2pt]
    \hline
    \hline
    \end{tabular}
    \caption{Radius-0 basis operators for each choice of $\vec{r}, \alpha$, shown in the first row.}
    \label{tab:hex_basis_0th}
\end{table}

\begin{table}[t!]
    \centering
    \begin{tabular}{c  @{\hspace{12pt}} c  @{\hspace{12pt}} c  @{\hspace{12pt}} c}
    \hline
    \hline
      \multicolumn{2}{c}{ $(\vec{r}$, A, $\uparrow/\downarrow)$} & \multicolumn{2}{c}{$(\vec{r}$, B,  $\uparrow/\downarrow)$} \\
        \hline
        $\vec{u}=0$ & $\vec{u}=\vec{e}_1,\vec{e}_2$ & $\vec{u}=0$ &  $\vec{u}=-\vec{e}_1,-\vec{e}_2$ \\
    \hline    
     $\hat{c}^{\dagger}_{\vec{r}, A \uparrow}\hat{c}^{\dagger}_{\vec{r}, B \uparrow}$ &   $\hat{c}^{\dagger}_{\vec{r}, A \uparrow}\hat{c}^{\dagger}_{\vec{r}+\vec{u}, B \uparrow}$ & $\hat{c}^{\dagger}_{\vec{r}, B \uparrow}\hat{c}^{\dagger}_{\vec{r}, A \uparrow}$ & $\hat{c}^{\dagger}_{\vec{r}, B \uparrow}\hat{c}^{\dagger}_{\vec{r}+\vec{u}, A \uparrow}$  \\
         $\hat{c}^{\dagger}_{\vec{r}, A \uparrow}\hat{c}^{\dagger}_{\vec{r}, B \downarrow}$  &  $\hat{c}^{\dagger}_{\vec{r}, A \uparrow}\hat{c}^{\dagger}_{\vec{r}+\vec{u}, B \downarrow}$& $\hat{c}^{\dagger}_{\vec{r}, B \uparrow}\hat{c}^{\dagger}_{\vec{r}, A \downarrow}$ & $\hat{c}^{\dagger}_{\vec{r}, B \uparrow}\hat{c}^{\dagger}_{\vec{r}+\vec{u}, A \downarrow}$  \\
         $\hat{c}^{\dagger}_{\vec{r}, A \downarrow}\hat{c}^{\dagger}_{\vec{r}, B \uparrow}$  &  $\hat{c}^{\dagger}_{\vec{r}, A \downarrow}\hat{c}^{\dagger}_{\vec{r}+\vec{u}, B \uparrow}$ & $\hat{c}^{\dagger}_{\vec{r}, B \downarrow}\hat{c}^{\dagger}_{\vec{r}, A \uparrow}$& $\hat{c}^{\dagger}_{\vec{r}, B \downarrow}\hat{c}^{\dagger}_{\vec{r}+\vec{u}, A \uparrow}$  \\
         $\hat{c}^{\dagger}_{\vec{r}, A \downarrow}\hat{c}^{\dagger}_{\vec{r}, B \downarrow}$ & $\hat{c}^{\dagger}_{\vec{r}, A \downarrow}\hat{c}^{\dagger}_{\vec{r}+\vec{u}, B \downarrow}$ & $\hat{c}^{\dagger}_{\vec{r}, B \downarrow}\hat{c}^{\dagger}_{\vec{r}, A \downarrow}$ &
         $\hat{c}^{\dagger}_{\vec{r}, B \downarrow}\hat{c}^{\dagger}_{\vec{r}+\vec{u}, A \downarrow}$   \\[2pt]
    \hline
    \hline
    \end{tabular}
    \caption{Radius-1 basis operators for each choice of $\vec{r}, \alpha$, shown in the first row. For compactness, we listed in a unique column operators stemming from $\alpha$ that only differ by spin, as the spin degrees of freedom does not influence the position of the electronic operators.}
    \label{tab:hex_basis_1st}
\end{table}

\subsection{Green's functions in the flattened Hamiltonian limit}\label{appendix:flat H}
In this section we discuss in more detail how to obtain the spectrum of $\Go$ and $\Gt$ at zero frequency, in the limiting case of spectrally flattened Hamiltonian, see Sec.~\ref{sec:flat H}.
We consider once more the spectrally flattened Hamiltonian introduced in Eq.~\eqref{eq:flattened Hamiltonian}.
As in Sec.~\ref{sec:flat H}, we consider the local Hilbert space of each unit cell at position $\vec{r}$ to be composed by the states of three orbitals, each one containing two states paired by TRS. We denote the single particle states by $1, 2, 3$ and their TRS partners by $\bar{1}, \Bar{2}, \Bar{3}$, see Eq.~\eqref{eq:TRS of 1,2,3 orbitals}.
We consider the AL and MAL ground states defined in Eqs.~\eqref{eq:AL gs flat limit} and~\eqref{eq:MAL gs flat limit}.
\\
We build a basis for $\Go$ and $\Gt$ out of electronic operators that create or annihilate electrons in one of the six single-particle states of each unit cell, namely $\hat{c}^{\dagger}_{\vec{r},j}, \, \hat{c}_{\vec{r},j}$ ($j=1, 2, 3, \Bar{1}, \Bar{2}, \Bar{3}$).

As in Sec.~\ref{sec:flat H}, we consider as basis for $\Go$ the set of $\hat{c}^{\dagger}_{\Vec{r}, j}$ operators, and for $\Gt$ we consider any (antisymmetrized) combination of two such operators. Without antisymmetrization the resulting basis is not orthonormal, and can be taken care of by introducing a multiplication factor of $1/2$ in the definition of $\Gt$.

The one-particle retarded Green's function with the Hamiltonian defined as in~\eqref{eq:flattened Hamiltonian} becomes
\begin{equation}
\begin{split}
   g^{(1)}_{\vec{r}_1, i, \vec{r}_2, j}  & = - \frac{1}{\Delta} \expval{\hat{c}_{\vec{r}_1, i} \Big[\sum_{\substack{n \in \mathscr{H}_{N+1}}}\ket{n}\bra{n}\Big] \hat{c}^{\dagger}_{\vec{r}_2, j}}{\mathrm{GS}}  + \frac{1}{\Delta} \expval{\hat{c}^{\dagger}_{\vec{r}_2, j} \Big[\sum_{\substack{n \in \mathscr{H}_{N-1}}}\ket{n}\bra{n}\Big] \hat{c}_{\vec{r}_1, i}}{\mathrm{GS}}= \frac{1}{\Delta} \delta_{i j} \delta_{\vec{r}_1, \vec{r}_2} \lambda_1
\end{split}
\end{equation}
For the AL state, the eigenvalues and eigenstates of $\Go$ are
\begin{equation}
    \begin{split}
        \lambda_1 = \frac{1}{\Delta} &: \quad \hat{c}^{\dagger}_{\Vec{r}, 1}, \hat{c}^{\dagger}_{\Vec{r}, \bar{1}} \quad \text{(occupied)}, \\
         \lambda_1 = -\frac{1}{\Delta} &: \quad \hat{c}^{\dagger}_{\Vec{r}, 2}, \hat{c}^{\dagger}_{\Vec{r}, \bar{2}}, \hat{c}^{\dagger}_{\Vec{r}, 2}, \hat{c}^{\dagger}_{\Vec{r}, \bar{2}} \quad \text{(non-occupied)},
    \end{split}
\end{equation}
and for the MAL state
\begin{equation}
    \begin{split}
        \lambda_1 = 0 &:\quad \hat{c}^{\dagger}_{\Vec{r}, 1}, \hat{c}^{\dagger}_{\Vec{r}, \bar{1}} , \hat{c}^{\dagger}_{\Vec{r}, 2}, \hat{c}^{\dagger}_{\Vec{r}, \bar{2}} \quad \text{(half-occupied)}, \\
         \lambda_1 = -\frac{1}{\Delta} &: \quad \hat{c}^{\dagger}_{\Vec{r}, 2}, \hat{c}^{\dagger}_{\Vec{r}, \bar{2}} \quad \text{(non-occupied)}.
    \end{split}
\end{equation}

With the flattened Hamiltonian in Eq.~\eqref{eq:flattened Hamiltonian}, the two-particle retarded Green's function becomes
\begin{equation}\label{eq:G2 flat limit SI}
\begin{split}
   g^{(2)}_{\vec{r}_1, i, j, \Vec{u};  \vec{r}_2, m, l, \Vec{v}} =& \frac{1}{\Delta}\expval{\hat{c}_{\vec{r}_1 + \Vec{u}, j} \hat{c}_{\vec{r}_1, i} \Big[\sum_{n \in \mathscr{H}_{N+2}}\ket{n}\bra{n}\Big] \hat{c}^{\dagger}_{\vec{r}_2, l} \hat{c}^{\dagger}_{\vec{r}_2+\Vec{v}, m}}{\mathrm{GS}}  \\
   &\quad + \frac{1}{\Delta} \expval{\hat{c}^{\dagger}_{\vec{r}_2+\Vec{v}, l} \hat{c}^{\dagger}_{\vec{r}_2, m} \Big[\sum_{n \in \mathscr{H}_{N+2}}\ket{n}\bra{n}\Big] \hat{c}_{\vec{r}_1+\Vec{u}, j} \hat{c}_{\vec{r}_1, i}}{\mathrm{GS}}
\end{split}
\end{equation}
For the AL state, the set of eigenvalues and eigenstate is
\begin{table}[ht!]
    \centering
    \begin{tabular}{c@{\hspace{12pt}}c@{\hspace{12pt}}c}
    \hline
    \hline
      $\lambda_2 \Delta$   &  Eigenstates \\
      \hline
      && \\[-8pt]
      \multirow{2}{*}{$1$}   &  $ \hat{c}^{\dagger}_{\Vec{r}, 1}\hat{c}^{\dagger}_{\Vec{r}', \bar{1}}$ & occupied \\[8pt]
       & $\hat{c}^{\dagger}_{\Vec{r}, 2}\hat{c}^{\dagger}_{\Vec{r}', \bar{2}}, \hat{c}^{\dagger}_{\Vec{r}, 3}\hat{c}^{\dagger}_{\Vec{r}', \bar{3}}, \hat{c}^{\dagger}_{\Vec{r}, 2}\hat{c}^{\dagger}_{\Vec{r}', \bar{3}}, \hat{c}^{\dagger}_{\Vec{r}, 3}\hat{c}^{\dagger}_{\Vec{r}', \bar{2}}, 
        \hat{c}^{\dagger}_{\Vec{r}, 2}\hat{c}^{\dagger}_{\Vec{r}', 3},   \hat{c}^{\dagger}_{\Vec{r}, \bar{2}}\hat{c}^{\dagger}_{\Vec{r}', \bar{3}}$ & non-occupied\\[8pt]
        \hline
        && \\[-8pt]
        $0$ & $\hat{c}^{\dagger}_{\Vec{r}, 1}\hat{c}^{\dagger}_{\Vec{r}', j} \ (j = 2, \Bar{2}, 3, \Bar{3}), \quad \hat{c}^{\dagger}_{\Vec{r}, \bar{1}}\hat{c}^{\dagger}_{\Vec{r}', j} \ (j = 2, \Bar{2}, 3, \Bar{3})$ & mixed\\
         && \\[-8pt]
    \hline
    \hline
    \end{tabular}
    \label{tab:AL_G2_Eigenstates_flatH}
\end{table}
for any $\Vec{r}, \vec{r}' \in \Lambda$, and for the MAL state
\begin{table}[ht!]
    \centering
    \begin{tabular}{c@{\hspace{12pt}}c@{\hspace{12pt}}c}
    \hline
    \hline
      $\lambda_2 \Delta$   &  Eigenstates & \\
      \hline
      && \\[-8pt]
      $2$ & $\frac{1}{\sqrt{2}}(\hat{c}^{\dagger}_{\Vec{r}, 1}\hat{c}^{\dagger}_{\Vec{r}, \bar{2}} - \hat{c}^{\dagger}_{\Vec{r}, \bar{1}}\hat{c}^{\dagger}_{\Vec{r}, 2})$ & occupied \\[8pt]
      \hline
      $1$ & $\hat{c}^{\dagger}_{\Vec{r}, 3}\hat{c}^{\dagger}_{\Vec{r}', \bar{3}}$ & non-occupied \\[8pt]
      \hline
      && \\[-8pt]
      \multirow{2}{*}{$\frac{1}{2}$}   &  $\hat{c}^{\dagger}_{\Vec{r}, 1}\hat{c}^{\dagger}_{\Vec{r}', j}, \ \hat{c}^{\dagger}_{\Vec{r}, \bar{1}}\hat{c}^{\dagger}_{\vec{r}', j}, \ \hat{c}^{\dagger}_{\Vec{r}, 2}\hat{c}^{\dagger}_{\Vec{r}', j}, \ \hat{c}^{\dagger}_{\Vec{r}, \bar{2}}\hat{c}^{\dagger}_{\Vec{r}', j} \ (j = 3, \Bar{3})$ & mixed\\[8pt]
      & $\frac{1}{\sqrt{2}}(\hat{c}^{\dagger}_{\Vec{r}, 1} \hat{c}^{\dagger}_{\Vec{r}', \bar{2}} - \hat{c}^{\dagger}_{\Vec{r}', \bar{1}}\hat{c}^{\dagger}_{\Vec{r}, 2}) \ \ \text{if} \ \Vec{r}\neq \Vec{r}'$\\[8pt]
      \hline
      && \\[-8pt]
    \multirow{2}{*}{$0$}  &  $\hat{c}^{\dagger}_{\Vec{r}, 1}\hat{c}^{\dagger}_{\Vec{r}', \bar{1}},\ \hat{c}^{\dagger}_{\Vec{r}, 2}\hat{c}^{\dagger}_{\Vec{r}', \bar{2}},\ \hat{c}^{\dagger}_{\Vec{r}, 1}\hat{c}^{\dagger}_{\Vec{r}', 2},\ \hat{c}^{\dagger}_{\Vec{r}, \bar{1}}\hat{c}^{\dagger}_{\Vec{r}, \bar{2}}$\\[8pt]
         & $\frac{1}{\sqrt{2}}(\hat{c}^{\dagger}_{\Vec{r}, \bar{2}} + \hat{c}^{\dagger}_{\Vec{r}', \bar{1}}\hat{c}^{\dagger}_{\Vec{r}, 2}) \ \ \text{if} \  \Vec{r}\neq \Vec{r}'$ \\
    && \\[-8pt]
    \hline
    \hline
    \end{tabular}
    \label{tab:MAL_G2_Eigenstates_flatH}
\end{table}
In the following, we rely on the notion of these expectation values to explicitly compute the spectra of $\Gt$ in momentum space in the flattened Hamiltonian limit.

\paragraph{1D example: MAL ground state}
Let us consider a one-dimensional chain, with two spinful orbitals per unit cell, which we label by $a$ and $b$. In the following, we denote $\hat{a}_{r\sigma}, \hat{a}^{\dagger}_{r\sigma}$ ($\hat{b}_{r\sigma}, \hat{b}^{\dagger}_{r\sigma}$) as the annihilation and creation operators for the orbital $a$ ($b$), with spin $\sigma$ and position corresponding to the unit cell $r$.
We consider the MAL ground state
\begin{equation}\label{eq:MAL GS flat limit appendix}
   \ket{\mathrm{GS}} = \prod_r \frac{1}{\sqrt{2}}(\hat{a}^{\dagger}_{r\uparrow}\hat{b}^{\dagger}_{r\downarrow} - \hat{a}^{\dagger}_{r\downarrow}\hat{b}^{\dagger}_{r\uparrow}) \ket{0} = \prod_r \hat{O}^{\dagger}_{r, \xi}\ket{0},
\end{equation}
where we assume TRS $\mathcal{T}$, hence the two pairs of operators at each position $r$ should be TRS related: $\mathcal{T}\hat{a}^{\dagger}_{r\uparrow}\hat{b}^{\dagger}_{r\downarrow} = -\hat{a}^{\dagger}_{r\downarrow}\hat{b}^{\dagger}_{r\uparrow}$.
In addition, we constrain the radius of the correlations probed by $\Gt$ to be $u_{\max}=1$.
The basis elements $\hat{v}_i$, $i=1,\cdots, 29$ are listed in Tab.~\ref{tab:basis}.
\begin{table}[ht!]
    \centering
    \begin{tabular}{cc @{\hspace{12pt}}cc}
    \hline
    \hline
    \multicolumn{4}{c}{$\hat{v}_{i}$ for $u=0$}\\
   \hline
        (1) &\quad $\hat{a}^{\dagger}_{r\uparrow} \hat{a}^{\dagger}_{r\downarrow} $ & (9)  &\quad $\hat{b}^{\dagger}_{r\uparrow} \hat{b}^{\dagger}_{r\uparrow} $ \\
        (2) &\quad $\hat{a}^{\dagger}_{r\uparrow} \hat{b}^{\dagger}_{r\uparrow} $ & (10)  &\quad $\hat{b}^{\dagger}_{r\uparrow} \hat{b}^{\dagger}_{r\downarrow}$\\
        (3) &\quad $\hat{a}^{\dagger}_{r\uparrow} \hat{b}^{\dagger}_{r\downarrow} $& (11)  &\quad $\hat{b}^{\dagger}_{r\downarrow} \hat{a}^{\dagger}_{r\uparrow} $\\
        (4) &\quad $\hat{a}^{\dagger}_{r\downarrow} \hat{a}^{\dagger}_{r\uparrow} $ &(12)  &\quad $\hat{b}^{\dagger}_{r\downarrow} \hat{a}^{\dagger}_{r\downarrow} $\\
        (5) &\quad $\hat{a}^{\dagger}_{r\downarrow} \hat{b}^{\dagger}_{r\uparrow} $ & (13)  &\quad $\hat{b}^{\dagger}_{r\downarrow} \hat{b}^{\dagger}_{r\uparrow} $\\
        (6) &\quad $\hat{a}^{\dagger}_{r\downarrow} \hat{b}^{\dagger}_{r\downarrow} $&&\\
        (7)  &\quad $\hat{b}^{\dagger}_{r\uparrow} \hat{a}^{\dagger}_{r\uparrow} $&&\\
        (8)  &\quad $\hat{b}^{\dagger}_{r\uparrow} \hat{a}^{\dagger}_{r\downarrow} $&&\\[3pt]
         \hline
       \end{tabular} \hspace{0.5cm}
        \begin{tabular}{cc@{\hspace{12pt}}cc}
        \hline
        \hline
        \multicolumn{4}{c}{$\hat{v}_{i}$ for $u=\pm1$} \\
        \hline
        (14) &\quad $\hat{a}^{\dagger}_{r\uparrow} \hat{a}^{\dagger}_{r+u\uparrow} $& (22) &\quad $\hat{b}^{\dagger}_{r\uparrow} \hat{a}^{\dagger}_{r+u\uparrow} $\\
        (15) &\quad $\hat{a}^{\dagger}_{r\uparrow} \hat{a}^{\dagger}_{r+u\downarrow} $&(23) &\quad $\hat{b}^{\dagger}_{r\uparrow} \hat{a}^{\dagger}_{r+u\downarrow} $\\
        (16) &\quad $\hat{a}^{\dagger}_{r\uparrow} \hat{b}^{\dagger}_{r+u\uparrow} $ & (24) &\quad $\hat{b}^{\dagger}_{r\uparrow} \hat{b}^{\dagger}_{r+u\uparrow} $\\
        (17) &\quad $\hat{c}^{\dagger}_{r, 1a, \uparrow}
        \hat{b}^{\dagger}_{r+u\downarrow} $ & (25) &\quad $\hat{b}^{\dagger}_{r\uparrow} \hat{b}^{\dagger}_{r+u\downarrow} $\\
        (18) &\quad $\hat{a}^{\dagger}_{r\downarrow} \hat{a}^{\dagger}_{r+u\uparrow} $&(26) &\quad $\hat{b}^{\dagger}_{r\downarrow} \hat{a}^{\dagger}_{r+u\uparrow} $\\
        (19) &\quad $\hat{a}^{\dagger}_{r\downarrow} \hat{a}^{\dagger}_{r+u\downarrow} $ & (27) &\quad $\hat{b}^{\dagger}_{r\downarrow}
        \hat{a}^{\dagger}_{r+u\downarrow} $\\
        (20) &\quad $\hat{a}^{\dagger}_{r\downarrow} \hat{b}^{\dagger}_{r+u\uparrow} $ & (28) &\quad $\hat{b}^{\dagger}_{r\downarrow} \hat{b}^{\dagger}_{r+u\uparrow} $\\
        (21) &\quad $\hat{a}^{\dagger}_{r\downarrow} \hat{b}^{\dagger}_{r+u\downarrow} $ & (29) &\quad $\hat{b}^{\dagger}_{r\downarrow} \hat{b}^{\dagger}_{r+u\downarrow} $\\[3pt]
        \hline
    \end{tabular}
    \caption{Basis for the Green's function operators in the $u=0$, $u=\pm 1$ sectors.}
    \label{tab:basis}
\end{table}
Note that this basis is not anti-symmetrized, hence the eigenvalues resulting from the diagonalization in this basis should be multiplied by a factor of $1/2$ to take into account the anti-symmetrization factor.

We now want to compute the values of all the blocks entering in the matrix
\begin{equation}\label{eq:G2rr' general form}
     g^{(2)}_{r r'} = \frac{1}{2} \begin{pmatrix}
   G^{0, 0}_{r r'} & G^{0, 1}_{r r'} & G^{0, -1}_{r r'} \\
    G^{1, 0}_{r r'} & G^{1, 1}_{r r'} & G^{1, -1}_{r r'} \\
     G^{-1, 0}_{r r'} & G^{-1, 1}_{r r'} & G^{-1, -1}_{r r'}
    \end{pmatrix}
\end{equation}
where we use the notation $G^{u_1, u_2}_{r r'}$, with $u_1, u_2 \in \{0, 1, -1\}$, with the index $u_1$ ($u_2$) referred to $r$ ($r'$), and the $\frac{1}{2}$ factor accounts for the antisymmetrization.

For the case of $r=r'$, we find
\begin{equation}\label{eq:non-zero G2rr' flat MAL}
\begin{split}
    &G^{u_1=0, u_2=0}_{r,r} = \mathbf{A} = \begin{pmatrix} \mathbf{M} & \mathbf{0}_{d_{V_a} \times d_{V_b}} \\
    \mathbf{0}_{d_{V_b} \times d_{V_a}} & \mathbf{0}_{d_{V_b} \times d_{V_b}}
    \end{pmatrix}, \quad \mathbf{M} = \frac{1}{2\Delta}\begin{pmatrix}
     +1 & +1 & -1 & -1\\
     +1 & +1 & -1 & -1\\
     -1 & -1 & +1 & +1\\
     -1 & -1 & +1 & +1\\
    \end{pmatrix}_{V_a},\\
    & G^{u_1=1, u_2=1}_{r,r} =  G^{u_1=-1, u_2=-1}_{r,r} =\mathbf{B} = \frac{1}{4\Delta} \mathbbm{1}_{16\times 16}
\end{split}
\end{equation}
where $V_a$ indicates the subspace containing basis elements $\{\hat{v}_3, \hat{v}_5, \hat{v}_8, \hat{v}_{11}\}$, and $V_b$ all the remaining elements with $u=0$. We indicated with $\mathbf{0}_{n \times m}$ the null matrix of dimension $(n, m)$, and $d_{V_a}, d_{V_b}$ indicate the dimension of the two subspaces. All the blocks not listed in~\eqref{eq:non-zero G2rr' flat MAL} have all entries equal to zero.
For $g^{(2)}_{r r'}$, with $r\neq r'$, the only non-zero contributions are
\begin{equation}
    G^{u_1=1, u_2=-1}_{r, r+1} = -\mathbf{B}, \qquad G^{u_1=-1, u_2=1}_{r, r-1} = -\mathbf{B}.
\end{equation}
We now compute the explicit form of $\Gt(q)$, starting from the $g^{(2)}_{rr'}$ derived above. The definition reads
\begin{equation}
    \Gt(q) = \frac{1}{N} \sum_{r, r'} e^{\mathrm{i}(r-r')q} g^{(2)}_{r r'}.
\end{equation}
The only non-zero contributions in the sum correspond to terms with $r'=r$, $r'=r+1$ and $r'=r-1$.
Hence,
\begin{equation}
\begin{split}
    \Gt(q) &=\frac{1}{N} \sum_{r=1}^N g^{(2)}_{r r} + \frac{1}{N} \sum_{r, r, r\neq r'}  e^{\mathrm{i}q (r-r')} g^{(2)}_{r r'} =\frac{1}{N} \sum_{r=1}^N g^{(2)}_{r r} + \frac{1}{N} \sum_{r=1}^N \sum_{r'=r+1, r-1}  e^{\mathrm{i}q (r-r')} g^{(2)}_{r r'} \\
    &= \frac{1}{N} \sum_{r=1}^N g^{(2)}_{r r} + \frac{1}{N} \sum_{r=1}^N (e^{\mathrm{i}q} g^{(2)}_{r,r-1} + e^{-\mathrm{i}q} g^{(2)}_{r,r+1}).
\end{split}
\end{equation}

Inserting the explicit form of the matrices $g^{(2)}_{r r'}$ found above, we get
\begin{equation}\label{eq:Flat H Gq MAL}
\begin{split}
    \Gt(q) &= \begin{pmatrix}
   \mathbf{A} & 0 & 0 \\
   0 & \mathbf{B} & -\mathbf{B}e^{-\mathrm{i}q} \\
   0& -\mathbf{B}e^{\mathrm{i}q} & \mathbf{B}
    \end{pmatrix} := \begin{pmatrix}
    \mathbf{A} & 0 \\
     0 & \tilde{\mathbf{B}} \\
    \end{pmatrix}.
\end{split}
\end{equation}
After diagonalization, we obtain
\begin{equation}\label{eq:Gq diagonal MAL}
    \begin{split}
    U \Gt(q) U^{\dagger} = \begin{pmatrix}
     \Lambda_{\mathbf{A}} & 0 \\
     0& \Lambda_{\tilde{\mathbf{B}}} \\
     \end{pmatrix}, \qquad \Lambda_{\mathbf{A}} = \text{diag}\big(\frac{2}{\Delta}, \underbrace{0,\cdots, 0}_{d_{V_b}}\big), \quad \Lambda_{\tilde{\mathbf{B}}} = \text{diag}\big(\underbrace{\frac{1}{\Delta}, \frac{1}{\Delta}, \cdots}_{8},\underbrace{0, \cdots, 0}_{8}\big),
    \end{split}
\end{equation}
with $ \Lambda_{\mathbf{A}}$ and $\Lambda_{\mathbf{B}}$ diagonal matrices, which we indicate by diag($\cdots$) and in parenthesis we list the diagonal entries. The eigenvalues obtained above are the same for any value of the momentum $q$.
The single eigenvalue below the gap at a fixed $q$ with eigenvalue $\lambda_2 = 2/\Delta$ corresponds to the eigenstate
\begin{equation}\label{eq:lowest eigenvalue MAL flatH}
   \hat{O}^{\dagger}_{q} = \sum_k \frac{1}{\sqrt{2}}(\hat{a}^{\dagger}_{k, \uparrow}\hat{b}^{\dagger}_{-k+q, \downarrow } - \hat{a}^{\dagger}_{k, \downarrow}  \hat{b}^{\dagger}_{-k+q, \uparrow })
\end{equation}
which is the real-space MAL operator appearing in the ground state of Eq.~\eqref{eq:MAL GS flat limit appendix}, transformed to momentum space.
In conclusion, the inverse spectrum of $\Gt$ is composed by a set of $q$-independent bands, one of which has $\lambda^{-1}_2=\Delta/2<\Delta$, and all the remaining ones have eigenvalues $\lambda^{-1}_2 \geq \Delta$, with $\Delta$ the many-body energy gap.

Note that~\eqref{eq:lowest eigenvalue MAL flatH} also proves that, for an MAL ground state and in the limit of flattened Hamiltonian, the lowest lying eigenvalue corresponds to an eigenvalue that transforms in the same representation as the MAL operator out of which the ground state is constructed.

\vspace{10 pt}
\paragraph{1D example: AL ground state}
Let us again consider a one-dimensional chain, with two spinful orbitals per unit cell, which we label by $a$ and $b$.
We now consider the AL ground state
\begin{equation}
  \ket{\mathrm{GS}} = \prod_r \hat{a}^{\dagger}_{r\uparrow}\hat{a}^{\dagger}_{r\downarrow}\ket{0}
\end{equation}
where we assume TRS $\mathcal{T}$ of the ground state.
We want to compute all the $g^{(2)}_{r r'}$'s, as defined in in Eq.~\eqref{eq:G2rr' general form} and we consider the basis already defined in Tab.~\ref{tab:basis}.
For the case of $r=r'$, the non-zero blocks are
For the diagonal block $g^{(2)}_{r r}$, we find
\begin{equation}
    g^{u_1=0, u_2=0}_{r,r} = \mathbf{C} = \begin{pmatrix}
    \mathbf{P} & \mathbf{0}_{d_{V_p} \times d_{V_d}}\\
    \mathbf{0}_{d_{V_d} \times d_{V_p}} & \mathbf{0}_{d_{V_d} \times d_{V_d}}
    \end{pmatrix}, \quad \mathbf{P} = \frac{1}{2\Delta} \begin{pmatrix}
    +1 & -1 & 0 & 0 \\
    -1 & +1 & 0 & 0 \\
    0 & 0 & +1 & -1 \\
    0 & 0 & -1 & +1 
    \end{pmatrix}_{V_c}
\end{equation}
with $V_p = \text{span}(\{\hat{v}_1, \hat{v}_4, \hat{v}_{10}, \hat{v}_{13}\})$ and $V_d$ the obtained from the remaining states in the $u=0$ sector.
The remaining non-zero blocks are
\begin{equation}
    g^{u_1=1, u_2=1}_{rr} = g^{u_1=-1,u_2=-1}_{rr} = \mathbf{D} = \frac{1}{2\Delta} \text{diag}(1, 1, 0, 0, 1, 1, 0, 0, 0, 0, 1, 1, 0, 0, 1, 1).
\end{equation}
where the eigenvalue is $1$ if the two operators are both occupied or both non-occupied in the ground state, and $0$ else.
Hence, 
\begin{equation}
    g^{(2)}_{r r} = \begin{pmatrix}
     \mathbf{C} & 0 & 0 \\
     0 &  \mathbf{D} & 0 \\
     0 & 0 &  \mathbf{D}
    \end{pmatrix}.
\end{equation}
For the matrix $g_{r, r'=r\pm1}$, we find
\begin{equation}
     g_{r,r+1} = \begin{pmatrix}
     0 & 0 & 0 \\
     0 &  0 & -\mathbf{D} \\
     0 & 0 & 0
    \end{pmatrix}, \quad  g_{r,r-1} = \begin{pmatrix}
     0 & 0 & 0 \\
     0 &  0 & 0 \\
     0 & -\mathbf{D} & 0
    \end{pmatrix}.
\end{equation}
The Fourier transform finally gives
\begin{equation}\label{eq:Flat H Gq AL}
\Gt(q) = \begin{pmatrix}
\mathbf{C} & 0 & 0 \\
0 & \mathbf{D} & -e^{-\mathrm{i}q}\mathbf{D} \\
0 & -e^{\mathrm{i}q}\mathbf{D} & \mathbf{D}
\end{pmatrix}= \begin{pmatrix}
\mathbf{C} & 0 \\
0 & \tilde{\mathbf{D}}
\end{pmatrix}.
\end{equation}
and from the diagonalization one obtains
\begin{equation}\label{eq:Gq diagonal AL}
    U \Gt(q) U^{\dagger} = \begin{pmatrix}
     \Lambda_{\mathbf{C}} & 0 \\
     0& \Lambda_{\tilde{\mathbf{D}}} \\
     \end{pmatrix}, \qquad \Lambda_{\mathbf{C}} = \text{diag}\big(\frac{1}{\Delta},\frac{1}{\Delta}, 0,\cdots, 0\big), \quad \Lambda_{\tilde{\mathbf{D}}} = \text{diag}\big(\frac{1}{\Delta}, \frac{1}{\Delta}, \cdots, 0, \cdots, 0\big).
\end{equation}
The latter result implies that in the inverse $\Gt(q)$ spectra originating from a AL ground state, there are only flat bands at $\lambda^{-1}_2=\Delta$ or at infinity.

\vspace{10pt}
From the previous considerations, one can extrapolate the general behavior for the case of $u_{\max}>1$.
If one considers $u_{\max}>1$, the set of basis functions listed in Tab.~\ref{tab:basis} has to be enlarged, and it should include operators with $u=0, \pm 1, \pm 2,..., \pm u_{\text{max}}$.
By repeating calculation of $\Gt$ as the ones outlined above, one obtains a block-matrix structure analogous to the one in Eq.~\eqref{eq:Flat H Gq MAL} and~\eqref{eq:Flat H Gq AL}:
for example, for $u_{\max}=3$, we have and the AL ground state $\Gt(q)$ becomes
\begin{equation}\label{eq:larger cluster Gq flatH}
    \Gt(q)= \begin{pmatrix}
    \mathbf{C} & 0 & 0 & 0 & 0 & 0 & 0 \\
    0 & \mathbf{D} &  \mathbf{D} e^{\mathrm{i} q}  & 0 & 0 & 0 & 0 \\
    0 &  \mathbf{D} e^{-\mathrm{i} q}  &  \mathbf{D}  & 0 & 0 & 0 & 0 \\
    0 & 0 & 0 & \mathbf{D}  & \mathbf{D} e^{\mathrm{i}2 q}  & 0 & 0 \\
    0 & 0 & 0 & \mathbf{D} e^{-\mathrm{i} 2q}  & \mathbf{D}  & 0 & 0 \\
    0 & 0 & 0 & 0 & 0 & \mathbf{D}  &  e^{\mathrm{i}3 q}\mathbf{D}  \\
    0 & 0 & 0 & 0 & 0 & \mathbf{D} e^{-\mathrm{i} 3q}  & \mathbf{D}  \\
    \end{pmatrix}
\end{equation}
where we ordered the basis blocks according to the list $u=(0, 1, -1, 2, -2, 3, -3)$.
Now,~\eqref{eq:larger cluster Gq flatH} has eigenvalues $\Lambda_{\mathbf{C}}$ and $\bigoplus_{i=1}^3\Lambda_{\mathbf{D}}$, with the $\Lambda_{\mathbf{C}/\mathbf{D}}$ sets as defined in~\eqref{eq:Gq diagonal AL}. A similar result holds for MAL states, upon replacing $\mathbf{C}, \mathbf{D}$ by $\mathbf{A}, \mathbf{B}$ respectively.

\vspace{10pt}
\paragraph{Stacking property in the flattened Hamiltonian limit}
As an additional remark, note that in this limit one can see the property of stacking of MAL's and AL's, which results into `stacking' the flat bands in $\Gt(q)$ obtained from the individual spectra of the single non-overlapping MALs and ALs.
For example, considering now four spinful orbitals in the unit cell, with creation operators labeled by $a, b, c, d$ and subscript indicating the position and spin $\sigma$, and for a ground state of the type
\begin{equation}
    \ket{\mathrm{GS}} =\prod_{\vec{r}} \hat{O}^{\dagger}_{\vec{r}, \xi_1} \hat{O}^{\dagger}_{\vec{r}, \xi_2} \ket{0} , \qquad \hat{O}^{\dagger}_{\vec{r}, \xi_1} =  \frac{1}{\sqrt{2}}(\hat{a}^{\dagger}_{r\uparrow}\hat{b}^{\dagger}_{r\downarrow} - \hat{a}^{\dagger}_{r\downarrow}\hat{b}^{\dagger}_{r\uparrow}) , \quad \hat{O}^{\dagger}_{\vec{r}, \xi_2} =  \frac{1}{\sqrt{2}}(\hat{c}^{\dagger}_{r\uparrow}\hat{d}^{\dagger}_{r\downarrow} - \hat{c}^{\dagger}_{r\downarrow}\hat{d}^{\dagger}_{r\uparrow})
\end{equation}
the inverse spectra of $\Gt$ has two bands with eigenvalue $\lambda^{-1}_2=\Delta/2$, with eigenstates equivalent to $\hat{O}^{\dagger}_{\vec{r}, \xi_1}$ and $\hat{O}^{\dagger}_{\vec{r}, \xi_2}$.

\subsection{Particle-hole Green's functions}\label{Appendix:particle-hole G2}
As mentioned in the main text, the formalism developed in this paper is suitable to be extended to other kinds of correlations functions, other than the particle-particle retarded Green's functions. In particular, for the case $n=2$ one can define another kind of correlation function, namely the retarded Green's function of the particle-hole type, which we indicate by $\Gt_{\mathrm{ph}}$, defined as
\begin{equation}\label{eq:particle-hole Green's function}
    g^{(2)}_{\mathrm{ph} \ i j, l m}(\omega) = \mathrm{FT}_{\omega}[- \Theta(t)\expval{[\hat{c}^{\dagger}_{j}(t) \hat{c}_i(t), \hat{c}^{\dagger}_{l}(0) \hat{c}_m(0)]}_{\mathrm{GS}}].
\end{equation} 
This correlation function, re-expressed as a matrix and taken at imaginary frequency $\omega=0$, is positive semidefinite.

The spectrum of $\Gt_{\mathrm{ph}}$ can be used in diagnosing MAL states. 
To show this, we compute the spectrum of the particle-hole Green's function with an MAL ground state, in the limiting case of a flattened Hamiltonian (see Sec.~\ref{sec:flat H}).
Let us consider as an example of MAL state
\begin{equation}\label{eq:MAL gs for nn G2}
    \ket{\mathrm{MAL}} = \prod_{\vec{r}\in\Lambda}\frac{1}{\sqrt{2}}(\hat{c}^{\dagger}_{\vec{r}, 1}\hat{c}^{\dagger}_{\vec{r} , \bar{2}} - \hat{c}^{\dagger}_{\vec{r}, \bar{1}}\hat{c}^{\dagger}_{\vec{r}, 2}) \ket{0},
\end{equation}
where $\hat{c}^{\dagger}_{\vec{r}, j}$ and $\hat{c}^{\dagger}_{\vec{r}, \bar{j}}$, for $j=1, 2$, are the creation operators for orthogonal states related pairwise by TRS, see Eq.~\eqref{eq:TRS of 1,2,3 orbitals}.
For each term in the product~\eqref{eq:MAL gs for nn G2}, we can re-write the operator in parenthesis in four possible ways
\begin{equation}\label{eq:decomposition MAL in particle-hole app}
    \begin{split}
 \prod_{\vec{r}\in\Lambda}\frac{1}{\sqrt{2}}(\hat{c}^{\dagger}_{\vec{r}, 1}\hat{c}^{\dagger}_{\vec{r} , \bar{2}} - \hat{c}^{\dagger}_{\vec{r}, \bar{1}}\hat{c}^{\dagger}_{\vec{r}, 2}) \ket{0} = \prod_{\vec{r}\in\Lambda}
   \begin{cases}  
       & \frac{1}{\sqrt{2}}(\hat{c}^{\dagger}_{\vec{r},1} \hat{c}_{\vec{r},\bar{1}} - \hat{c}^{\dagger}_{\vec{r}, 2} \hat{c}_{\vec{r}, \bar{2}} ) \hat{c}^{\dagger}_{\vec{r}, \bar{1}}\hat{c}^{\dagger}_{\vec{r}, \bar{2}}\ket{0}\\[6pt]
       & \frac{1}{\sqrt{2}}(\hat{c}^{\dagger}_{\vec{r}, 1} \hat{c}_{\vec{r}, 2} + \hat{c}^{\dagger}_{\vec{r}, \bar{1}} \hat{c}_{\vec{r}, \bar{2}}) \hat{c}^{\dagger}_{\vec{r}, 2} \hat{c}^{\dagger}_{\vec{r}, \bar{2}}\ket{0}\\[6pt]
       & \frac{1}{\sqrt{2}}(-\hat{c}^{\dagger}_{\vec{r}, \bar{2}} \hat{c}_{\vec{r}, \bar{1}} - \hat{c}^{\dagger}_{\vec{r}, 2} \hat{c}_{\vec{r}, 1} ) \hat{c}^{\dagger}_{\vec{r}, \bar{1}}\hat{c}^{\dagger}_{\vec{r}, 1}\ket{0}\\[6pt]
       & \frac{1}{\sqrt{2}}(-\hat{c}^{\dagger}_{\vec{r}, \bar{2}} \hat{c}_{\vec{r}, 2} + \hat{c}^{\dagger}_{\vec{r}, \bar{1}} \hat{c}_{\vec{r}, 1} ) \hat{c}^{\dagger}_{\vec{r}, 2}\hat{c}^{\dagger}_{\vec{r}, 1}\ket{0}
       \end{cases}
    \end{split}
\end{equation}
One can repeat the calculation of App.~\ref{appendix:flat H}, but now for $\Gt_{\mathrm{ph}}$. The first difference as compared to the case of $\Gt$ is that the states appearing in the Lehmann decomposition of $\Gt_{\mathrm{ph}}$ belong to the $N$-particle sector, as the operators acting on the ground state are particle-number conserving. This results into terms whose denominator diverges due to the $1/(E_{m}-E_0)$ factor, but these terms also carry a trivial representation. Therefore, we can discard them by subtracting from $\Gt_{\mathrm{ph}}$ the divergent part.
Analogously to the case of $\Gt_{\mathrm{ph}}$, entangled operators in the ground state, such as the terms in parenthesis in each line of Eq.~\eqref{eq:decomposition MAL in particle-hole app}

The spectrum of $\Gt_{\mathrm{ph}}$ for the state in Eq.~\eqref{eq:decomposition MAL in particle-hole app} has several eigenvalues below the gap, i.\,e., with eigenvalue $\lambda^{-1}_{2, \mathrm{ph}}=\Delta/2$, whose eigenstates are listed in Tab.~\ref{tab:App eval below gap flat H G densitydensity}. There is a divergent eigenvalue, with $\Delta/\lambda_2\rightarrow0$ arising from the operators that when acting on the ground state result into a non-zero overlap with the ground state itself. There are in total six eigenvalues below the gap, four of them with eigenstates corresponding to the operators enclosed in parenthesis in each line of Eq.~\eqref{eq:decomposition MAL in particle-hole app}, in analogy to the lowest-lying eigenvalue in $\Gt$, and two additional eigenvalues derive from the density contribution of $\Gt_{\mathrm{ph}}$. Note that two of the eigenstates listed in Tab.~\ref{tab:App eval below gap flat H G densitydensity} are related by TRS.

\begin{table}[t!]
    \centering
    \begin{tabular}{c@{\hspace{12pt}}c}
    \hline
    \hline
   $\lambda_{2, \mathrm{ph}}^{-1}/\Delta$ & Eigenstate \\
 \hline
 $1/2$ & $\frac{1}{\sqrt{2}}(\hat{c}^{\dagger}_{\vec{r},1} \hat{c}_{\vec{r},\bar{1}} - \hat{c}^{\dagger}_{\vec{r}, 2} \hat{c}_{\vec{r}, \bar{2}} ) $  \\
 $1/2$ & $\frac{1}{\sqrt{2}}(\hat{c}^{\dagger}_{\vec{r}, 1} \hat{c}_{\vec{r}, 2} + \hat{c}^{\dagger}_{\vec{r}, \bar{1}} \hat{c}_{\vec{r}, \bar{2}}) $ \\
 $1/2$ & $\frac{1}{\sqrt{2}}(-\hat{c}^{\dagger}_{\vec{r}, \bar{2}} \hat{c}_{\vec{r}, \bar{1}} - \hat{c}^{\dagger}_{\vec{r}, 2} \hat{c}_{\vec{r}, 1} )$ \\
 $1/2$ & $\frac{1}{\sqrt{2}}(-\hat{c}^{\dagger}_{\vec{r}, \bar{2}} \hat{c}_{\vec{r}, 2} + \hat{c}^{\dagger}_{\vec{r}, \bar{1}} \hat{c}_{\vec{r}, 1} )$ \\
  $1/2$ & $\frac{1}{2}(\hat{c}_{\vec{r}, 1}^{\dagger}  \hat{c}_{\vec{r}, 1} -  \hat{c}_{\vec{r}, \bar{1}}^{\dagger}  \hat{c}_{\vec{r}, \bar{1}} - \hat{c}_{\vec{r}, 2}^{\dagger}  \hat{c}_{\vec{r}, 2}+  \hat{c}_{\vec{r}, \bar{2}}^{\dagger}  \hat{c}_{\vec{r}, \bar{2}})$\\
 0 & $\frac{1}{2}(\hat{c}_{\vec{r}, 1}^{\dagger}  \hat{c}_{\vec{r}, 1}+  \hat{c}_{\vec{r}, \bar{1}}^{\dagger}  \hat{c}_{\vec{r}, \bar{1}} + \hat{c}_{\vec{r}, 2}^{\dagger}  \hat{c}_{\vec{r}, 2}+  \hat{c}_{\vec{r}, \bar{2}}^{\dagger}  \hat{c}_{\vec{r}, \bar{2}})$ \\
    \hline
    \hline
    \end{tabular}
    \caption{Inverted eigenvalues of $\Gt_{\mathrm{ph}}$ below the gap $\Delta$ (fist column) with their respective eigenstates (second column) in the flattened Hamiltonian limit, for the ground state in Eq.~\eqref{eq:MAL gs for nn G2}.}
    \label{tab:App eval below gap flat H G densitydensity}
\end{table}

Following the same procedure adopted to fill the entries in Tab.~\ref{tab:App eval below gap flat H G densitydensity}, we compiled the list of irreps appearing below the particle-hole gap in the spectrum of $\Gt_{\mathrm{ph}}$ for each type of MAL entering in the 1D classification of Tab.~\ref{tab:List of AL and MAL in 1D}. As all the entries are distinguished, the spectrum of $\Gt_{\mathrm{ph}}$ provides an additional mean to diagnose MAL states.

\begin{figure}[t]
    \centering
    \includegraphics[width=0.5\textwidth]{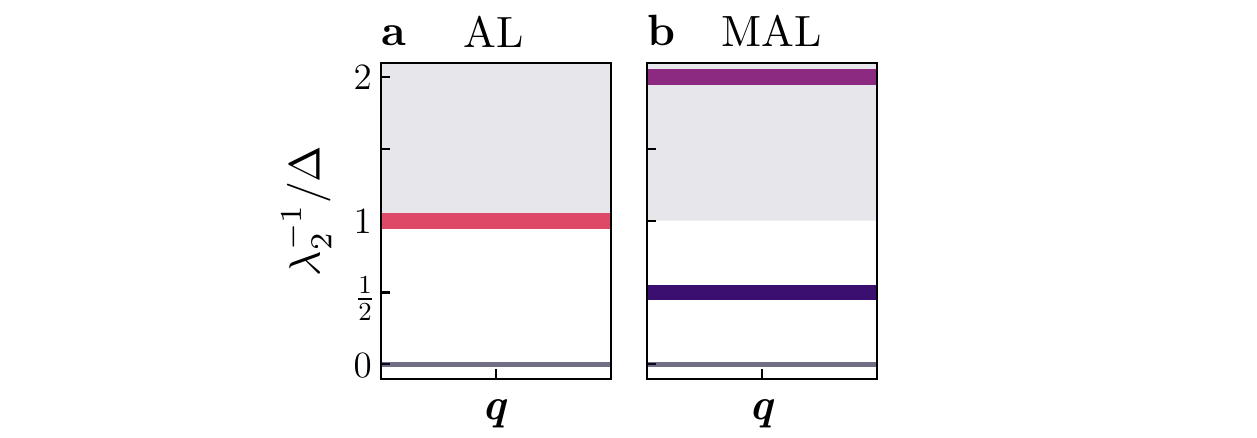}
    \caption{Inverted $\Gt_{\mathrm{ph}}$ spectrum for the \textbf{a} AL and \textbf{b} MAL ground states. The shaded region corresponds to the spectrum above the noninteracting bound. Thin lines correspond to singly degenerate eigenvalues.}
    \label{fig:flatH_gph}
\end{figure}

Figure~\ref{fig:flatH_gph} shows the spectrum of $\Gt_{\mathrm{ph}}$ for the AL and MAL ground state of Sec.~\ref{sec:flat H}.

\section{Examples of classifications of MAL-induced band representation in 2D}
\label{app:2D_Examples}

\begin{figure}[t!]
    \centering
    \includegraphics[width=0.5\textwidth]{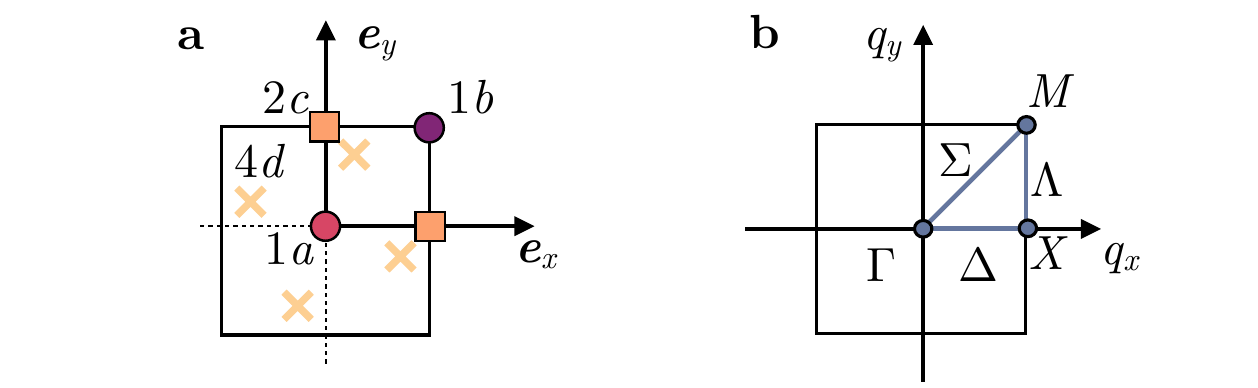}
    \caption{\textbf{a} Wyckoff positions and \textbf{b} Brillouin zone of a square lattice.}
    \label{fig:squarelattice}
\end{figure}
We consider as an explicit example the double point-group $C^D_{4v}$ ($4mm$ in Bilbao server notation \cite{BilbaoElcoro:ks5574}). 
The space group is $P4mm$ (No. 99), and has three maximal Wyckoff positions: $1a$, $1b$, $2c$, see Fig.~\ref{fig:squarelattice}\textcolor{red}{a}. The symmetries of this group are \cite{CrystTables}
\begin{equation}
    (1)\ \mathbbm{1} \quad (2) \ C_4 \quad (3) \  C^{-1}_4 \quad (4) \ C_2 \quad (5) \ m_{01} \quad (6) \ m_{10} \quad (7) \ m_{1 \bar{1}} \quad (8) \ m_{\bar{1} 1} 
\end{equation}
We focus on Wyckoff positions $1a$ and $1b$, as it is not possible to define a single MAL out of two orbitals placed at the $2c$ positions without the breaking of some of the lattice symmetries.

The Brillouin zone is characterized by four maximal momenta in the Brillouin zone, denoted as $\Gamma, X, M$ (Fig.~\ref{fig:squarelattice}).
The co-little groups at $\Gamma$ and $M$ are isomorphic to $C^D_{4v}$ (double point-group $4mm$ No. 13), while at $X$ is isomorphic to $C^D_{2v}$ (double point-group $mm2$ No. 7).
We list the character table for these two groups in Tabs.~\ref{tab:point-group of C4v at gamma and M} and~\ref{tab:point-group of C2v at X}. 
\begin{table}[ht]
    \centering
    \begin{tabular}{ c c | c c c c c c c}
    \hline
    \hline
     
   (a) &  (b) & $E$ & $C_4$ & $C_2$  & $ \sigma_v$ & $\sigma_d$ & $\bar{E}$ & $\bar{C}_4$ \\
    
         \hline
 $A_1$ & $\Gamma_1$ & 1 & 1 & 1 & 1 & 1 & 1 & 1 \\
 $B_1$ & $\Gamma_2$ &1 & -1 & 1 & 1 & -1 & 1 & -1 \\
 $B_2$ &  $\Gamma_3$ & 1 & -1 & 1 & -1 & 1 & 1 & -1 \\
 $A_2$ &$\Gamma_4$ &  1 & 1 & 1 & -1 & -1 & 1 & 1 \\
 $E$ &$\Gamma_5$ &  2 & 0 & -2 & 0 & 0 & 2 & 0 \\
 $\bar E_2$ &  $\bar\Gamma_6$ & 2 & -$\sqrt{2}$ & 0 & 0 & 0 & -2 & $\sqrt{2}$ \\
 $\bar E_1$ &$\bar\Gamma_7$ &  2 & $\sqrt{2}$ & 0 & 0 & 0 & -2 & -$\sqrt{2}$ \\
\hline
\hline
    \end{tabular}
    \caption{Character table of the little group $G_{\Gamma} \equiv C_{4v}$~\cite{BilbaoElcoro:ks5574}. The character table is the same for $G_{M} \sim C_{4v}$. In the text we use notation (b). For the point $M$ we use exactly the same representations with label $M$ instead of $\Gamma$.}
    \label{tab:point-group of C4v at gamma and M}
\end{table}

\begin{table}[ht]
    \centering
    \begin{tabular}{ c c|c c c c c }
    \hline
    \hline
    
     (a) & (b) & $E$ & $C_2$ & $\sigma_v(01)$ & $\sigma_v(10)$ & $\bar{E}$ \\
     
         \hline
  $A_1$ & $\Gamma_1$ & 1 & 1 & 1 & 1 & 1  \\
  $A_2$ & $\Gamma_2$ & 1 & 1 & -1 & -1 & 1  \\
  $B_2$ & $\Gamma_3$ &  1 & -1 & -1 & 1 & 1  \\
  $B_1$ & $\Gamma_4$ &   1 & -1 & 1 & -1 & 1  \\
  $\bar{E}$ &$\bar \Gamma_5$ &  2 & 0 & 0 & 0 & -2  \\
\hline
\hline
    \end{tabular}
    \caption{Character table of the little group $G_{X} \equiv C_{2v}$~\cite{BilbaoElcoro:ks5574}. In the text we use notation (b) where we replace $\Gamma$ by $X$.}
    \label{tab:point-group of C2v at X}
\end{table}

\begin{table}[ht]
    \centering
    \begin{tabular}{cc|cccc}
    \hline
    \hline
    
     (a) & (b) & $E$ & $\sigma_v(01)$ & $\bar{E}$ & $\bar{\sigma}_v(01)$ \\ 
    
     \hline
       $A'$  & $\Gamma_1$  & $1$ & 1 & 1 & 1  \\
        $A''$ &  $\Gamma_2$   & $1$ & $-1$ & 1 & $-1$ \\
        $^2\bar{E}$ & $\bar{\Gamma}_3$  & 1 & $-\mathrm{i}$ & $-1$ &  $\mathrm{i}$  \\
         $^1\bar{E}$ &  $\bar{\Gamma}_4$  & 1 &  $\mathrm{i}$ & $-1$ &  $-\mathrm{i}$ \\
         \hline
         \hline
    \end{tabular}
    \caption{Character table of the little co-group of $C_s$~\cite{BilbaoElcoro:ks5574}.}
    \label{tab:character table Cs}
\end{table}

To list all the possible MALs compatible with $C^D_{4v}$ and with clusters entangling electrons within the same Wyckoff position, we consider all the irreps of the allowed orbitals at $1a$ and $1b$. For each site-symmetry group, we write the list of allowed orbital representations $\rho_{\alpha}$ and the representation induced in momentum space by the site-symmetry representation $\rho_{\Vec{q}^W}$. Then, two particle representations can be obtained by combining the orbital representations
\begin{equation}\label{eq:square lattice irrep decomposition}
    \rho_{\bm{k}, \alpha,W} = \rho_{\alpha} \otimes  \rho_{\bm{k}}^W.
\end{equation}
When the decomposition in Eq.~\eqref{eq:square lattice irrep decomposition} is allowed, the contribution coming from the internal degrees of freedom is factored out from the sum over momenta
\begin{equation}\label{eq:square lattice rho2}
\begin{split}
    \rho^{(2)}_{\bm{q}, \alpha, W_1, \beta, W_2}
    &=  \rho_{\alpha} \otimes \rho_{\beta} \otimes \rho^W_{\Vec{q}}.
\end{split}
\end{equation}
By computing the tensor product in~\eqref{eq:square lattice rho2}, one can derive the full list of allowed irreps for two-particle operators.
Finally, we subduce the representations in Eq.~\eqref{eq:square lattice rho2} to the little groups of the maximal momentum points in the Brillouin zone.

For orbitals placed at Wyckoff positions $1a$ and $1b$, we evaluate the form of each $\rho^W_{\bm{k}}$ for a subset of elements of $C_{4v}$: $\{C_4, C_2, m_{01}, m_{1\bar{1}}\}$.
We have
\begin{equation}\label{eq:C4v represent momentum at 1a}
    \rho^{1a}_{\bm{q}}(\{S|00\}) = 1, \quad \rho^{1a}_{\bm{q}}(\mathcal{T}) = \mathcal{K} \qquad \Rightarrow \qquad \rho^{1a}_{\bm{q}}
\end{equation}
where $S$ indicates any of the symmetry elements in $C^D_{4v}$ and $\mathcal{T}$ TRS. For the Wyckoff position $1b$, the representation of the elements of $C_{4v}$ is
\begin{align}\label{eq: square lattice rep at 1b}
     & \rho^{1b}_{\bm{q}}(\{C_4|00\}) = e^{\mathrm{i}\bm{q}\cdot \bm{e}_x}, &
    &\rho^{1b}_{\bm{q}}(\{C_2|00\}) = e^{\mathrm{i}\bm{q}\cdot (\bm{e}_x+\bm{e}_y)} &    &\rho^{1b}_{\bm{q}}(\mathcal{T}) = \mathcal{K} \\
    &\rho^{1b}_{\bm{q}}(\{m_{01}|00\}) = e^{\mathrm{i}\bm{q}\cdot \bm{e}_y} & &\rho^{1b}_{\bm{q}}(\{m_{1\bar{1}}|00\}) = 1 \nonumber.
\end{align}
The representation under which the orbitals transform, the $\rho_{\alpha}$ in Eq.~\eqref{eq:square lattice irrep decomposition}, corresponds to one of the spinful irreps of $C^D_{4v}$. 

\subsection{Example for \textit{1b} Wyckoff position}
We now carry out the explicit calculation of the allowed two-particle representations in the $\Gti$ spectrum for a square lattice, for some combinations of orbital representations and Wyckoff positions. We will go through the steps shortly outlined in the above paragraph, while following the methods developed in the TQC approach (see Supplementary Information of Ref.~\cite{Bradlyn2017}, Sec. III).
As a specific example, we consider two spinful orbitals, each one transforming in the $\Bar{E}_1$ representation of the point-group, both placed at the Wyckoff position $1b$.
The two electron states in a single $\Bar{E}_1$-orbital are $\ket{\Bar{E}_1, \uparrow}, \ket{\Bar{E}_1, \downarrow}$, and they transform in the representation $\rho_{\Bar{E}_1}$, given by
\begin{align}\label{eq:repr spin bar gamma_7}
    &\rho_{\Bar{E}_1}(\{C_4|00\}) = e^{\mathrm{i} \frac{\pi}{4} s_z} &
    &\rho_{\Bar{E}_1}(\{C_2|00\}) = \mathrm{i}s_z &    &\rho_{\Bar{E}_1}(\mathcal{T}) = \mathrm{i}s_y \\
    &\rho_{\Bar{E}_1}(\{m_{01}|00\}) = \mathrm{i} s_x &  &\rho_{\Bar{E}_1}(\{m_{1\bar{1}}|00\}) = \mathrm{i} s_x, \nonumber
\end{align}
where $s_i$ indicate Pauli matrices acting on the spin degrees of freedom. Comparing Eq.~\eqref{eq:repr spin bar gamma_7} with Tabs.~\ref{tab:point-group of C4v at gamma and M} and~\ref{tab:point-group of C2v at X}, we see that $(\rho_{\Bar{E}_1}\downarrow G_{\Gamma}) \equiv \bar{\Gamma}_7 \otimes \Gamma_1 = \bar{\Gamma}_7$ at the $k$-point $\Gamma$, $(\rho_{\Bar{E}_1}\downarrow G_{M}) \equiv \bar{M}_7 \otimes M_1 = \bar{M}_7$ at the $M$ point, and $(\rho_{\Bar{E}_1}\downarrow G_{X}) \equiv \bar{X}_5$ at the $X$ point in the Brillouin Zone.

Eq.~\eqref{eq:square lattice rho2} becomes
\begin{equation}
    \rho^{(2)}_{\bm{q}, \Bar{E}_1\oplus\Bar{E}_1, 1b} = \rho_{\Bar{E}_1} \otimes \rho_{\Bar{E}_1}\otimes \rho^{1b}_{\bm{q}}.
\end{equation}
Using Eqs.~\eqref{eq: square lattice rep at 1b} and~\eqref{eq:repr spin bar gamma_7}, we find the two-particle representation
\begin{align}\label{eq:rho 2 part 2s 1b}
   & \rho^{(2)}_{\bm{q}, 2\Bar{E}_1, 1b}(C_4) = \big( e^{\mathrm{i}\frac{\pi}{4} s_z} \otimes e^{\mathrm{i} \frac{\pi}{4} s_z} \big) e^{\mathrm{i}q_x}
&   & \rho^{(2)}_{\bm{q}, 2\Bar{E}_1, 1b}(C_2) = \mathrm{i}^2 (s_z \otimes s_z) e^{\mathrm{i}(q_x + q_y)} \\
    & \rho^{(2)}_{\bm{q}, 2\Bar{E}_1, 1b}(m_{01}) = \mathrm{i}^2 (s_x \otimes s_x) e^{\mathrm{i} q_y} & &\rho^{(2)}_{\bm{q}, 2\Bar{E}_1, 1b}(m_{1\bar 1}) = \mathrm{i}^2 (s_x \otimes s_x) \nonumber \\
    & \rho^{(2)}_{\bm{q}, 2\Bar{E}_1, 1b}(\mathcal{T}) = -s_y \otimes s_y \mathcal{K} \nonumber
\end{align}
Computing the trace of these expression at maximal momenta then leads to the decomposition in terms of irreps
\begin{equation}\label{eq:square lattice 2s1b irreps}
    \rho^{(2)}_{\Gamma, 2\Bar{E}_1, 1b} \equiv \Gamma_5 \oplus \Gamma_1 \oplus \Gamma_4, \quad \rho^{(2)}_{M, 2\Bar{E}_1, 1b} \equiv M_5 \oplus M_2 \oplus M_3,
    \quad \rho^{(2)}_{X, 2\Bar{E}_1, 1b} \equiv \rho^{(2)}_{Y, 2\Bar{E}_1, 1b} \equiv X_1 \oplus X_2 \oplus X_3 \oplus X_4
\end{equation}
A MAL operator composed by two $\bar{E}_1$-orbitals placed at $1b$ will result in a set of eigenvalues below the gap and at maximal points of $\bm{q}$ transforming in the same irreps as the ones in Eq.~\eqref{eq:square lattice 2s1b irreps}.

We are left with the task of computing the connectivity of these representations through compatibility relations.
We first consider the line $\Sigma$ connecting $\Gamma$ and $M$. This line is left invariant by $m_{1 \bar{1}}$, and the point-group is $C_s$ (see Tab.~\ref{tab:character table Cs}).
The compatibility relations for the $\Gamma$ point are
\begin{equation}\label{eq:square lattice connectivity rel gamma-M}
    \begin{split}
        \Gamma_5 \downarrow G_{\Sigma} = \Sigma_1 \oplus \Sigma_2, \quad
        \Gamma_1 \downarrow G_{\Sigma} = \Sigma_1 \quad
        \Gamma_4 \downarrow G_{\Sigma} = \Sigma_2
    \end{split}
\end{equation}
and for the $M$ point
\begin{equation}
    \begin{split}
        M_5 \downarrow G_{\Sigma} = \Sigma_1 \oplus \Sigma_2, \quad
        M_2 \downarrow G_{\Sigma} = \Sigma_2 \quad
        M_3 \downarrow G_{\Sigma} = \Sigma_1.
    \end{split}
\end{equation}
The line $\Lambda$ connecting $M$ and $X$ is invariant under symmetry operation $m_{10}$, and again the symmetry group is $C_s$.
The compatibility relations between $X$ and $\Lambda$ are
\begin{equation}
    X_1 \downarrow G_{\Lambda} = \Lambda_1, \quad 
    X_2 \downarrow G_{\Lambda} = \Lambda_2, \quad
    X_3 \downarrow G_{\Lambda} = \Lambda_1, \quad
    X_4 \downarrow G_{\Lambda} = \Lambda_2.
\end{equation}
while for $M$ and $\Lambda$ 
\begin{equation}\label{eq:square lattice connectivity rel M-X}
    \begin{split}
        M_5 \downarrow G_{\Lambda} = \Lambda_1 \oplus \Lambda_2, \quad
        M_2 \downarrow G_{\Lambda} = \Lambda_1 \quad
        M_3 \downarrow G_{\Lambda} = \Lambda_2.
    \end{split}
\end{equation}
The remaining line $\Delta$ connects $\Gamma$ and $X$ and is invariant under $m_{01}$, with symmetry group $C_s$.
The connectivity relations between $\Gamma$ and $\Delta$ are 
\begin{equation}
    \begin{split}
        \Gamma_5 \downarrow G_{\Delta} = \Delta_1 \oplus \Delta_2, \quad
        \Gamma_1 \downarrow G_{\Delta} = \Delta_1 \quad
        \Gamma_4 \downarrow G_{\Delta} = \Delta_2,
    \end{split}
\end{equation}
while between $X$ and $\Delta$ they are
\begin{equation}
    X_1 \downarrow G_{\Delta} = \Delta_1, \quad 
    X_2 \downarrow G_{\Delta} = \Delta_2, \quad
    X_3 \downarrow G_{\Delta} = \Delta_2, \quad
    X_4 \downarrow G_{\Delta} = \Delta_1.
\end{equation}

\begin{figure}[t]
    \centering
    \includegraphics[width=1\textwidth]{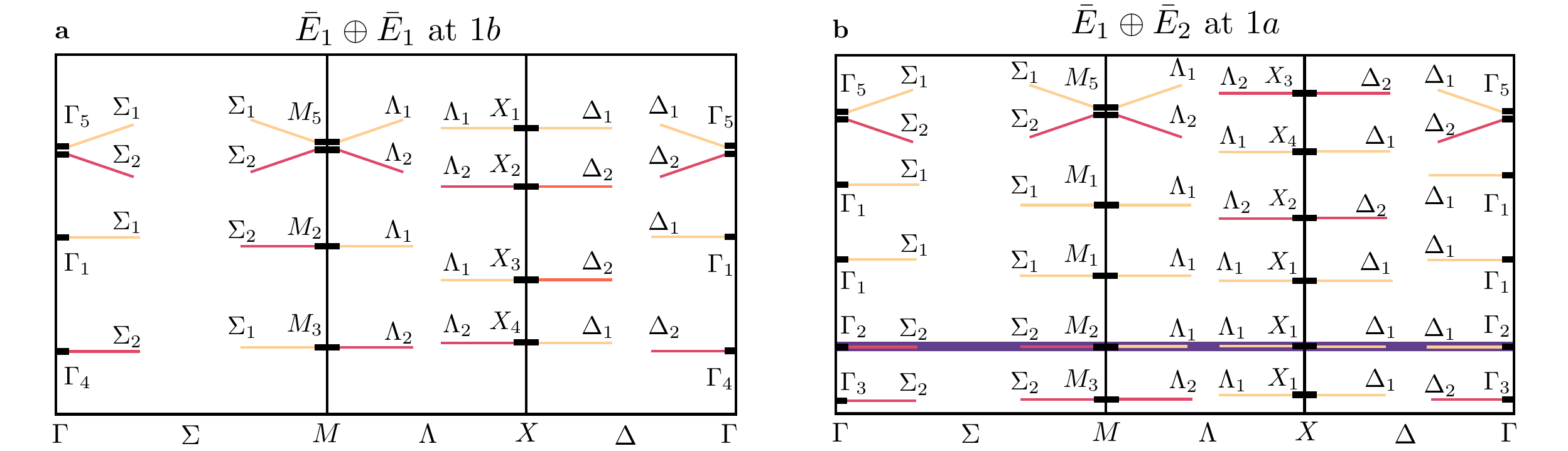}
    \caption{Representations of two-particle Green's function eigenvalues stemming from \textbf{a} two $\bar{E}_1$-orbitals placed at Wyckoff position $1b$ and \textbf{b} a $\bar{E}_1\oplus \Bar{E}_2$-orbitals placed at Wyckoff position $1a$. Representations are labeled according to notation (b). In \textbf{b}, the single band appearing in the $\Gt$ spectra of the diamond square lattice model in the regime $t_1 \ll t_3$ is highlighted in blue.}
    \label{fig:G2Bands_2Dexample}
\end{figure}

\paragraph{Basis functions}
For completeness, we write the explicit basis functions corresponding to the irreps of the eigenvalue structure in Fig.~\ref{fig:G2Bands_2Dexample}\textcolor{red}{a}.
Consider the representation of two $\Bar{E}_1$ orbital electrons, for now without momentum quantum number. At the maximally symmetric positions, this corresponds to the Kronecker product
\begin{equation}\label{eq:decomposition gm1 bargm7 bargm7}
    (A_1 \otimes \bar{E}_1)_a \otimes (A_1 \otimes \bar{E}_1)_b = E_{a b} \oplus A_{1, a b}  \oplus A_{2,  a b},
\end{equation}
where $a, b$ are indexes for the two $s$ orbitals. The basis function for each spinful orbital is 
\begin{equation}
    (A_1 \otimes \bar{E}_1)_i : \quad (\hat{c}^{\dagger}_{1b, \bar{E}_1^i, \downarrow}, \hat{c}^{\dagger}_{1b, \bar{E}_1^i, \uparrow}), \quad i=a, b.
\end{equation}
The basis functions of the irreps appearing in the decomposition~\eqref{eq:decomposition gm1 bargm7 bargm7} can be obtained using the Clebsch-Gordan decomposition coefficients
\begin{equation}\label{eq:basis functions for two spin square lattice 2s at 1b}
    \begin{split}
        A_1: & \quad \hat{O}^{\dagger}_{A_1} = \frac{1}{\sqrt{2}}(\hat{c}^{\dagger}_{1b, \bar{E}_1^a, \uparrow} \hat{c}^{\dagger}_{1b, \bar{E}_1^b, \downarrow} - \hat{c}^{\dagger}_{1b, \bar{E}_1^a, \downarrow} \hat{c}^{\dagger}_{1b, \bar{E}_1^b,  \uparrow})\\
        A_2: & \quad \hat{O}^{\dagger}_{A_2} = -\frac{\mathrm{i}}{\sqrt{2}}(\hat{c}^{\dagger}_{1b, \bar{E}_1^a, \uparrow} \hat{c}^{\dagger}_{1b, \bar{E}_1^b, \downarrow} + \hat{c}^{\dagger}_{1b, \bar{E}_1^a, \downarrow} \hat{c}^{\dagger}_{1b, \bar{E}_1^b,  \uparrow})\\
        E: & \quad \{  \hat{O}^{\dagger}_{E, 1} = \frac{\mathrm{i}}{\sqrt{2}}(\hat{c}^{\dagger}_{1b, \bar{E}_1^a, \uparrow} \hat{c}^{\dagger}_{1b, \bar{E}_1^b, \uparrow} - \hat{c}^{\dagger}_{1b, \bar{E}_1^a, \downarrow} \hat{c}^{\dagger}_{1b, \bar{E}_1^b, \downarrow}), \\
        & \quad \hat{O}^{\dagger}_{E, 2} = \frac{1}{\sqrt{2}}(\hat{c}^{\dagger}_{1b, \bar{E}_1^a, \uparrow} \hat{c}^{\dagger}_{1b, \bar{E}_1^b, \uparrow} + \hat{c}^{\dagger}_{1b, \bar{E}_1^a, \downarrow} \hat{c}^{\dagger}_{1b, \bar{E}_1^b, \downarrow}) \}
    \end{split}
\end{equation}
At a specific value of the momentum $\bm{q}$, these basis functions are refined by the introduction of a momentum label, which enriches their transformation properties by the relevant representation at the specific value of  $\bm{q}$. 
For example, the $\Gamma_1$ basis function at $\Gamma$ becomes
\begin{equation}\label{eq:basis function with q dependence}
    \hat{O}^{\dagger}_{1b, A_1, \bm{q}} = \frac{1}{\sqrt{N}} \sum^N_{\bm{k}=1} \frac{1}{\sqrt{2}} (\hat{c}^{\dagger}_{1b, \bar{E}_1^a, \uparrow, \bm{k}} \hat{c}^{\dagger}_{1b, \bar{E}_1^b, \downarrow, -\bm{k}+\bm{q}} - \hat{c}^{\dagger}_{1b, \bar{E}_1^a, \downarrow, \bm{k}} \hat{c}^{\dagger}_{1b, \bar{E}_1^b,  \uparrow, -\bm{k}+\bm{q}})
\end{equation}
At non maximally symmetric points, so for $\bm{q} \neq \Gamma, M$, the $\Gamma_5$ representation splits into two one dimensional irreps. This can also be seen by observing that the two basis elements $\hat{O}^{1, 2}_5$ transform with opposite mirror eigenvalue.

\subsection{Example for \textit{1a} Wyckoff position}\label{subsec:App 2D examples E1 E2 at 1B}
Here we consider the two orbitals corresponding to the $\Gamma_5$ representation placed at $1a$.
The single electron state transforms in the representation
\begin{equation}\label{eq:repr spinful p orbitals}
(E \otimes SU(2) \downarrow C^D_{4v}) = \Bar{E}_1 \oplus \Bar{E}_2
\end{equation}
The basis functions for the irreps on the left hand side of~\eqref{eq:repr spinful p orbitals} are
\begin{equation}
\{(\uparrow, \ p_+ ), (\downarrow, \ p_-)\} \in \bar{E}_1, \quad \{(\uparrow, \ p_-), (\downarrow, \ p_+)\} \in \bar{E}_2, \qquad  p_{\pm} =\frac{\mathrm{i}}{\sqrt{2}} (p_x \pm \mathrm{i} p_y), \quad \mathcal{T}p_{\pm} = - p_{\mp}.
\end{equation}
The momentum representation at the Wyckoff position $1a$ is trivial ($A_1$), as shown in Eq.~\eqref{eq:C4v represent momentum at 1a}. Hence, the two-particle representations, after taking into account the site-symmetry contribution, becomes
\begin{equation}\label{eq:C4v 2p orbitals at 1a 2 particle repr decompositon}
(\bar{E}_1 \oplus \bar{E}_1)_a \otimes  (\bar{E}_1 \oplus \bar{E}_2)_b = (A_1 \oplus A_2 \oplus E)^2_{ab} \oplus (B_1 \oplus B_2 \oplus E)^2_{ab}
\end{equation}
and at momentum $\Gamma, M$ becomes
\begin{equation}
\begin{split}
    (\bar{\Gamma}_7 \oplus \bar{\Gamma}_6)_a \otimes  (\bar{\Gamma}_7 \oplus \bar{\Gamma}_6)_b &= (\Gamma_1 \oplus \Gamma_4 \oplus \Gamma_5)^2_{ab} \oplus (\Gamma_2 \oplus \Gamma_3 \oplus \Gamma_5)^2_{ab},\\
     (\bar{M}_7 \oplus \bar{M}_6)_a \otimes (\bar{M}_7 \oplus \bar{M}_6)_b &= (M_1 \oplus M_4 \oplus M_5)^2_{ab} \oplus (M_2 \oplus M_3 \oplus M_5)^2_{ab}.
\end{split}
\end{equation}
If there is a single pair of $(\Bar{E}_1\oplus\Bar{E}_2) $-orbitals ($a=b$), the list of representations appearing in~\eqref{eq:C4v 2p orbitals at 1a 2 particle repr decompositon} is reduced by Pauli exclusion principle
\begin{equation}\label{eq:C4v 2p orb at 1a irrep reduced by Pauli}
    (A_1 \oplus A_2 \oplus E)^2 \oplus (B_1 \oplus B_2 \oplus E)^2 \rightarrow A_1 \oplus A_1 \oplus (B_1 \oplus B_2 \oplus E).
\end{equation}
The basis functions for the first two $\Gamma_1$ factors in~\eqref{eq:C4v 2p orb at 1a irrep reduced by Pauli} are $\hat{c}^{\dagger}_{1a,p_+, \uparrow} \hat{c}^{\dagger}_{1a,p_-, \downarrow} \in A_1$ and  $\hat{c}^{\dagger}_{1a,p_-, \uparrow} \hat{c}^{\dagger}_{1a,p_+, \downarrow} \in A_1$, and the remaining ones
\begin{equation}
\begin{split}
    B_1:& \quad  \hat{O}^{\dagger}_{B_1}=\frac{1}{\sqrt{2}}(-\hat{c}^{\dagger}_{1a, p_+, \uparrow} \hat{c}^{\dagger}_{1a, p_+, \downarrow} + \hat{c}^{\dagger}_{1a, p_-, \downarrow} \hat{c}^{\dagger}_{1a, p_-, \uparrow}), \\
    B_2:& \quad \hat{O}^{\dagger}_{B_2}= \frac{\mathrm{i}}{\sqrt{2}}(\hat{c}^{\dagger}_{1a, p_+, \uparrow} \hat{c}^{\dagger}_{1a, p_+, \downarrow} + \hat{c}^{\dagger}_{1a, p_-, \downarrow} \hat{c}^{\dagger}_{1a, p_-, \uparrow}), \\
    E:& \quad \{ \hat{O}^{\dagger}_{E, 1}= \frac{1}{\sqrt{2}}(\hat{c}^{\dagger}_{1a, p_+, \uparrow} \hat{c}^{\dagger}_{1a, p_-, \uparrow} + \hat{c}^{\dagger}_{1a, p_-, \downarrow} \hat{c}^{\dagger}_{1a, p_+, \downarrow}),  \\
    & \quad \hat{O}^{\dagger}_{E, 2}=  \frac{\mathrm{i}}{\sqrt{2}}(-\hat{c}^{\dagger}_{1a, p_+, \uparrow} \hat{c}^{\dagger}_{1a, p_-, \uparrow} + \hat{c}^{\dagger}_{1a, p_+, \downarrow} \hat{a}^{\dagger}_{p_-, \downarrow})\}.
\end{split}
\end{equation}
In writing these expressions we omitted the $\bm{q}$ dependence, which can be retrieved as in Eq.~\eqref{eq:basis function with q dependence}.
When two sets of $(\Bar{E}_1\oplus\Bar{E}_2)$ orbitals are considered ($a \neq b$), all the representations in Eq.~\eqref{eq:C4v 2p orbitals at 1a 2 particle repr decompositon} are allowed. Fig.~\ref{fig:G2Bands_2Dexample}\textcolor{red}{b} shows the resulting ``band structure" of the two-particle Green's function for a single pair of $(\Bar{E}_1\oplus\Bar{E}_2)$-orbitals placed at $1a$.

\section{Numerical calculations}\label{app:Numerics}
\subsection{Quantum Monte Carlo}
\label{sec:App_QMC}
The square lattice with only nearest-neighbor hoppings and the diamond chain model at $t_2 = 0$ are {\it bipartite lattices}  and can therefore be studied in absence of a sign problem within the determinant Quantum Monte Carlo (DQMC) method. In this work, we employ a high-precision 4-valued Hubbard-Stratonovich decomposition~\cite{PhysRevB.56.15001}. The method allows one to exactly simulate the system at temperature $T = 1/\beta$, which is described by the partition function $\mathcal{Z} = \mbox{Tr}\,e^{-\beta \hat{H}}.$ In this work, we consider temperatures sufficiently low, i.\,e., $\beta \Delta \gg 1$, with $\Delta$ being the system's many-body gap. To this end, we employ $\beta t = 10 L$. Throughout the numerical simulations, the Hamiltonian is defined such that the ground state energy $E_0$ is zero and, with this choice, $\mathcal{Z} = 1$ with exponential precision. 

\subsubsection{Measurement of Green's functions}

Within DQMC, one has access to thermal Matsubara-separated averages of the form
\begin{equation}
    \label{eq:qmc_corr}
    \langle \hat{B}(\tau) \hat{A}(0)\rangle_{\beta} = \frac{1}{\mathcal{Z}} \mbox{Tr}\big( e^{-\beta \hat{H}} e^{+ t \hat{H}} \hat{B} e^{-\tau \hat{H}} \hat{A}\big)
\end{equation}
at any time separation $0 \leqslant \tau < \beta$, where the $\hat{A}$ and $\hat{B}$ operators are arbitrary strings of fermionic creation and annihilation operators~\cite{PhysRevB.56.15001}. This allows one to compute the expectations 
\begin{equation}
    g^{(1/2)}_{\hat{A},\,\hat{B}} (\omega = 0) = \left< \hat{B} \frac{1}{-\hat{H}} \hat{A} \right> \pm \left< \hat{A} \frac{1}{+ \hat{H}} \hat{B} \right>,
\end{equation}
corresponding to single- and two-particle Green's functions defined in the main text at zero frequency. To compute $g^{(1/2)}_{\hat{A},\,\hat{B}}$ using the Matsubara-separated thermal averages, we numerically compute the time-separation integrals
\begin{gather}
    \int\limits_{0}^{\beta / 2} d\tau \langle \hat{B}(\tau) \hat{A}(0)\rangle_{\beta} \pm \int\limits_{\beta / 2}^{\beta} d\tau\langle \hat{B}(\tau) \hat{A}(0)\rangle_{\beta} = \sum\limits_n e^{-\beta \epsilon_n} \int \limits_{0}^{\beta / 2} d\tau e^{+\tau \epsilon_n} \langle n | \hat{B} e^{-\tau \hat{H}} \hat{A}|n\rangle \pm \sum\limits_n e^{-\beta \epsilon_n} \int \limits_{\beta / 2}^{\beta} d\tau e^{+\tau \epsilon_n} \langle n | \hat{B} e^{-\tau \hat{H}} \hat{A}|n\rangle  \nonumber \\ 
    = \sum\limits_n  e^{-\beta \epsilon_n} \left< n \middle| \hat{B} \frac{1}{\epsilon_n - \hat{H}} \left( e^{\beta (\epsilon_n - \hat{H})} - e^{\beta (\epsilon_n - \hat{H}) / 2}\right) \hat{A} \middle| n\right> \pm \sum\limits_n  e^{-\beta \epsilon_n} \left< n \middle| \hat{B} \frac{1}{\epsilon_n - \hat{H}} \left( e^{\beta (\epsilon_n - \hat{H}) / 2} - \mathbb{1} \right) \hat{A} \middle| n\right>  \\ 
    = \sum\limits_{nm} \frac{\langle n | \hat{B} | m \rangle \langle m | \hat{A} | n \rangle}{\epsilon_n - \epsilon_m} \left(\left[e^{- \beta \epsilon_m} - e^{-\beta (\epsilon_n + \epsilon_m) / 2} \right] \pm \left[e^{-\beta (\epsilon_n + \epsilon_m) / 2} - e^{-\beta \epsilon_n} \right]\right).\nonumber
\end{gather}

Here, the sums run over the full set of the Hamiltonian eigenstates. We note that, since $\hat{A} = c^{\dag}_i c^{\dag}_j$ for $\Gt$ and $\hat{A} = c^{\dag}_i$ for $\Go$, $|\epsilon_n - \epsilon_m| > 0$, i.\,e., strictly greater than zero, bounded from below by $E(N + 2) - E(N)$ in the former and by $E(N + 1) - E(N)$ in the latter case, where $N$ is the ground state filling. Since also the ground state energy is taken to be zero, exponents of the form $e^{-\beta (\epsilon_n + \epsilon_m) / 2}$ vanish at zero temperature. The remaining exponents $e^{- \beta \epsilon_m}$ and $e^{- \beta \epsilon_n}$ only survive if, correspondingly, $m = 0$ or $n = 0$. Thus, the resulting expression reads 
\begin{gather}
    \int\limits_{0}^{\beta / 2} d\tau \langle \hat{B}(\tau) \hat{A}(0)\rangle_{\beta} \pm \int\limits_{\beta / 2}^{\beta} d\tau\langle \hat{B}(\tau) \hat{A}(0)\rangle_{\beta} = \sum\limits_{n} \frac{\langle n|\hat{B}|0\rangle \langle 0 | \hat{A} | n \rangle}{\epsilon_n} \mp \sum\limits_{m} \frac{\langle 0|\hat{B}|m\rangle \langle m | \hat{A} | 0 \rangle}{-\epsilon_m} = \\ = \langle 0| \hat{A} \hat{H}^{-1} \hat{B} |0\rangle \mp \langle 0| \hat{B} (-\hat{H})^{-1} \hat{A} |0\rangle = G^{(1/2)}_{\hat{A},\,\hat{B}} (\omega = 0).\nonumber
\end{gather}

\begin{figure}[t!]
    \centering
    \includegraphics[width=\columnwidth]{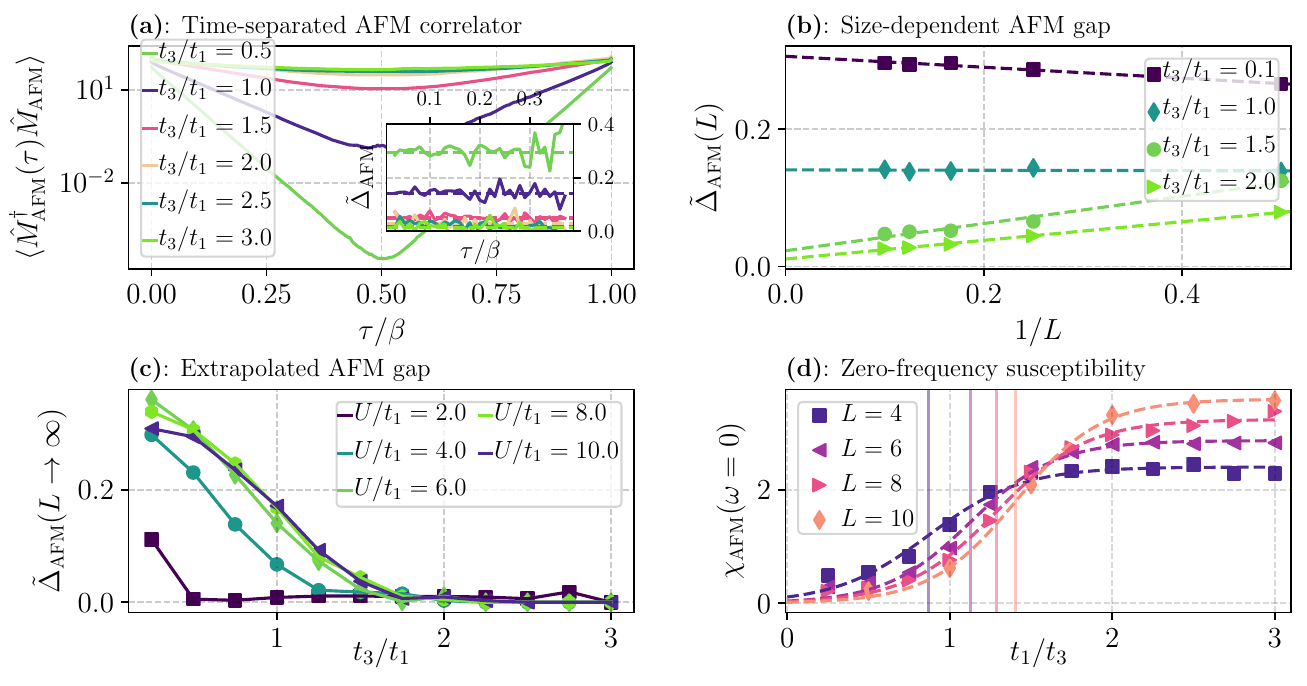}
    \caption{\textbf{Diamond chain} {\bf (a)} Imaginary time-dependent AFM operator correlator obtained within QMC simulations of the $L = 10$ diamond chain at $U / t_1 = 6$. The inset illustrates the behaviour of the time-dependent AFM gap-estimator $\tilde{\Delta}(\tau)$ defined in Eq.\,\eqref{eq:estimator} in a region of $\tau / \beta$ chosen such that the estimator is constant. The horizontal lines in the inset show constant fits. {\bf (b)} Thermodynamic limit $L \to \infty$ extrapolations of the AFM gap at $U / t_1 = 6.0$. The dashed lines show the $A + B / L$ fits. {\bf (c)} Thermodynamic limit extrapolated gap estimator as a function of $t_3 / t_1$. Vanishing gap indicates establishment of the AFM order. The points are connected to guide the eye. {\bf (d)} Zero-frequency AFM susceptibility as a function of $t_3 / t_1$ at $U / t_1 = 6.0$. The vertical lines show the inflection point in the hyperbolic tangent fit. Inset: A/B bipartition of the diamond chain lattice.}
    \label{fig:DC_illustration}
\end{figure}

\subsubsection{Interacting single- and two-particle gaps}

The DQMC approach allows one to extract single- or two-particle gaps $\Delta_1 = E(N + 1) - E(N)$ and $\Delta_2 = E(N + 2) - E(N),$ as well as ground state energies in the Hilbert space half-filled sectors with defined quantum numbers such as momentum. First, to obtain $\Delta_1$ and $\Delta_2$, consider an operator $\hat{\mathcal{O}}$ superposing strings of one or two fermionic creation operators. Assuming, as above, $\beta \Delta \gg 1$, for $\tau < \beta / 2$ within DQMC we measure 
\begin{gather}
C_{\hat{\mathcal{O}}}(\tau) = \langle \hat{\mathcal{O}}^{\dag}(\tau) \hat{\mathcal{O}}(0) \rangle_{\beta} = \sum\limits_{nm} e^{-\beta \epsilon_n + \tau \epsilon_n - \tau \epsilon_m} \langle n | \hat{\mathcal{O}}^{\dag} |m \rangle \langle m | \hat{\mathcal{O}} | n \rangle \to \sum\limits_m |\langle m | \hat{\mathcal{O}} | 0 \rangle|^2,
\end{gather}
where we used the fact that at $n \neq 0$ the quantity vanishes exponentiallly fast due to the low temperature and the restriction $\tau < \beta / 2$. In the resulting expression, in the case of one and two-fermion operators $\hat{\mathcal{O}}$, at sufficiently large $\tau$ the correlator behaves as $C(\tau) = e^{-\Delta_{1/2} \tau} |\langle 0_{1/2}| \hat{\mathcal{O}} |0 \rangle|^2|$ with $0_{1/2}$ being the ground state in the $N + 1$ or $N + 2$ particle sectors. To extract $\Delta_{1/2}$ from the measured $C(\tau)$, we consider a window of $\tau$ with $\tau$ being large enough to neglect higher-energy $\epsilon_m$ contributions to $C(\tau)$, but not too large (since exponentially-decaying $C(\tau)$ becomes increasingly complicated to measure as $\tau \to \beta / 2$). 

Within this work, we consider a set of operators $\hat{\mathcal{O}}_{\alpha} = c^{\dag}_{\alpha}$ in case of $\Delta_1$ and $\hat{\mathcal{O}}_{\alpha,\,\beta} = c^{\dag}_{\alpha} c^{\dag}_{\beta}$ for $\Delta_2$. Here, $\alpha$ and $\beta$ are composite indices enumerating lattice sites and spins. We then measure $C^{\alpha}_{\alpha'}(\tau) = \langle \hat{\mathcal{O}}^{\dag}_{\alpha}(\tau) \hat{\mathcal{O}}_{\alpha'} \rangle_{\beta}$ or $C^{\alpha,\beta}_{\alpha',\beta'}(\tau) = \langle \hat{\mathcal{O}}^{\dag}_{\alpha,\,\beta}(\tau) \hat{\mathcal{O}}_{\alpha'\,\beta'}\rangle_{\beta}$ in those cases, respectively. Treating $(\alpha,\,\beta)$ in the latter case as a composite index, we diagonalize the correlator at zero time separation $C(\tau = 0) = U^{\dag} \Lambda(\tau = 0) U$ and then apply the same transformation for all $\tau:$ $\Lambda(\tau) = U C(\tau) U^{\dag}$. The unitary transformation $U$ builds new operators $\hat{\mathcal{O}}'_{(\boldsymbol k, \mathcal{Q})}$ superposing one- or two-creation strings to have defined momentum $\boldsymbol k$ and other quantum numbers $\mathcal{Q}$ which are, in case of a diamond chain, $M_x$ and $M_y$ mirrors and also include rotation in the case of the square lattice. Such an operator has only overlap with $|m \rangle$ states in the $(\boldsymbol k, \mathcal{Q})$ symmetry sector. As a result, the respective entry of the correlator $\Lambda(\tau)_{(\boldsymbol k, \mathcal{Q})}$ reaches the single-exponential behavior regime more quickly, as the operator only induces an overlap with the states with a defined set of quantum numbers.

\begin{figure}[t!]
    \centering
    \includegraphics[width=\columnwidth]{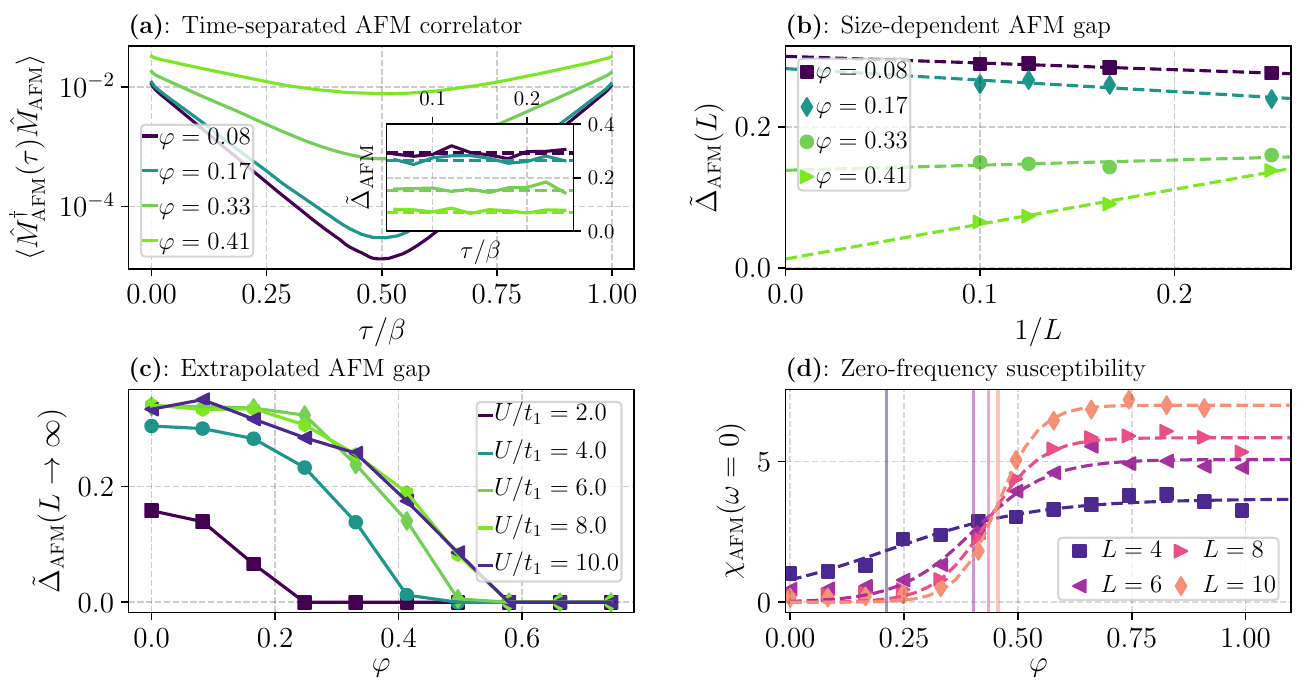}
    \caption{\textbf{Checkerboard lattice of Hubbard squares} {\bf (a)} Imaginary time-dependent AFM operator correlator obtained within QMC simulations of the $L = 10$ diamond chain at $U / t = 4$. The inset illustrates the behaviour of the time-dependent AFM gap-estimator $\tilde{\Delta}(\tau)$ defined in Eq.\,\eqref{eq:estimator} in a region of $\tau / \beta$ chosen such that the estimator is constant. The horizontal lines in the inset show constant fits. {\bf (b)} Thermodynamic limit $L \to \infty$ extrapolations of the AFM gap at $U / t = 4.0$. The dashed lines show the $A + B / L$ fits. {\bf (c)} Thermodynamic limit extrapolated gap estimator as a function of $\varphi$. Vanishing gap indicates establishment of the AFM order. The points are connected to guide the eye. {\bf (d)} Zero-frequency AFM susceptibility as a function of $\varphi$ at $U / t = 4.0$. The vertical lines show the inflection point in the hyperbolic tangent fit.}
    \label{fig:SQUARE_illustration}
\end{figure}

Having obtained diagonal responses matrix $\Lambda(\tau)$, we employ the two estimators for the gap:
\begin{gather}
    \label{eq:estimator}
    \log \Lambda_{(\boldsymbol k, \mathcal{Q})}(\tau) = A - \Delta_{(\boldsymbol k, \mathcal{Q})} \tau,\\
    \Delta_{(\boldsymbol k, \mathcal{Q})} = -\frac{1}{\delta \tau} \frac{1}{N_w} \sum\limits_{\tau = \tau_0}^{\tau_0 + N_w} \log \frac{\Lambda_{(\boldsymbol k, \mathcal{Q})}(\tau + \delta \tau)}{\Lambda_{(\boldsymbol k, \mathcal{Q})}(\tau)},
\end{gather}
where $\delta \tau$ is the Suzuki-Trotter discretization employed in the computation. The first estimator fits logarithmic decay of the correlator, while the second estimator averages the logarithmic derivative over a window of $\tau$. Those two estimators usually agree within $5\,\%$ and we take average of the two for a more robust result. 

\subsubsection{Antiferromagnetic order within QMC}
Time-separated correlator Eq.\,\eqref{eq:qmc_corr} can also be used to obtain zero-frequency susceptibilities signaling tendencies towards establishment of a symmetry-breaking order. It can also be used to estimate the gap $\Delta_{\boldsymbol k, \mathcal{Q}}(N)$ between the ground state energy in the Hilbert space sector at half-filling with the quantum numbers $(\boldsymbol k, \mathcal{Q})$, and the global ground state at half-filling.

To study both antiferromagnetic (AFM) susceptibility and the respective gap $\Delta_{\text{AFM}}$ between the global ground state and the ground state in the $S = 1$, $\boldsymbol{k} = \pi$ sector for the diamond chain, $S = 1$, $\boldsymbol{k} = (\pi, \pi)$ sector for the square lattice, we define the AFM operator as 
\begin{gather}
\hat{M}_{\text{AFM}} = \frac{1}{N_{\text{sites}}}\sum_i (\hat{n}_{i,\,\uparrow} - \hat{n}_{i,\,\downarrow}) (-1)^{z_i},
\end{gather}
where $z_i = \pm 1$ on A and B sublattices, respectively, and $i$ runs over the whole lattice. The A and B sublattices arrangement for the cases of the diamond chain is shown in the inset of Fig.~\ref{fig:DC_illustration}\textcolor{red}{d}. The A/B bipartition is the standard checkerboard separation in the case of the square lattice. Thus, the operator $\hat{M}_{\text{AFM}}$, acting on a ground state, generates overlap with the required symmetry sector. To extract the AFM gap $\Delta_{\text{AFM}}$, we study the behavior of the time-separated correlator 
\begin{gather}
  \chi_{\text{AFM}}(\tau) = \langle \hat{M}_{\text{AFM}}^{\dagger}(\tau)  \hat{M}_{\text{AFM}}(0) \rangle.
\end{gather}

Such correlator is shown in the figures Fig.~\ref{fig:DC_illustration}\textcolor{red}{a} for the diamond chain lattice and in Fig.~\ref{fig:SQUARE_illustration}\textcolor{red}{a} for the square lattice, respectively. On the logarithmic scale, one can clearly see the region of linear decay with $\tau$. In the insets, we show the gap estimators obtained within the averaging (second) technique Eq.\,\eqref{eq:estimator}. We see, modulo Monte Carlo noise, robust constant estimators, signalling that the $\chi_{\text{AFM}}(\tau)$ correlator's decay is dominated by a single exponential. 

As the next step, in Fig.~\ref{fig:DC_illustration}\textcolor{red}{b} and Fig.~\ref{fig:SQUARE_illustration}\textcolor{red}{b}, we show extrapolations of the $\tilde \Delta_{\text{AFM}}(L)$ estimator to the thermodynamic limit $L \to \infty$ at different values of $t_3 / t_1$ (diamond chain) and $\varphi$ (square lattice), respectively. We observe that the gap extrapolation follows a clear $1 / L$ trend. Applying the $A + B / L$ fit, we extract the thermodynamic-limit extrapolated gap $\tilde \Delta_{\text{AFM}} (L \to \infty)$. Lastly, in Fig.~\ref{fig:DC_illustration}\textcolor{red}{c} and Fig.~\ref{fig:SQUARE_illustration}\textcolor{red}{c}, we show the extrapolated gap value $\tilde \Delta_{\text{AFM}} (L \to \infty)$ as a function of $t_3 / t_1$ and $\varphi$, respectively. Vanishing gap opens prospect to the spontaneous breaking of translational and $SU(2)$ symmetries. The position of vanishing gap is used in the main text as one of the phase boundary criteria.

To obtain the susceptibility signalling such symmetry breaking, we compute the zero-frequency response 
\begin{gather} 
\chi_{\text{AFM}}(\omega = 0) = \int\limits_{0}^{\beta} \text{d} \tau \chi_{\text{AFM}}(\tau).
\end{gather}

The results are shown in Fig.~\ref{fig:DC_illustration}\textcolor{red}{d} and in Fig.~\ref{fig:SQUARE_illustration}\textcolor{red}{d}. The data is fitted with the ansatz $A (1 + \tanh[(r - r_0)/\delta])$. The location of the inflection point $r_0 = (t_3 / t_1)^*$, or $\varphi^*$ at $L = 10$ is used in the main text to pin down the phase transition between the AFM and FMI MAL phases.

\subsection{Exact diagonalization data}
\label{sec:App_ED}
\begin{figure}[t!]
    \centering
    \includegraphics[width=\columnwidth]{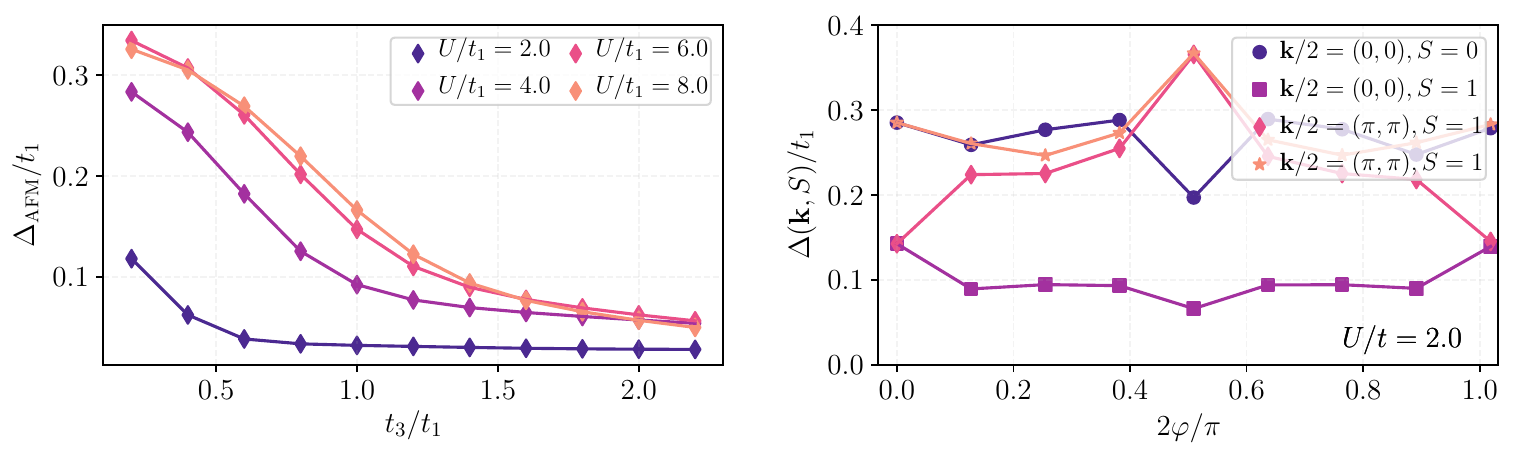}
    \caption{{\bf (a)} Gap $\Delta_{\text{AFM}}$ between the ground state and the $\boldsymbol k = \pi, S = 1$ state obtained from exact diagonalization of the $L = 4$ diamond chain lattice at $U / t_1 = 2.0,\,4.0,\,6.0$ and $8.0$. The horizontal dashed line is an arbitrary threshold. {\bf (b)} Gaps between the ground state and lowest excited states with momentum $\boldsymbol k$ and spin $S$ as a function of $\varphi$ computed within ED of the $4 \times 4$ checkerboard lattice at $U / t = 2.0$.}
    \label{fig:ED_illustration}
\end{figure}

In Fig.~\ref{fig:ED_illustration}, we present spectral gaps obtained on the $L = 4$ (16-sites) diamond chain and $4 \times 4$ square systems. In Fig.~\ref{fig:ED_illustration}\textcolor{red}{a}, we show the AFM gap $\Delta_{\text{AFM}}$ as a function of $t_3 / t_1$ for a set of $U / t_1$ within the diamond chain lattice. The figure confirms the two tendencies observed within the QMC simulation. Namely, (i) increasing $t_3 / t_1$ leads to the shrinking AFM gap, while (ii) growing $U / t_1$ increases the transition point $(t_3 / t_1)^*$ between the MAL and AFM phases.

In Fig.~\ref{fig:ED_illustration}\textcolor{red}{b}, in the case of the $4 \times 4$, due to the explicit translational symmetry breaking at $t_3 \neq t_1$ ($t_1 = t \cos \varphi$, $t_3 = t \sin \varphi$), within ED one can only fix the half-momentum $\boldsymbol{k} / 2$. For instance, $\boldsymbol{k} / 2 = (0, 0)$ corresponds to the standard AFM momentum $\boldsymbol{k} = (\pi, \pi)$ at $t_1 = t_3$. Therefore, in the figure we observe closing of the AFM gap, as well as sharp reordering of the excited levels structure in the region $0.4 \leqslant 2 \varphi / \pi\leqslant 0.6$, signaling transition from the FMI MAL phase to the AFM phase.

\subsection{Exact diagonalization for the Star of David}\label{app:DavidStar_ED}
\begin{figure}[t]
    \centering   \includegraphics[width=\textwidth]{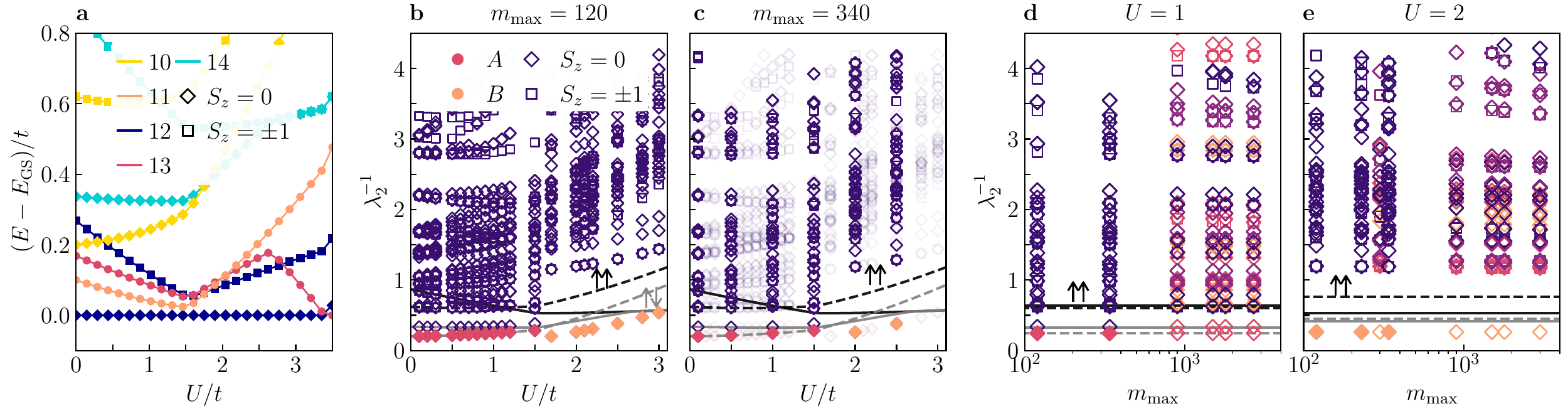}
    \caption{\textbf{Numerical calculations for the Star of David.} \textbf{a} Energies for the $N=10, 11, 12, 13, 14$ particle sectors as a function of $U/t$. \textbf{b}-\textbf{c} Spectrum of $\Gt$ as a function of $U$ for \textbf{b} $m_{\max}=120$, \textbf{c} $m_{\max}=340$ and the $\ket{m}$ states selected by spin sector $S_z$, but not by $C_6$ sector. In \textbf{c}, the transparent points correspond to the ones of panel \textbf{b}. The lowest inverse eigenvalues at each value of $U/t$ are colored according to the irrep of $C_6^D$ in which they transform, either $A$ or $B$ (see the character table in Tab.~\ref{tab:character table C6}).
    \textbf{d}-\textbf{e} Comparison of inverted $\Gt$ spectra at fixed $U=1, 2$ (for  \textbf{d} and \textbf{e}, respectively) as a function of $m_{\max}$. The spectra at fixed $m_{\max}$ with the lowest eigenvalue filled by a color correspond to numerical calculations where the $\{\ket{m}\}$ states are taken from the appropriate $S_z$ symmetry sector, while the spectra at fixed $m_{\max}$ with no eigenvalue filled by color correspond to computations where the $\{\ket{m}\}$ states are taken from the appropriate $S_z$ and $C_6$ symmetry sectors. For the latter case, the value of the $m_{\max}$ states taken in each sector with fixed eigenvalue of $S_z, C_6$ is multiplied by a factor of $6$, to compare the resulting spectrum with the case of no $C_6$ symmetry implemented (see the legend in Fig.~\ref{fig:star_of_david}\textcolor{red}{d} in the main text for these points).
    In \textbf{b}--\textbf{e} the gray (black) line mark the $S_z=0$ ($S_z=\pm1$) two-particle gap, and continuous (dashed) lines correspond to the $\mathscr{H}_{N-2}$ ($\mathscr{H}_{N+2}$) sector.
    In all of the panels, the model parameters are chosen to be $\mu_*/t=1.5$, $\mu/t=-0.7$, $t'/t=1$, $t''/t=0.4$, corresponding to the dashed line in the phase diagram of Fig.~\ref{fig:star_of_david}\textcolor{red}{b}. }
    \label{fig:panel_david_star_SM}
\end{figure}

For the model presented in Sec.~\ref{sec:star of David}, we performed ED calculations.
The phase diagram of Fig.~\ref{fig:star_of_david}\textcolor{red}{b} is computed in the full Hilbert space of $N=12$ electrons. To ensure that the many-body ground state lies in the twelve-electron sector, and for $\mu_*=1.5$ and $\mu=-0.7$, we computed the ground state in the $N$, $N\pm1$ and $N\pm2$ particle sectors, see Fig.~\ref{fig:panel_david_star_SM}\textcolor{red}{a}.

To compute the spectrum of $\Gt$, we approximate the Hamiltonian in Eq.~\eqref{eq:G2 with O operators} by decomposing it in terms of the first $m_{\max}$ lowest-energy states
\begin{equation}\label{eq:approx of H with mmax}
    [\hat{H}- \mathbb{1} E_{\mathrm{GS}}]^{-1} \approx \sum_{m=0}^{m_{\max}} \frac{\ket{m}\bra{m}}{E_m - E_{\mathrm{GS}}},
\end{equation}
with $m \in \mathscr{H}_{N\pm2}$ depending on which term of Eq.~\eqref{eq:G2 with O operators} is being calculated.
As we are mainly interested in the low-lying inverse spectrum of $\Gt$, it is sufficient to consider the first $m_{\max}$ states with the cutoff $m_{\max}\sim10^2$. 
In fact, we expect the lowest $\lambda_2^{-1}$ to be well approximated by considering the Hamiltonian's low-lying energy states: in order to obtain a low $\lambda_2^{-1}$, the states obtained by acting with the eigenstate  $\hat{O}^{\dagger}$ and  $\hat{O}$  on the ground state need to mainly overlap with states $\ket{m}$ characterized by small energies $E_{m}$'s, so that the denominator in Eq.~\eqref{eq:approx of H with mmax} is large
\begin{equation}\label{eq:estimate of lambda_2 ED cutoff}
    \lambda^{-1}_2 \approx \sum_{\substack{m = 1,\\ m \in \mathscr{H}_{N+2}}}^{m_{\max}} \frac{|\bra{m}\hat{O}^{\dagger} \ket{\mathrm{GS}}|^2}{E_m - E_{\mathrm{GS}}} + \sum_{\substack{m = 1,\\ m \in \mathscr{H}_{N-2}}}^{m_{\max}} \frac{|\bra{m}\hat{O} \ket{\mathrm{GS}}|^2}{E_m - E_{\mathrm{GS}}}.
\end{equation}
To improve the quality of each overlap in the summations of Eq.~\eqref{eq:estimate of lambda_2 ED cutoff}, the set of $m_{\max}$ lowest-energy $\ket{m}$ states can be chosen to fall within the Hilbert space of the states with quantum numbers equal to those of the states $\hat{O}^{\dagger}\ket{\mathrm{GS}}$, for $\{\ket{m}\} \in \mathscr
H_{N+2}$, and $\hat{O}\ket{\mathrm{GS}}$, for $\{\ket{m}\} \in \mathscr
H_{N-2}$. For this model, we ran calculations with $\ket{m}$ states that fall within the appropriate $S_z$ symmetry sector (see Fig.~\textcolor{red}{8}\textcolor{red}{b}-\textcolor{red}{c}), as well as calculations where both the $S_z$ and $C_6$ sectors are implemented (as in Fig.~\ref{fig:star_of_david}\textcolor{red}{d} in the main text, as well as for the vertical lines where the lowest eigenvalues have an empty marker in Fig.~\ref{fig:panel_david_star_SM}\textcolor{red}{d}-\textcolor{red}{e}).

Figure~\textcolor{red}{8}\textcolor{red}{b}--\textcolor{red}{d} shows the low-lying inverted $\Gt$ spectra as a function of $U$ for increasing value of $m_{\max}$. There, one sees that the low-lying spectrum remains remains unchanged as $m_{\max}$ increases. There, one sees that the spectrum between the values of $\lambda_2^{-1} \in [0, 1]$ is fully converged already at values of $m_{\max} \approx 150$ and with $C_6$ symmetry implemented. For Fig.~\textcolor{red}{8}\textcolor{red}{d} in the main text, we choose $m_{\max}=300$ and $C_6$ symmetry implemented as it gives a correct description of the low-lying spectra of $\Gt$ in both phases.

\begin{table}[t!]
    \centering
    \begin{tabular}{c@{\hspace{10pt}}c@{\hspace{10pt}}|@{\hspace{10pt}}c@{\hspace{10pt}}c@{\hspace{10pt}}c@{\hspace{10pt}}c@{\hspace{10pt}}c@{\hspace{10pt}}c@{\hspace{10pt}}c@{\hspace{10pt}}c@{\hspace{10pt}}c@{\hspace{10pt}}c@{\hspace{10pt}}c@{\hspace{10pt}}c}
    \hline
    \hline
    
     (a) & (b) & $E$ & $C_3$ & $C^{-1}_3$ & $C_2$ & $C_6^{-1}$ & $C_6$ & $\Bar{E}$ & $\bar{C}_3$ & $\bar{C}^{-1}_3$ & $\bar{C}_2$ & $\bar{C}^{-1}_6$ & $\bar{C}_6$\\
     
    \hline
        $A$ & $\Gamma_1$ & $1$ & $1$ & $1$ & $1$ & $1$ & $1$ & $1$ & $1$ & $1$ & $1$ & $1$ & $1$ \\
        $B$ & $\Gamma_2$ & $1$ & $1$ & $1$ & $-1$ & $-1$ & $-1$ & $1$ & $1$ & $1$ & $-1$ & $-1$ & $-1$ \\
        $^2E_1$ & $\Gamma_3$ & $1$ & $\omega^2$ & $-\omega$ & $1$ & $\omega^2$ & $-\omega$ & $1$ & $\omega^2$ & $-\omega$ & $1$ & $\omega^2$ & $-\omega$ \\
        $^2E_2$ & $\Gamma_4$ & $1$ & $\omega^2$ & $-\omega$ & $-1$ & $-\omega^2$ & $\omega$ & $1$ & $\omega^2$ & $-\omega$ & $-1$ & $-\omega^2$ & $\omega$ \\
        $^1E_1$ & $\Gamma_5$ & $1$ & $-\omega$ & $\omega^2$ & $1$ & $-\omega$ & $\omega^2$ & $1$ & $-\omega$ & $\omega^2$ & $1$ & $-\omega$ & $\omega^2$ \\
        $^1E_2$ & $\Gamma_6$ & $1$ & $-\omega$ & $\omega^2$ & $-1$ & $\omega$ & $-\omega^2$ & $1$ & $-\omega$ & $\omega^2$ & $-1$ & $\omega$ & $-\omega^2$ \\
        $^2\bar{E}_1$ & $\bar{\Gamma}_7$ & $1$ & $-1$ & $-1$ & $-\mathrm{i}$ & $\mathrm{i}$ & $-\mathrm{i}$ & $-1$ & $1$ & $1$ & $\mathrm{i}$ & $-\mathrm{i}$ & $\mathrm{i}$ \\
        $^1\bar{E}_1$ & $\bar{\Gamma}_8$ & $1$ & $-1$ & $-1$ & $\mathrm{i}$ & $-\mathrm{i}$ & $\mathrm{i}$ & $-1$ & $1$ & $1$ & $-\mathrm{i}$ & $\mathrm{i}$ & $-\mathrm{i}$ \\
        
        $^2\bar{E}_2$ & $\bar{\Gamma}_9$ & $1$ & $-\omega^2$ & $\omega$ & $-\mathrm{i}$ & $\mathrm{i}\omega^2$ & $\mathrm{i}\omega$ & $-1$ & $-\mathrm{i}\omega$ & $\mathrm{i}\omega^2$ & $\mathrm{i}$ & $-\mathrm{i}\omega^2$ & $-\mathrm{i}\omega$ \\
        $^1\bar{E}_3$ & $\bar{\Gamma}_{10}$ & $1$ & $-\omega^2$ & $\omega$ & $\mathrm{i}$ & $-\mathrm{i}\omega^2$ & $-\mathrm{i}\omega$ & $-1$ & $-\mathrm{i}\omega$ & $\mathrm{i}\omega^2$ & $-\mathrm{i}$ & $\mathrm{i}\omega^2$ & $\mathrm{i}\omega$ \\
        $^2\bar{E}_3$ & $\bar{\Gamma}_{11}$ & $1$ & $\omega$ & $-\omega^2$ & $-\mathrm{i}$ & $-\mathrm{i}\omega$ & $-\mathrm{i}\omega^2$ & $-1$ & $\mathrm{i}\omega^2$ & $-\mathrm{i}\omega$ & $\mathrm{i}$ & $\mathrm{i}\omega$ & $\mathrm{i}\omega^2$ \\
        $^1\bar{E}_2$ & $\bar{\Gamma}_{12}$ & $1$ & $\omega$ & $-\omega^2$ & $\mathrm{i}$ & $\mathrm{i}\omega$ & $\mathrm{i}\omega^2$ & $-1$ & $\mathrm{i}\omega^2$ & $-\mathrm{i}\omega$ & $-\mathrm{i}$ & $-\mathrm{i}\omega$ & $-\mathrm{i}\omega^2$ \\
    \hline
    \hline
    \end{tabular}
    \caption{Character table of the group $C_6^D$, with $\omega = e^{\mathrm{i}\pi/3}$~\cite{BilbaoElcoro:ks5574}.}
    \label{tab:character table C6}
\end{table}

\section{Character tables and Note on the notation}
Throughout the main text and the Appendix, we used the same notations of the Bilbao Crystallographic server~\cite{BilbaoElcoro:ks5574} to label the irreps of different point-groups. 
In the character tables, the irrep labels listed under column (a) correspond to the notation of Ref.~\cite{BradleyBook2} and (b) to Ref.~\cite{CracknellBook} for the $\Gamma$ point, respectively appearing in the Bilbao Crystallographic server as (2) and (3).

\begin{table}[ht!]
    \centering
    \begin{tabular}{c@{\hspace{10pt}}c@{\hspace{10pt}}|@{\hspace{5pt}}c@{\hspace{10pt}}c@{\hspace{10pt}}c@{\hspace{10pt}}c@{\hspace{10pt}}c@{\hspace{10pt}}c@{\hspace{10pt}}c@{\hspace{10pt}}c@{\hspace{10pt}}c@{\hspace{10pt}}c}
    \hline
    \hline
    
     (a) & (b) & $E$ & $C_2(001)$ & $C_2(010)$ & $C_2(100)$ & $\bar{E}$ & $\mathcal{I}$ & $\sigma_v(001)$ & $\sigma_v(010)$ & $\sigma_v(100)$ & $\bar{\mathcal{I}}$ \\ 
     
     \hline
       $A_g$  & $\Gamma_1^+$  & $1$ & $1$ & $1$ & $1$ & $1$ & $1$ & $1$ & $1$ & $1$ & $1$  \\
        $A_u$  & $\Gamma_1^-$  & $1$ & $1$ & $1$ & $1$ & $1$ & $-1$ & $-1$ & $-1$ & $-1$ & $-1$  \\
        $B_{1g}$  & $\Gamma_2^+$  & $1$ & $1$ & $-1$ & $-1$ & $1$ & $1$ & $1$ & $-1$ & $-1$ & $1$  \\
        $B_{1u}$  & $\Gamma_2^-$  & $1$ & $1$ & $-1$ & $-1$ & $1$ & $-1$ & $-1$ & $1$ & $1$ & $-1$  \\
        $B_{3g}$  & $\Gamma_3^+$  & $1$ & $-1$ & $-1$ & $1$ & $1$ & $1$ & $-1$ & $-1$ & $1$ & $1$  \\
        $B_{3u}$  & $\Gamma_3^-$  & $1$ & $-1$ & $-1$ & $1$ & $1$ & $-1$ & $1$ & $1$ & $-1$ & $-1$  \\
        $B_{2g}$  & $\Gamma_4^+$  & $1$ & $-1$ & $1$ & $-1$ & $1$ & $1$ & $-1$ & $1$ & $-1$ & $1$  \\
        $B_{2u}$  & $\Gamma_4^-$  & $1$ & $-1$ & $1$ & $-1$ & $1$ & $-1$ & $1$ & $-1$ & $1$ & $-1$  \\
        $\bar{E}_g$  & $\bar{\Gamma}_5$  & $2$ & $0$ & $0$ & $0$ & $-2$ & $2$ & $0$ & $0$ & $0$ & $-2$  \\
        $\bar{E}_u$  & $\bar{\Gamma}_6$  & $2$ & $0$ & $0$ & $0$ & $-2$ & $-2$ & $0$ & $0$ & $0$ & $2$  \\
        \hline
         \hline
    \end{tabular}
    \caption{Character table of the double point-group $D_{2h}$~\cite{BilbaoElcoro:ks5574}.}
    \label{tab:character table D2h}
\end{table}

\begin{table}[ht!]
    \centering
    \begin{tabular}{cc|ccccccccc}
    \hline
    \hline
    
     (a) & (b) & $E$ & $C_3$ & $C_2$ & $C_6$ & $\sigma_v(10)$ & $\sigma_v(1\bar{1})$ & $\bar{E}$ & $\bar{C}_{3}$ & $\bar{C}_6$ \\ 
     
     \hline
        $A_1$  & $\Gamma_1$  & $1$ & $1$ & $1$ & $1$  & $1$ & $1$ & $1$  & $1$ & $1$   \\
        $A_2$  & $\Gamma_2$  & $1$ & $1$ & $1$ & $1$  & $-1$ & $-1$ & $1$  & $1$ & $1$   \\
        $B_2$  & $\Gamma_3$  & $1$ & $1$ & $-1$ & $-1$  & $-1$ & $1$ & $1$  & $1$ & $-1$   \\
        $B_1$  & $\Gamma_4$  & $1$ & $1$ & $-1$ & $-1$  & $1$ & $-1$ & $1$  & $1$ & $-1$   \\
        $E_2$  & $\Gamma_5$  & $2$ & $-1$ & $2$ & $-1$  & $0$ & $0$ & $2$  & $-1$ & $-1$   \\
        $E_1$  & $\Gamma_6$  & $2$ & $-1$ & $-2$ & $1$  & $0$ & $0$ & $2$  & $-1$ & $1$   \\
        $\bar{E}_3$  & $\bar{\Gamma}_7$  & $2$ & $-2$ & $0$ & $0$  & $0$ & $0$ & $-2$  & $2$ & $0$   \\
        $\bar{E}_2$  & $\bar{\Gamma}_8$  & $2$ & $1$ & $0$ & $-\sqrt{3}$  & $0$ & $0$ & $-2$  & $-1$ & $\sqrt{3}$   \\
        $\bar{E}_1$  & $\bar{\Gamma}_9$  & $2$ & $1$ & $0$ & $\sqrt{3}$  & $0$ & $0$ & $-2$  & $-1$ & $-\sqrt{3}$   \\
         \hline
         \hline
    \end{tabular}
    \caption{Character table of the double point-group $C_{6v}$~\cite{BilbaoElcoro:ks5574}.}
    \label{tab:character table C6v}
\end{table}
\clearpage
\newpage
\end{appendix}

\end{document}